\documentclass[aps, prd, reprint,10pt, notitlepage, a4paper,floats, floaqutfix,amsmath, amssymb, amsfonts,superscriptaddress,showpacs, showkeys,nofootinbib,longbibliography]{revtex4-2}

\pdfoutput=1

\usepackage{epsfig}
\usepackage{graphics}
\usepackage{graphicx}
\usepackage{amsmath,amsfonts,amssymb,amsthm}
\usepackage{amsfonts}
\usepackage[usenames,dvipsnames]{xcolor}
\usepackage{wasysym}
\usepackage{times}
\usepackage{mathptmx}
\usepackage{gensymb}
\usepackage{appendix}
\usepackage{listings}
\usepackage{url}
\usepackage[normalem]{ulem}
\usepackage{alltt}
\usepackage[colorlinks]{hyperref}
\usepackage{cleveref}
\usepackage{longtable}
\usepackage{enumitem}
\setlist{nosep}
\usepackage{color}
\usepackage{calc}
\usepackage{tensor}
\usepackage{bm}
\usepackage{times}
\usepackage{multirow}
\usepackage[varg]{txfonts}
\usepackage{float}
\usepackage{dcolumn}
\usepackage[nolist,nohyperlinks]{acronym}
\usepackage{xspace}
\usepackage[english]{babel}
\usepackage[abs]{overpic}
\usepackage{pict2e}
\usepackage[caption=false]{subfig} 
\allowdisplaybreaks[1]
\usepackage[utf8]{inputenc}
\usepackage{gensymb}
\usepackage{bm}
\usepackage{stackengine}
\usepackage{boldline,multirow}
\usepackage{braket}
\usepackage{longtable}
\usepackage{bm} 
\usepackage{makecell}

\usepackage{tabularx}
\usepackage{rotating}

\DeclareMathAlphabet{\mathcalstd}{OMS}{cmsy}{m}{n}
\DeclareMathAlphabet{\mathpzc}{OT1}{pzc}{m}{it}

\newcommand{\AEI}{\affiliation{Max Planck Institute for Gravitational Physics (Albert Einstein Institute), Am M\"uhlenberg 1, Potsdam 14476, Germany}}
\newcommand{\Maryland}{\affiliation{Department of Physics, University of Maryland, College Park, MD 20742, USA}}
\newcommand{\PI}{\affiliation{Perimeter Institute for Theoretical Physics, 31 Caroline Street North, Waterloo, ON N2L 2Y5, Canada}}
\newcommand{\Stanford}{\affiliation{Kavli Institute for Particle Astrophysics and Cosmology and Department of Physics,\\ Stanford University, 382 Via Pueblo Mall, Stanford, CA 94305, USA
}}

\definecolor{dodgerblue}{HTML}{1E90FF}
\definecolor{viennared}{HTML}{DA0A14}

\hypersetup{citecolor=dodgerblue}

\newcommand{\seobfivehm}{{\texttt{SEOBNRv5HM}}}
\newcommand{\seobfivephm}{{{\texttt{SEOBNRv5PHM}}}}
\newcommand{\seobfourphm}{{\texttt{SEOBNRv4PHM}}}

\newcommand{\xphm}{{\texttt{IMRPhenomXPHM}}}
\newcommand{\tphm}{{\texttt{IMRPhenomTPHM}}}
\newcommand{\teob}{{\texttt{TEOBResumS-GIOTTO}}}
\newcommand{\nrsur}{{\texttt{NRSur7dq4}}}

\def\mr{\mathrm}

\def\cross{\times}

\def\Lhat{\bm{l}}
\def\LNhat{\bm{l}_N}
\def\dotLNhat{\dot{\bm{l}}_N}

\def\lamNhat{\bm{\lambda}_N}
\def\xhat{\hat{\bm{e}}_{1}}
\def\yhat{\hat{\bm{e}}_{2}}
\def\zhat{\hat{\bm{e}}_{3}}

\def\J2P{J\rightarrow P}
\def\I2J{I\rightarrow J}

\acrodef{PN}{post-Newtonian}
\acrodef{EOB}{effective-one-body}
\acrodef{NR}{numerical relativity}
\acrodef{GW}{gravitational wave}
\acrodef{BBH}{binary black hole}
\acrodef{BH}{black hole}
\acrodef{BNS}{binary neutron star}
\acrodef{NSBH}{neutron star-black hole}
\acrodef{SNR}{signal-to-noise ratio}
\acrodef{aLIGO}{Advanced LIGO}
\acrodef{AdV}{Advanced Virgo}

\begin{document}



\title{\texttt{SEOBNRv5PHM}: Next generation of accurate and efficient multipolar precessing-spin effective-one-body waveforms for binary black holes}

\author{Antoni Ramos-Buades}\email{antoni.ramos.buades@aei.mpg.de}\AEI
\author{Alessandra Buonanno}\AEI\Maryland
\author{H\'ector Estell\'es}\AEI
\author{Mohammed Khalil}\PI\AEI\Maryland
\author{\\Deyan P. Mihaylov}\AEI
\author{Serguei Ossokine}\AEI
\author{Lorenzo Pompili}\AEI
\author{Mahlet Shiferaw}\AEI \Stanford

\date{\today}

\begin{abstract}
Spin precession is one of the key physical effects that could unveil the origin of the compact binaries detected by ground- and space-based gravitational-wave (GW) detectors, and shed light on their possible formation channels.
Efficiently and accurately modeling the GW signals emitted by these systems is crucial to extract their properties. Here, we present \seobfivephm, a multipolar precessing-spin waveform model within the effective-one-body (EOB) formalism for the full signal (i.e. inspiral, merger and ringdown) of binary black holes (BBHs).
In the non-precessing limit, the model reduces to \seobfivehm, which is calibrated to $442$ numerical-relativity (NR) simulations, $13$ waveforms from BH perturbation theory, and non-spinning energy flux from second-order gravitational self-force theory. We remark that \seobfivephm~is not calibrated to precessing-spin NR waveforms from the Simulating eXtreme Spacetimes Collaboration. We validate \seobfivephm~by computing the unfaithfulness against 1543 precessing-spin NR waveforms, and find that for $99.8 \%$ $(84.4\%)$ of the cases, the maximum value, in the total mass range $20-300 M_\odot$, is below $3\%$ $(1\%)$. These numbers reduce to $95.3\%$ $(60.8\%)$ when using the previous version of the \texttt{SEOBNR} family, \seobfourphm, and  to $78.2\%$ $(38.3 \%)$ when using the state-of-the-art frequency-domain multipolar precessing-spin phenomenological \xphm~model.
Due to much better computational efficiency of \seobfivephm~compared to \seobfourphm,~we are  also able to  perform extensive Bayesian parameter estimation on synthetic signals and GW events observed by LIGO-Virgo detectors. We show that \seobfivephm~can be used as a standard tool for inference analyses to extract astrophysical and cosmological information of large catalogues of BBHs.
\end{abstract}

\maketitle


\section{Introduction}\label{sec:intro}
Since the first detection of a gravitational-wave (GW) signal in 2015
\cite{LIGOScientific:2016aoc}, GW astronomy has quickly transitioned
from a dozen of events observed in the first and second observing runs
\cite{LIGOScientific:2018mvr,Venumadhav:2019lyq} of the LIGO and Virgo
GW ground-based detectors
\cite{TheLIGOScientific:2014jea,TheVirgo:2014hva} to more than a
hundred of GW event candidates in the latest observing run of the LIGO, Virgo and KAGRA detectors
\cite{LIGOScientific:2020ibl,LIGOScientific:2021usb,LIGOScientific:2021djp,Nitz:2021uxj,Olsen:2022pin,KAGRA:2018plz,KAGRA:2020tym}. With
the upcoming upgrades of the existing ground-based detectors, as well
as the planned next-generation GW detectors, such as the
ground-based Einstein Telescope~\cite{Punturo:2010zz} and Cosmic
Explorer~\cite{Reitze:2019iox,Evans:2021gyd}, or the space-based Laser Interferometer
Space Antenna (LISA) \cite{amaro2017laser}, it is expected an
increasing rate of detected mergers of compact binaries. In order to
maximize the science output of such experiments, it is essential to
accurately model the GWs emitted from binary systems.

One of the most active research areas in the field of GW source modeling
concerns with the accurate description of the two-body motion when spins are
misaligned with respect to the orbital angular momentum of the system. In this situation, both
the spins and the orbital angular momentum precess around the direction of the
total angular momentum~\cite{Apostolatos:1994mx}. In addition to spin precession, asymmetries
in the masses of the binary components excite multipoles beyond the
quadrupolar order \cite{Blanchet:2013haa} which induce a rich
structure in the GW signal, and complicate substantially its
modeling. Measurements of spin precession and higher multipoles can
provide key information about the formation channels of the observed
systems~\cite{Rodriguez:2016vmx,Stevenson:2017dlk,Talbot:2017yur,Zhu:2017znf,Kimball:2020opk,Gompertz:2021xub}
and break degeneracies among parameters
\cite{Vecchio:2003tn,Lang:2006bsg,Chatziioannou:2014coa,Graff:2015bba,Pratten:2020igi,Johnson-McDaniel:2021rvv,Krishnendu:2021cyi,Biscoveanu:2021nvg,Steinle:2022rhj},
allowing high precision GW astronomy and accurate measurements of
cosmological parameters
\cite{CalderonBustillo:2020kcg,Ezquiaga:2021ayr,LIGOScientific:2021aug}, as well as unique
tests of General Relativity (GR)
\cite{Huwyler:2011iq,LIGOScientific:2020tif,LIGOScientific:2021sio}.

Accurate models for precessing-spin binary black holes have been developed within different modeling frameworks: the phenomenological approach, the numerical relativity (NR) surrogate models and the effective-one-body (EOB) formalism.

Phenomenological
models~\cite{Pan:2007nw,Ajith:2007qp,Ajith:2009bn,Santamaria:2010yb,Hannam:2013oca,Husa:2015iqa,Khan:2015jqa,London:2017bcn,Khan:2018fmp,Khan:2019kot,Dietrich:2019kaq,Pratten:2020fqn,Pratten:2020ceb,Garcia-Quiros:2020qpx,Estelles:2020osj,Estelles:2020twz,Estelles:2021gvs,Hamilton:2021pkf}
are built upon ans\"atze based on post-Newtonian (PN) and EOB theory during
the inspiral, and functional forms of the waveform in the intermediate
and merger-ringdown parts, which are calibrated to EOB and NR
waveforms. Recently, there has been efforts to include calibration to
precessing-spin NR waveforms \cite{Hamilton:2021pkf}, and there is
ongoing work to include these improvements in the latest
frequency-domain precessing-spin \xphm~\cite{Pratten:2020ceb} model,
which we use throughout this paper. Within the IMRPhenom family we also
employ the time-domain \tphm~model
\cite{Estelles:2020osj,Estelles:2020twz,Estelles:2021gvs}, which
includes an improved description of the spin precession with respect
to the \xphm~model.

The surrogate
models~\cite{Blackman:2015pia,Blackman:2017dfb,Blackman:2017pcm,Varma:2018mmi,Varma:2019csw,Williams:2019vub,Rifat:2019ltp,Islam:2021mha,Islam:2022laz}
interpolate NR waveforms, and they have been proven the most accurate
method to produce models for higher multipoles \cite{Varma:2018mmi}
and spin-precession \cite{Blackman:2017pcm,Varma:2019csw}. However,
these models are limited to the region in parameter space where NR simulations
are available, and are restricted to the length of NR waveforms,
unless they are hybridized with EOB waveforms \cite{Varma:2018mmi,Yoo:2022erv}. In this paper, we consider the state-of-the-art
surrogate waveform model, \nrsur~\cite{Varma:2019csw}, which includes
spin-precession, all the multipoles in the co-precessing frame up to
$l=4$, mass ratios $q\in[1-4]$, dimensionless spins up to 0.8 and
binary total masses $ \gtrsim 60 M_\odot$ at a starting frequency of 20 Hz.

The EOB formalism~\cite{Buonanno:1998gg,Buonanno:2000ef,Damour:2000we,Damour:2001tu,Buonanno:2005xu}
combines information from several analytical methods, such as
post-Newtonian (PN) and small mass-ratio approximations, with results
from NR simulations.  The EOB waveform models consist of three main
building blocks: 1) the Hamiltonian, which describes the conservative
dynamics, 2) the radiation-reaction (RR) force, which accounts for the
energy and angular momentum losses due to GW emission, and 3) the
inspiral-merger-ringdown waveform modes, built upon improved PN resummations
for the inspiral part, and functional forms calibrated to NR waveforms
in the merger-ringdown.  EOB waveform models have been constructed for
quasi-circular non-spinning~\cite{Buonanno:2000ef,Damour:2000we,Buonanno:2006ui,Buonanno:2007pf,Damour:2007yf,Damour:2008gu,Buonanno:2009qa,
  Pan:2011gk,Damour:2012ky,Damour:2015isa,Nagar:2019wds} and
spinning~\cite{Damour:2001tu,Buonanno:2005xu,Damour:2007vq,
  Damour:2008qf,Pan:2009wj,Damour:2008te,Barausse:2009xi,Barausse:2011ys,Nagar:2011fx,Damour:2014sva,Balmelli:2015zsa,Khalil:2020mmr,Taracchini:2012ig,Taracchini:2013rva,Bohe:2016gbl,Cotesta:2018fcv,Pan:2013rra,Babak:2016tgq,Ossokine:2020kjp,Nagar:2018plt,Nagar:2018zoe,Akcay:2020qrj,Gamba:2021ydi}
binaries. Furthermore, orbital
eccentricity~\cite{Bini:2012ji,Hinderer:2017jcs,Chiaramello:2020ehz,Nagar:2021gss,Khalil:2021txt,Ramos-Buades:2021adz,Albanesi:2022xge}
and
matter~\cite{Bernuzzi:2014owa,Hinderer:2016eia,Steinhoff:2016rfi,Akcay:2018yyh,Steinhoff:2021dsn,Matas:2020wab,Gonzalez:2022prs}
effects, as well as information from
post-Minkowskian~\cite{Damour:2016gwp,Damour:2017zjx,Antonelli:2019ytb,Damgaard:2021rnk,Khalil:2022ylj,Damour:2022ybd}
and small mass-ratio
approximations~\cite{Damour:2009sm,Yunes:2009ef,Yunes:2010zj,Barausse:2011dq,Akcay:2012ea,Antonelli:2019fmq,Nagar:2022fep}
have been also incorporated in EOB models. To increase the
computational efficiency of the EOB waveforms, reduced-order
frequency-domain or surrogate models have been
developed~\cite{Field:2013cfa,Purrer:2014fza,Purrer:2015tud,Lackey:2016krb,Lackey:2018zvw,Cotesta:2020qhw,Gadre:2022sed,Tissino:2022thn,Khan:2020fso,Thomas:2022rmc}.

In the EOB formalism two main waveform families exist:
\texttt{SEOBNR}~\cite{Bohe:2016gbl,Cotesta:2018fcv,Ossokine:2020kjp}
and
\texttt{TEOBResumS}~\cite{Nagar:2018zoe,Nagar:2020pcj,Gamba:2021ydi}.
Within the \texttt{SEOBNR} family, here we present a new multipolar
precessing-spin waveform model, \seobfivephm\footnote{\seobfivephm~is publicly available through the python package \texttt{pySEOBNR} \href{https://git.ligo.org/waveforms/software/pyseobnr}{\texttt{git.ligo.org/waveforms/software/pyseobnr}}. Stable versions of \texttt{pySEOBNR} are published through the Python Package Index (PyPI), and can be installed via ~\texttt{pip install pyseobnr}.}, for quasi-circular
binary black holes (BBHs). Precessing-spin waveforms can be
constructed from an aligned-spin waveform in the co-precessing frame,
in which the BBH is viewed from the maximum radiation axis and the GW
signal resembles a non-precessing one, by applying a time-dependent
rotation to the inertial
frame~\cite{Apostolatos:1994mx,Buonanno:2002fy,Schmidt:2010it,Boyle:2011gg,OShaughnessy:2011pmr,Schmidt:2012rh}.
The precessing-spin \texttt{SEOBNRv3} \cite{Pan:2013rra,Babak:2016tgq}
and \seobfourphm~\cite{Ossokine:2020kjp} models employ a full EOB
precessing-spin Hamiltonian~\cite{Barausse:2009xi,Barausse:2011ys} to
evolve the dynamics in the co-precessing frame. To improve the
computational efficiency, the time-domain phenomenological
\tphm~\cite{Estelles:2020osj,Estelles:2021gvs} model builds the
precessing waveform employing a purely aligned-spin
dynamics. Similarly, the precessing-spin \texttt{TEOBResumS} model,
\texttt{TEOBResumS-GIOTTO}~\cite{Akcay:2020qrj,Gamba:2021ydi} builds
computational efficient precessing-spin waveforms evolving an
aligned-spin EOB Hamiltonian in the co-precessing frame.

To increase computational efficiency, \seobfivephm~follows a similar
approach as in Refs.~\cite{Estelles:2020osj,Akcay:2020qrj,Gamba:2021ydi}, and
decouples the evolution of the spins from the orbital dynamics by
using orbit-averaged, PN-expanded spin-precession
equations~\cite{SpinTaylorNotes,Estelles:2020twz,Akcay:2020qrj,Gamba:2021ydi}.
The latter, in \seobfivephm, includes higher PN orders and is derived from the
full-precessing spin \texttt{SEOBNRv5} Hamiltonian \cite{Balmelli:2015zsa,Khalil:2020mmr,Khalilv5}.
The \seobfivephm~model is built in the co-precessing frame upon the accurate multipolar aligned-spin
\seobfivehm~model \cite{Pompiliv5}, which is calibrated to 442 NR
simulations~\cite{Mroue:2013xna,Boyle:2019kee}, $13$ waveforms from BH
perturbation theory~\cite{Barausse:2011kb,Taracchini:2014zpa}, and nonspinning energy flux from second-order gravitational self-force theory~\cite{VandeMeentv5,Warburton:2021kwk,Wardell:2021fyy}.
The model includes the $(l,m)=\{(2,\pm 2),(2,\pm 1),(3,\pm 3),(3, \pm 2),(4,\pm 4),(4,\pm 3),(5,\pm 5)\}$
multipoles. We remark that the \seobfivephm~model is not calibrated to
precessing-spin NR simulations.

The standard way of validating waveform models is by comparing them with numerical solutions of the Einstein equations, i.e., NR waveforms. However, the high computational cost of producing NR simulations poses a challenge to finely populate the large dimensionality of the parameter space of quasi-circular precessing-spin BBHs (mass ratio and the six spin degrees of freedom). As a consequence, NR simulations of BBHs have been largely limited to mass ratios $q\leq 4 $ and dimensionless spins up to $0.8$, and length of 15-20 orbital cycles before merger~\cite{Gonzalez:2008bi,Buchman:2012dw,Chu:2009md,Mroue:2013xna,Jani:2016wkt,Foucart:2018lhe,Boyle:2019kee,Healy:2017psd,Hinder:2018fsy,Healy:2019jyf,Ossokine:2020kjp,Healy:2022wdn}. Here, we validate the new EOB precessing-spin waveform model, by comparing it to 1425 simulations from the public Simulating eXtreme Spacetimes (SXS) catalogue \cite{Boyle:2019kee}, as well as 118 NR simulations presented in Ref.~\cite{Ossokine:2020kjp}. When compared to NR simulations we find that \seobfivephm~provides  $99.8\%$ of cases with a maximum unfaithfulness, in total mass range $[20-300]M_\odot$, below $3\%$, while this number reduces to $95.3 \%$
for the previous generation of precessing-spin \texttt{SEOBNR} models, the \seobfourphm~model \cite{Ossokine:2020kjp}.

For the inspiral orbital dynamics \seobfivephm~uses the
post-adiabatic (PA) approximation~\cite{Nagar:2018gnk,Rettegno:2019tzh, Gamba:2021ydi,Mihaylov:2021bpf}.
This strategy for the evolution equations, combined with a new high-performance Python infrastructure \texttt{pySEOBNR} \cite{Mihaylovv5},
improves significantly the computational efficiency of the
\seobfivephm~model, and makes it comparable to the state-of-the-art
time-domain precessing-spin waveform models. The model is generally
$\sim 8-20$ times faster than \seobfourphm, which has been proven to
accurately describe quasi-circular precessing-spin binaries, and it
has been extensively employed to extract source properties of detected
GW signals \cite{LIGOScientific:2021usb,LIGOScientific:2021djp}. However, its high
computational cost requires the use of non-standard stochastic
sampling techniques for Bayesian inference studies, such as
\texttt{RIFT} \cite{Pankow:2015cra,Lange:2018pyp}, or machine learning
techniques such as \texttt{DINGO}
\cite{Green:2020dnx,Dax:2021tsq,Dax:2022pxd}. Here, we show that the
\seobfivephm~model can be employed with standard stochastic sampling
techniques due to its high computational efficiency. We perform
Bayesian inference studies with the \seobfivephm~model by
injecting synthetic NR signals into detector noise, and by reanalysing
GW events from previous observing runs. We find that the
\seobfivephm~model recovers accurately the injected synthetic NR
signals, as well a providing more constrained posterior distributions
in the analyzed GW events than the \seobfourphm~model.

This work is part of a series of articles
\cite{Khalilv5,VandeMeentv5,Pompiliv5,Mihaylovv5} describing the
\texttt{SEOBNRv5} family of models, and it is organized as follows. In
Secs. \ref{sec:EOBexpressions} and \ref{sec:EOBwaveforms} we develop
the multipolar EOB waveform model for precessing-spin BBHs,
\seobfivephm, and highlight improvements and differences with respect
to the previous generation of precessing-spin \texttt{SEOBNR} models. In
Sec. \ref{sec:WaveformValidation} we validate the accuracy of the
\seobfivephm~by comparing it to NR waveforms. We also compare the
performance of \seobfivephm~against other state-of-the-art
quasi-circular precessing-spin waveform models, notably \texttt{IMRPhenomXPHM}
and \texttt{TEOBResumS-GIOTTO}, and investigate in
which region of parameter space these models differ more from NR
waveforms and from each other. In Sec.~\ref{sec:PE}, we study the accuracy
of the precessing model using Bayesian inference analysis by injecting
synthetic NR waveforms in zero detector noise, and also by analysing
GW events detected in the latest observing runs of the LVK Collaboration. In Sec. \ref{sec:conclusions}, we summarize our
main conclusions and discuss future work. Finally, in Appendix
\ref{app:Ham} we provide the explicit expression of the Hamiltonian
used in the \seobfivephm~model~\cite{Khalilv5}, and in Appendix \ref{app:PA} we
specify the equations used to apply the PA approximation
in the \seobfivephm~model. In Appendix \ref{app:PhenomModels} we
compare the model with the state-of-the-art time-domain phenomenological model \tphm.

\section*{Notation}\label{sec:notation}
In this paper, we use geometric units, setting $G=c=1$ unless otherwise specified.

We consider a binary with masses $m_1$ and $m_2$, with $m_1 \geq m_2$, and spins $\bm{S}_1$ and $\bm{S}_2$. We define the following combinations of the masses:
\begin{equation}
\begin{gathered}
M \equiv m_1 + m_2, \quad \mu \equiv \frac{m_1m_2}{M}, \quad \nu \equiv \frac{\mu}{M},  \\
\delta \equiv\frac{m_1 - m_2}{M},  \quad q \equiv \frac{m_1}{m_2},
\end{gathered}
\end{equation}
where $\mr i = 1,2$. A relevant combination of masses for GW data analysis is the \textit{chirp mass} defined as \cite{Sathyaprakash:2009xs}
\begin{equation}
\mathcal{M} = \nu^{3/5} M.
\label{eq:chirpMass}
\end{equation}
We define the dimensionless spin vectors
\begin{gather}
\bm{\chi}_{\mr i} \equiv \frac{\bm{a}_{\mr i}}{m_{\mr i}} = \frac{\bm{S}_{\mr i}}{m_{\mr i}^2},
\end{gather}
along with the intermediate definition for $\bm{a}_{\mr i}$.
We also define the following combinations of the spins:
\begin{equation}
\bm{a}_\pm \equiv \bm{a}_1 \pm \bm{a}_2.
\end{equation}
The relative position and momentum vectors, in the binary's center-of-mass, are denoted $\bm{r}$ and $\bm{p}$, with
\begin{equation}
\bm p^2 = p_r^2 + \frac{L^2}{r^2}, \quad
p_r= \bm{n}\cdot\bm{p}, \quad
\bm{L}=\bm{r}\cross\bm{p},
\end{equation}
where $\bm{n}=\bm{r}/r$ and $\bm{L}$ is the orbital angular momentum with magnitude $L$. The direction of $\bm{L}$ is denoted as $\Lhat$. The total angular momentum is given by $\bm{J} = \bm{L} + \bm{S}_1 + \bm{S}_2$.
We express the precessing binary dynamics in an orthonormal frame $\{\LNhat,\bm{n},\lamNhat\}$, where $\LNhat$ is the direction of $\bm{L}_{\rm N}\equiv \mu \bm{r}\cross \dot{\bm{r}}$, and $\lamNhat \equiv \LNhat \cross \bm{n}$.
It is convenient to define the effective spin parameter $\chi_{\rm eff}$ \cite{Damour:2001tu,Racine:2008qv,Santamaria:2010yb},
\begin{equation}
\chi_{\rm eff} = \frac{1}{M}(\bm{a}_1 + \bm{a}_2)\cdot \LNhat,
\label{eq:chi_eff}
\end{equation}
and the effective precessing-spin parameter $\chi_p$ \cite{Schmidt:2014iyl},
\begin{equation}
\chi_{p} = \frac{1}{B_1 m_1^2}\max \left( B_1 m_1^2 \chi_{1,\perp}, B_2 m_2^2 \chi_{2,\perp}) \right),
\label{eq:chi_p}
\end{equation}
where $B_1 = 2+3m_2/(2 m_1)$, $B_2 = 2+3m_1/(2 m_2)$ and $\chi_{i,\perp}$ indicates the magnitude of the projection of the dimensionless spin vectors on the orbital plane.

\section{Effective-one-body dynamics of precessing-spin binary black holes}\label{sec:EOBexpressions}

For the two-body conservative dynamics, the EOB formalism relies on a Hamiltonian $H_{\rm EOB}$, constructed through an effective Hamiltonian $H_{\rm eff}$ of a test mass $\mu$ moving in a deformed Kerr spacetime of mass $M$ (the deformation parameter being $\nu$), and the following energy map connecting $H_{\rm eff}$ and $H_{\rm EOB}$
\begin{equation}
H_{\rm EOB} = M \sqrt{1+2 \nu \left(\frac{H_{\rm eff}}{\mu}-1 \right)}\,.
\label{eq:eq01}
\end{equation}
The deformation of the Kerr Hamiltonian is obtained by imposing that at each PN order, the PN-expanded EOB Hamiltonian agrees with a PN Hamiltonian through a canonical transformation.
In Ref.~\cite{Khalilv5}, an EOB Hamiltonian that includes all generic-spin information up to 4PN has been derived, while the non-spinning dynamics is incorporated up to 4PN with partial 5PN results. The dynamical variables of the generic EOB Hamiltonian are the orbital separation $\bm r$, the corresponding canonically conjugate momentum $\bm p$, and the spins $\bm{S}_{1,2}$.

For arbitrary orientations of the spins, both the orbital plane and the spins precess around the total angular momentum of the system $\bm{J}$. The equations of motion are as follows~\cite{Buonanno:2005xu}
\begin{equation}
\label{eq:fullSpin}
\begin{gathered}
\dot{\bm{r}} = \frac{\partial H^{\rm prec}_{\rm EOB}}{\partial \bm{p}}, \qquad
\dot{\bm{p}} = -\frac{\partial H^{\rm prec}_{\rm EOB}}{\partial \bm{r}} + \bm{\mathcal{F}}, \\
\dot{\bm{S}}_{1,2} = \frac{\partial H^{\rm prec}_{\rm EOB}}{\partial \bm{S}_{1,2}} \times \bm{S}_{1,2},
\end{gathered}
\end{equation}
where for \seobfivephm~the full precessing-spin Hamiltonian, $H^{\rm prec}_{\rm EOB}$, is given in Sec. II. D of Ref.~\cite{Khalilv5}, and it reduces as $\nu\to 0$ to the Kerr Hamiltonian for a test mass in a generic orbit.
Within the EOB formalism, the dissipative effects enter the dynamics through the RR force $\bm{\mathcal{F}}$, which is expressed in terms of the waveform modes \cite{Damour:2007xr,Damour:2008gu}.

It was shown in Refs.~\cite{Buonanno:2002fy,Schmidt:2010it,Boyle:2011gg,OShaughnessy:2011pmr,Schmidt:2012rh}
  that precessing-spin waveforms can be built starting from
  aligned-spin waveforms in the so-called co-precessing frame, in which the
  $z$-axis remains perpendicular to the instantaneous orbital plane,
  and then applying a suitable rotation to the inertial frame. The
  precessing-spin \texttt{SEOBNRv3} and \texttt{SEOBNRv4} models
  employed the full EOB precessing-spin
  Hamiltonian~\cite{Barausse:2009xi,Barausse:2011ys} to evolve the
  dynamics in the co-precessing frame. However, solving the EOB
  dynamics for generic spin configurations can be computationally
  expensive, as the EOB evolution equations \eqref{eq:fullSpin} lead to
  lengthy expressions~\cite{Knowles:2018hqq}. To build the
  precessing-spin \texttt{TEOBResumS} model and speed-up the
  computational time, Refs.~\cite{Akcay:2020qrj,Gamba:2021ydi} used an
  aligned-spin EOB Hamiltonian when evolving the equations in the
  co-precessing frame. Also, the \texttt{IMRPhenomT}
  model~\cite{Estelles:2020twz} was built using a purely aligned-spin
  dynamics in the co-precessing frame.

To build the computationally efficient precessing-spin dynamics
of \seobfivephm,  Ref.~\cite{Khalilv5} has leveraged the recent studies of Ref.~\cite{Estelles:2020twz,
Akcay:2020qrj,Gamba:2021ydi}, making some important modifications and improvements. In particular,
to enhance the accuracy in describing precessional effects, Ref.~\cite{Khalilv5} has found it
  important to incorporate at least partial precessing-spin
  information in the Hamiltonian used in the co-precessing frame. To
  achieve that, it has first obtained a precessing-spin
  Hamiltonian simpler than the full one, such that it reduces to the
  aligned-spin Hamiltonian in absence of spin precession, but only includes the in-plane spin components for
  circular orbits ($p_r=0$). Then, it has orbit averaged
  the in-plane spin components in the Hamiltonian, and used them
  when evolving the equations of motion involving the dynamical variables $r,p_r,\phi$ and
  $p_\phi$ in the co-precessing frame. Furthermore,  the evolution equations for the spin and angular
  momentum vectors are computed in a PN-expanded, orbit-averaged form
  for quasi-circular orbits, similarly to what was done in
  Refs.~\cite{SpinTaylorNotes,Estelles:2020twz,Akcay:2020qrj,Gamba:2021ydi},
  but, as we discuss below, Ref.~\cite{Khalilv5}, has included higher PN orders in the spin-spin sector, and has derived them from
  the \texttt{SEOBNRv5} EOB Hamiltonian, employing a different gauge
  and spin-supplementary condition with respect to Refs.~\cite{Estelles:2020twz,
Akcay:2020qrj,Gamba:2021ydi}.

Thus, in the \texttt{SEOBNRv5PHM} model, the equations of motion in the co-precessing frame read:
\begin{equation}
\label{eq:EOBEOMs}
\begin{aligned}
\dot{r}&=\xi(r) \frac{\partial H_{\rm EOB}^\text{pprec}}{\partial p_{r_*}}, \quad
&\dot{\phi} &=\frac{\partial H_{\rm EOB}^\text{pprec}}{\partial p_{\phi}},\\
\dot{p}_{r_*}&=-\xi(r)\frac{\partial H_{\rm EOB}^\text{pprec}}{\partial r} +\mathcal{F}_{r}, \quad
&\dot{p}_{\phi}&=\mathcal{F}_{\phi},
\end{aligned}
\end{equation}
where, as said, the Hamiltonian $H_{\rm EOB}^\text{pprec}$ reduces in the aligned-spin limit to the Hamiltonian used in  \seobfivehm~\cite{Pompiliv5}, while also including \emph{partial precessional} (pprec) effects.
Notably, the Hamiltonian incorporates orbit-averaged in-plane spin terms for circular orbits ($p_r=0$), while neglecting fourth order spin contributions (see Appendix~\ref{app:Ham} for the explicit expression of $H_{\rm EOB}^\text{pprec}$ and other details).

As in previous EOB models \cite{Taracchini:2013rva,Pan:2013rra,Cotesta:2018fcv,Ossokine:2020kjp}, the evolution of the radial momentum is performed using the tortoise-coordinate $p_{r_*} = p_r\xi(r)$, where $\xi(r) = dr/dr_*$.
The RR force is computed using~\cite{Buonanno:2005xu}
\begin{equation}
\mathcal{F_\phi}= -\frac{\Phi_E}{\Omega}, \qquad \mathcal{F}_r = \mathcal{F}_\phi \frac{p_{r}}{p_\phi},
\label{eq:RRforceAS}
\end{equation}
where $\Omega \equiv \dot{\phi}$ is the orbital frequency, and $\Phi_E$ is the energy flux for quasi-circular orbits, which can be written as~\cite{Damour:2007xr,Damour:2008gu}.
\begin{equation}
\Phi_E = \frac{\Omega^2}{16 \pi} \sum_{l=2}^8 \sum_{m=-l}^l m^2 |d_L h_{lm}|^2,
\label{eq:fluxAS}
\end{equation}
where $d_L$ is the luminosity distance from the binary to the observer, and $h_{lm}$ are the waveform modes.

In addition to the equations of motion~\eqref{eq:EOBEOMs}, the PN-expanded evolution equations for the spins and angular momentum, read:
\begin{subequations}
\label{eq:SLeqns}
\begin{align}
\dot{\bm S}_{\mr i} &= \bm{\Omega}_{S_{\mr i}}\times \bm{S}_{\mr i}, \label{eq:sdot}\\
\bm{L} &= \bm{L}(\LNhat,v,\bm{S}_{\mr i}),\\
\dotLNhat &= \dotLNhat(\LNhat,v,\bm{S}_{\mr i}), \label{eq:LNdot}
\end{align}
\end{subequations}
where $\bm{\Omega}_{S_{\mr i}}$ is the spin-precession frequency, $v \equiv (M \Omega_\text{PN})^{1/3}$ with $\Omega_\text{PN}$ being the PN-expanded orbital frequency (see below), and $\LNhat$ is the unit vector in the direction of $\bm{L}_N$. As said, these PN-expanded equations have been obtained in Ref.~\cite{Khalilv5} (consistently, from the {\tt SEOBNRv5} Hamiltonian and equations of motion) for precessing spins through an orbit-average procedure up to 4PN order, including spin-orbit (SO) contributions to next-to-next-to-leading order (NNLO), and spin-spin (SS) contributions to NNLO.
The spin-precession frequency is given by Eq.~(66) of Ref.~\cite{Khalilv5}, while $\bm{L}$ and $\dotLNhat$ are given there in Eqs.~(65) and (71).

We note that the SO and LO SS parts of the spin-precession frequency $\bm{\Omega}_{S_{\mr i}}$ agree with the orbit-averaged results given by Eqs. (1)-(5) of Refs.~\cite{Akcay:2020qrj,SpinTaylorNotes}, but the NLO and NNLO SS terms do not agree with Refs.~\cite{Bohe:2015ana,SpinTaylorNotes} because of the different gauge used for the {\tt SEOBNRv5} Hamiltonian.
Furthermore, our expressions for $\bm{L}(\LNhat,\Omega_{\rm PN},\bm{S}_{\mr i})$, and hence for $\dotLNhat$, differ at SO level from Ref.~\cite{Akcay:2020qrj} because of the different spin-supplementary condition used.

In practice, to solve the equations of motion, we first perform the PN-expanded evolution of the spin and angular momentum vectors using Eqs. \eqref{eq:SLeqns}, then we apply a subsequent EOB evolution using Eqs. \eqref{eq:EOBEOMs}, where the projections of the spins $\bm{S}_{1,2}$ onto $\LNhat$ and $\bm{L}(\LNhat)$ are updated at every timestep~\cite{Gamba:2021ydi}.
The solution of the PN-expanded equations~\eqref{eq:SLeqns} requires a prescription for the evolution of the orbital frequency, which we compute as follows
\begin{equation}
\dot{v} = \left [\frac{\dot{E}(v)}{dE(v)/dv} \right ]_{\rm PN-expanded},
\label{eq:vdot}
\end{equation}
where $E(v)$ is the binding energy of the binary, and  $\dot{E}(v)$ the circular-orbit PN-expanded energy flux.

The expression for $\dot{v}$ is given by Eq.~(69) of Ref.~\cite{Khalilv5}, which used the results of Ref.~\cite{Cho:2022syn} to obtain the NNLO SS contribution to the orbit-averaged energy flux. Our result for $\dot{v}$ agrees at the NNLO SO and LO SS with Eq.~(A1) of Ref.~\cite{Chatziioannou:2013dza}, but differs from it by including the NLO and NNLO SS contributions. Also, our PN-expanded equations are fully expanded in $v$.

The \seobfivephm~model employs the partial precessional Hamiltonian, $H^{\rm pprec}_{\rm EOB}$,
which reduces to the non-precessing \seobfivehm~ Hamiltonian in the
aligned-spin limit. This Hamiltonian contains parameters which feature
higher (yet unknown) PN orders and are calibrated to aligned-spin NR waveforms. These
calibration parameters are denoted by $a_6(\nu)$ and $d_{\rm SO}(\nu,
a_\pm)$ in Ref.~\cite{Pompiliv5}. From these two parameters only
$d_{\rm SO}$ contains a spin dependence, and thus, it is the only
calibration parameter affected by the variation of the spins with
time. In the \seobfivephm~we employ the projections of spins onto
$\LNhat$ to evaluate $d_{\rm SO}(\nu,\bm{a}_\pm \cdot \LNhat)$ at
every timestep of the evolution. The other calibration parameter
inherited from the underlying \seobfivehm~model is $\Delta
t^{22}_{\text{ISCO}}(\nu,a_\pm)$, which is a parameter determining the
time shift between the innermost stable circular orbit (ISCO) of the remnant Kerr BH,
and the time of the peak of the (2,2)-mode amplitude (see Sec. IV of
Ref.~\cite{Pompiliv5} for details). Here, we employ the projections of
the spins onto the Newtonian angular momentum evaluated at the time
the orbital separation $r$ crosses the ISCO\footnote{More specifically, the ISCO time is computed from the ISCO orbital
  separation $r_{\text{ISCO}} (\nu, a_\pm )$, which in the
  precessing-spin case depends on the values of the spins projected
  onto $\LNhat$ at a particular instant of time, which we decide to be
  $r=10M$, $r_{\rm ISCO}(\nu,\bm{a}_\pm \cdot \LNhat)|_{r=10M}$, for
  the reasons discussed in Sec.~\ref{sec:EOBwaveforms}.} to evaluate
the NR calibrated time shift, i.e., $\Delta
t^{22}_{\text{ISCO}}(\nu,\bm{a}_\pm \cdot \LNhat
)|_{t_{\text{ISCO}}}$.

Equations~\eqref{eq:EOBEOMs} have the same form of the evolution equations
in the aligned-spin \seobfivehm~model. This fact permits the use of
the PA approximation \cite{Nagar:2018gnk,Mihaylov:2021bpf}
in the precessing-spin \seobfivephm~model, as done in the underlying
aligned-spin \seobfivehm~ model \cite{Pompiliv5}. The use of the
PA approximation to evolve the EOB inspiral implies an
increase in speed and efficiency of the model as discussed in
Sec.~\ref{sec:benchmarks}, while the specific details of its
implementation are described in Appendix \ref{app:PA}. Furthermore,
the orbital frequency as computed in Eq.~\eqref{eq:vdot} allows an
adiabatic evolution, which permits to disentangle the starting
frequency of the EOB evolution with the reference frequency at which
the spins are specified, which introduces a novel feature in the
SEOBNR models\footnote{In the previous \seobfourphm~model, where
  Eqs.~\eqref{eq:fullSpin} are solved, the starting frequency and the
  reference frequency correspond to the same frequency. The
  specification of a reference frequency distinct from the starting
  frequency implies a backwards in time integration, which due to the
  RR force in the EOB dynamics would cause an increase
  of eccentricity in \seobfourphm, and thus it breaks the assumption of
  modeling quasi-circular binaries.} and highly benefits Bayesian
inference studies as shown in Sec.~\ref{sec:PE}.

In summary, our strategy to produce precessing-spin EOB waveforms shares common
aspects with the work developed in Refs.~\cite{Estelles:2020osj,
  Akcay:2020qrj, Gamba:2021ydi}, but it goes beyond them in several
aspects which we highlight again in the following. First, the
precessing-spin evolution equations, Eqs. \eqref{eq:SLeqns}, which are
implemented in \seobfivephm~and derived in Ref. \cite{Khalilv5},
include higher PN orders and are consistently derived from the generic
\texttt{SEOBNRv5} Hamiltonian. Then, the EOB dynamics is also improved
by including in the {\tt SEOBNRv5HM} Hamiltonian of
Ref.~\cite{Khalilv5,Pompiliv5} terms describing in-plane spin effects
and vanishing in the non-precessing limit. Moreover, all the spin
components entering into the Hamiltonian are used in the orbital
evolution (see Appendix \ref{app:Ham} for more details), instead of
just the projection onto $\LNhat$ as in
Refs. \cite{Akcay:2020qrj,Gamba:2021ydi}.

\section{Effective-one-body multipolar waveforms for precessing-spin binary black holes}\label{sec:EOBwaveforms}

In this section we describe the main building blocks to generate precessing-spin multipolar waveforms in the \seobfivephm~model.

\subsection{Inspiral-plunge waveforms}
\label{sec:InspiralWaveforms}

The construction of the inspiral-plunge waveforms follows a similar approach to
Ref.~\cite{Ossokine:2020kjp}, with the usage of the factorized, resummed version
\cite{Damour:2007xr,Pan:2010hz} of the frequency domain PN formulas of the modes
\cite{Arun:2008kb,Mishra:2016whh}. The factorized resummation has been developed
for non-precessing BBHs \cite{Damour:2008gu,Pan:2010hz,Cotesta:2018fcv,Khalilv5}
and it has been proven to improve the accuracy of the PN expressions in the
test-particle limit~\cite{Taracchini:2013wfa,Barausse:2011kb,Bernuzzi:2011aj,Harms:2015ixa}.

\begin{figure}[tbp!]
	\includegraphics[scale=0.2]{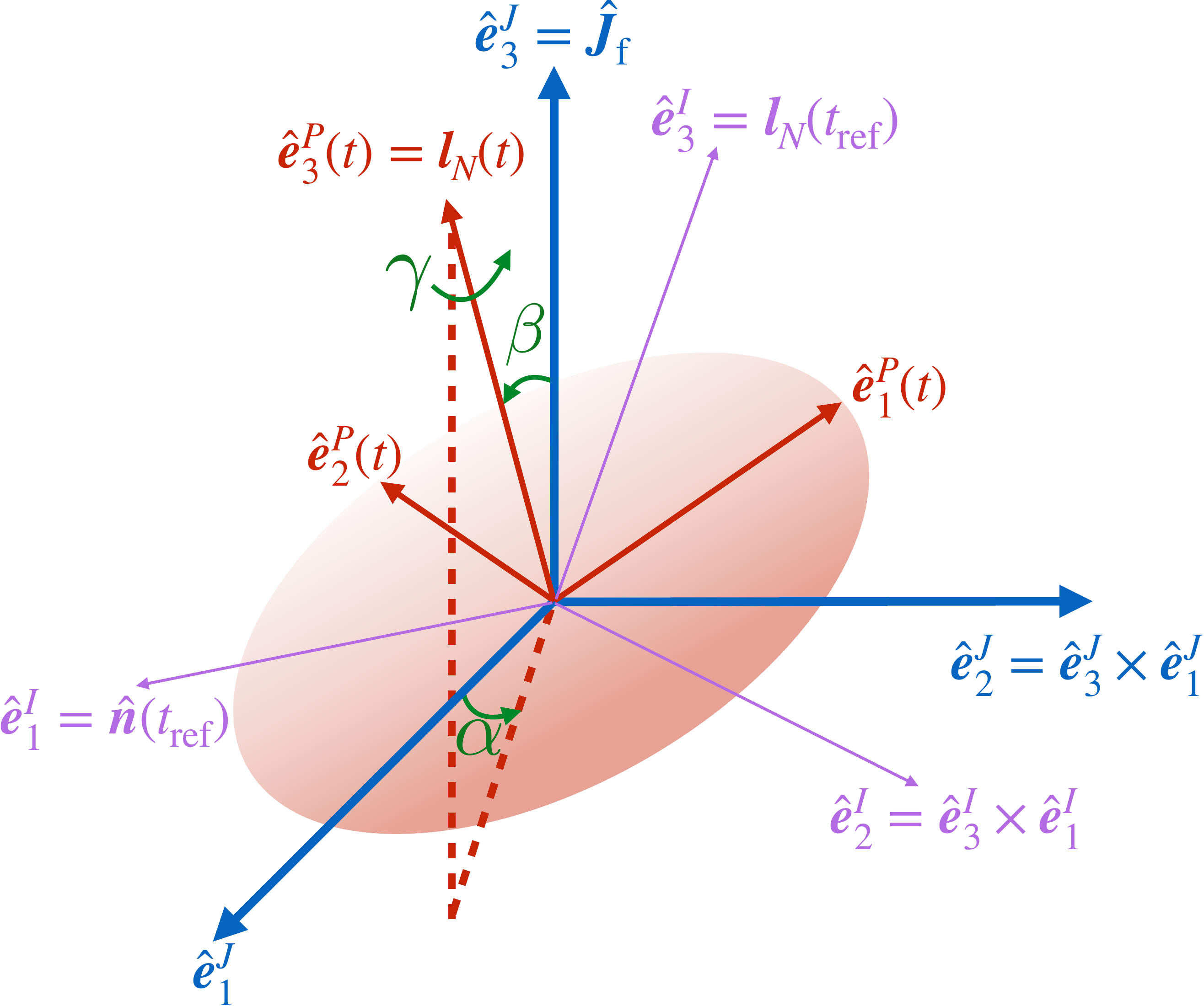}
	\caption{ Frames used in the construction of the \seobfivephm~model.
The co-precessing frame (red) is constructed such that its $z$-axis is instantaneously aligned with the Newtonian
angular momentum $\LNhat(t)$ and can be described by the Euler angles
$(\alpha,\beta,\gamma)$ with respect to $\bm{J}_{\rm f}$-frame (blue), while the
source frame (purple) corresponds to the inertial frame defined by the initial
Newtonian angular momentum $\LNhat(t_{\rm ref})$ and unit separation vector
$\hat{\bm{n}}(t_{\rm ref})$. For the \seobfivephm~model we adopt the convention that at $t_{\rm ref}$, the source and co-precessing
frames coincide. }
	\label{fig:frames}
\end{figure}

The components of the  RR force,  $\mathcal{F}_{r,\phi}$,  in Eq.
\eqref{eq:RRforceAS} depend on the amplitude of the individual GW modes
$|h_{lm}|$. In  \seobfivephm, the spins entering the GW modes (and  energy
flux) are projected onto the Newtonian orbital angular momentum,  $\bm{a}_{\pm}\cdot \LNhat$, since $\LNhat$
represents the direction perpendicular to the orbital plane (see Fig.
\ref{fig:frames}) and is provided by the PN-expanded EOB precessing-spin evolution equations\footnote{We note that in the \seobfourphm~model the spins were projected using $\Lhat$.}.

The GW polarizations in the inertial frame of the observer are required for data-analysis studies. As in Ref.~\cite{Ossokine:2020kjp}, the \seobfivephm~model
also defines three reference frames: 1)  the inertial frame of the observer (\textit{source
frame}) (whose quantities are indicated with a superscript $I$),  2) an inertial frame where
the $z$-axis is aligned with the final angular momentum of the system\footnote{This is computed as
the value of the solution of Eqs. \eqref{eq:SLeqns}  at the attachment point of
the merger-ringdown model.} (\textit{$\boldsymbol{J}_{\rm f}$-frame}), which helps with the construction of the merger ringdown,  (whose quantities are denoted with the
superscript $J$),  and finally 3) a non-inertial frame which tracks the
instantaneous motion of the orbital plane,  the \textit{co-precessing frame}
(whose quantities are denoted by the superscript P). The frames are depicted in Fig.~\ref{fig:frames} and described below\footnote{In Fig. \ref{fig:frames} we show the definition of the Euler angles between the co-precessing frame and the $\boldsymbol{J}_{\rm f}$-frame as a frame rotation. For time evolutions of the Euler angles, which qualitatively resemble the ones in \seobfivephm, we refer the reader to Refs. \cite{Buonanno:2002fy,Schmidt:2010it,Schmidt:2012rh,Boyle:2011gg,Babak:2016tgq,Estelles:2020osj,Akcay:2020qrj,Gamba:2021ydi}.}. 

The source frame is defined at a given reference frequency $f_{\rm ref}$
(corresponding to a reference time $t_{\rm ref}$) by the triad $\{\hat{\bm
e}^I_i\}$ $(i=1,2,3)$, where $\hat{\bm e}^I_1 = \bm{n}(t_{\rm ref})$,
$\hat{\bm e}^I_3 = \LNhat(t_{\rm ref})$, $\hat{\bm e}^I_2 = \hat{\bm e}^I_3
\times \hat{\bm e}^I_1$.  Meanwhile,  the $\bm{\hat{J}}_{\rm f}$-frame is constructed as
$\zhat^{J}=\hat{\bm{J}}_{\rm f}$,  $ \xhat^{J}=N[\xhat^{I}-(\xhat^{I}\cdot\zhat^{J})\zhat^{J}]$,  $\yhat^{J}=\zhat^{J}\times\xhat^{J}$ where the $N[]$ denotes normalization. The two frames are connected by a constant rotation given by:
\begin{equation}
\textbf{R}^{I \rightarrow J} =  \begin{pmatrix}
  \xhat^J\cdot\xhat^{I}  & \yhat^J\cdot\xhat^{I}   & \zhat^J\cdot\xhat^{I}\\[0.5em]
  \xhat^J\cdot\yhat^{I}  & \yhat^J\cdot\yhat^{I}   & \zhat^J\cdot\yhat^{I}\\[0.5em]
  \xhat^J\cdot\zhat^{I}  & \yhat^J\cdot\zhat^{I}   & \zhat^J\cdot\yhat^{I}\\
   \end{pmatrix}.
\label{eq:JframeMatrix}
\end{equation}
The rotation operation in Eq. \eqref{eq:JframeMatrix} can be also expressed as a
unit quaternion $q_{I \rightarrow J}$\footnote{To perform such a conversion, as
well as subsequent manipulations of quaternions (e.g., the enforcement of the
minimal rotation condition), we work with the \texttt{quaternion} Python package
\cite{mike_boyle_2022_6499564}.}.

Finally,  to construct the inertial GW modes $h^I_{lm}$ during the inspiral-plunge,  we
introduce the \textit{co-precessing frame},  which is defined by the triad
$\{\hat{\bm e}^P_i\}$ ($i=1,2,3$).  At every instant the $z$-axis of the
co-precessing frame is aligned with $\LNhat$ (i.e., $\hat{\bm e}^P_3(t)=\LNhat(t)$)
\footnote{Note that in Ref.~\cite{Ossokine:2020kjp}, the $z$-axis is aligned with
$\Lhat$ instead of $\LNhat$.}. In this frame,  the GW radiation resembles the
radiation from an aligned-spin binary
\cite{Buonanno:2002fy,Schmidt:2010it,Boyle:2011gg,OShaughnessy:2011pmr,Schmidt:2012rh}.
The other two axes lie in the orbital plane and are defined such that they
minimize precessional effects in the modes $h^P_{lm}$.  This is done by enforcing
the minimal rotation condition that relates the rotation from the $\bm{J}_{\rm
f}$-frame to the co-precessing frame~\cite{Boyle:2011gg}. This transformation is best parametrized by a unit quaternion that aligns the $z$-axis of the $\boldsymbol{J}_{\rm f}$-frame with $\LNhat$
\begin{equation}
q_{\J2P}(t) = \sqrt{-\LNhat(t)\zhat^{J}},
\end{equation}
and the minimal rotation condition is then simply $(\dot{q}\zhat^{J}\bar{q})_{0}=0$, where $(q)_{0}$ denotes taking the scalar part of the quaternion~\cite{Boyle:2011gg}, $\dot{q}$ denotes time derivative, and $\bar{q}$ denotes the conjugate of the quaternion (which is also its inverse). The minimal rotation condition has a residual freedom which corresponds to the integration constant~\cite{Boyle:2011gg}. We fix this freedom by demanding that at the reference time, the co-precessing frame and source frame coincide.

We calculate the co-precessing frame inspiral-plunge GW waveform modes by
evaluating the factorized,  resummed non-precessing modes along the EOB dynamics
described in Eqs. \eqref{eq:EOBEOMs}, with time-dependent projections of the
spins $\left\{\bm{a}_{\pm}\cdot \LNhat, \bm{a}_{\pm}\cdot\Lhat, \bm{a}_+\cdot \bm{a}_-, \bm{a}^2_{\pm}\right\}$. Following Ref.~\cite{Pompiliv5},
in which an EOB non-precessing multipolar waveform (\seobfivehm) calibrated
to NR non-precessing simulations was developed, we include in the co-precessing
frame of the \seobfivephm~model the $\{(2,\pm 2),(2,\pm 1),(3,\pm
3),(3,\pm 2),(4,\pm 4),(4,\pm 3), (5,\pm 5)\}$ modes, and make the assumption
$h^P_{l,-m}= (-1)^l h^{P *}_{l,m}$. As discussed in Sec. IIIB of
Ref.~\cite{Ossokine:2020kjp}, the inaccuracies due to neglecting mode asymmetries
should remain modest, and are expected to be at most comparable to other modeling
errors.

To assemble the inertial-frame modes, we first rotate $h_{\ell m}^{P}$ to the ${ \bm J }_{\rm f}$-frame using $\bar{q}_{\J2P}$,  and then from the ${\bm J }_{\rm f}$-frame to the source frame using $\bar{q}_{I \rightarrow J}$\footnote{We perform these rotations using  the
\texttt{scri}\cite{Boyle:2013nka,Boyle:2014ioa,Boyle:2015nqa} Python package.}.
To make contact with literature, it is useful to express these rotations in terms of Euler angles. Using the active ZYZ convention (see Fig.~\ref{fig:frames}),  the $J\rightarrow P$ rotation is given by
\begin{equation}
\label{eq:euler_convention}
q_{J\rightarrow P} = e^{\alpha \hat{\bm{z}}/2}e^{\beta \hat{\bm{y}}/2}e^{\gamma \hat{\bm{z}}/2}.
\end{equation}
In this formulation, the minimal rotation condition is given by $\dot{\gamma}=-\dot{\alpha}\cos\beta$ \cite{Boyle:2011gg}.

\subsection{Merger-ringdown waveforms}
\label{sec:MergerWaveforms}
After the coalescence, the description of a BBH system of two individual objects is no longer valid, and the EOB model builds the ringdown stage via a phenomenological model of the quasinormal modes (QNMs) of the remnant BHs, formed after the merger of the progenitors. The QNMs frequencies are tabulated functions of the final mass, $M_f$, and angular momentum $\bm{S}_f=M_f^2\bm{\chi}_f$  of the remnant BH \cite{Berti:2005ys}. The QNMs are defined with respect to the direction of the final spin, and thus, the description of the ringdown signal as a linear combination of QNMs, is formally valid only in an inertial frame with the z-axis parallel to $\bm{\chi}_f$.

Following Ref.~\cite{Ossokine:2020kjp}, in \seobfivephm~the attachment of the merger-rindown waveform is performed in the co-precessing frame. Therefore, we employ the merger-ringdown multipolar model developed for non-precessing BBHs (\seobfivehm) in  Ref.~\cite{Pompiliv5}.

The calculation of the waveform in the inertial observer's frame requires a description of the co-precessing frame Euler angles $\{\alpha(t),\beta(t),\gamma(t) \}$ which extends beyond merger. Here, we take advantage of a phenomenological prescription based on insights from NR simulations~\cite{OShaughnessy:2012iol}. More specifically, it was shown that the co-precessing frame continues to precess roughly around the direction of the final angular momentum with a precession frequency, $\omega_{\rm prec}$, proportional to the difference between the lowest overtone of the $(2,2)$ and $(2,1)$ QNM frequencies,  while the opening angle of the precessing cone,  $\beta$, tends to decrease at merger. This phenomenology translates into the following expressions for the merger-ringdown angles in \seobfivephm,
\begin{align}
\alpha^{\rm merger-RD} &= \alpha(t_{\rm match}) + \omega_{\rm prec}(t-t_{\rm match}),\label{eq:alphaRD}\\
\beta^{\rm merger-RD} &= \beta(t_{\rm match}),\label{eq:betaRD}\\
\gamma^{\rm merger-RD} &= \gamma(t_{\rm match}) - \omega_{\rm prec}(t-t_{\rm match}) \cos \beta(t_{\rm match}),
\label{eq:gammaRD}
\end{align}
where $t_{\rm match}$ is the time of attachment of the merger-ringdown model. We have also investigated non-constant post-merger extensions of the $\beta$ angle, such as the small opening angle approximation (see Eq. (24b) of Ref.~\cite{Estelles:2021gvs}), but we find that such an approximation may degrade the faithfulness of the model to NR for certain configurations.

The behavior noticed in Ref.~\cite{OShaughnessy:2012iol}, describes prograde configurations, were the remnant spin is positively aligned with the orbital angular momentum at merger. However, to keep the model generic and accurate in a wide parameter space of mass ratios and spins, we extend the prescription to the retrograde case (negative alignment of the final spin with respect to the angular momentum at merger), which is typical for high mass-ratio binaries, when the total angular momentum $\bm J$ is dominated by the primary spin $\bm{S}_1$ instead of $\bm{L}$. While we keep imposing simple precession around the final spin at a rate $\omega_{\rm prec} \geq 0$ in our model, we distinguish two cases depending on the direction of the total angular momentum at merger $\bm{\chi}_f \sim \bm{J}_{\rm f}$ with respect to the final orbital angular momentum $\bm{L}_f$,
\begin{equation}\label{eq:postmergerprec}
  \omega_{\rm prec}= \left\{\begin{aligned}
   &\omega^{\rm QNM}_{22}(\chi_{f})-\omega^{\rm QNM}_{21}(\chi_{f}) \quad \text{if} \quad \bm{\chi}_{f} \cdot \bm{L}_{f} > 0\\
   &\omega^{\rm QNM}_{2-1}(\chi_{f})-\omega^{\rm QNM}_{2-2}(\chi_{f}) \quad \text{if} \quad \bm{\chi}_{f} \cdot \bm{L}_{f} < 0
\end{aligned}\right.,
\end{equation}
where $\chi_f=|\bm{\chi}_f|$, and the QNM frequencies for negative $m$ are taken from the continous extension of the $m>0$, $\omega^{\rm QNM}_{lm}>0$ branch \cite{Berti:2005ys}. We stress that this prescription of the post-merger extension of the Euler angles for the retrograde case is much less tested than the prograde case due to the lack of NR simulations covering this region of parameter space, which also includes particular systems with transitional precession \cite{Apostolatos:1994mx}.

Following recent insights from NR of Ref.~\cite{Hamilton:2023znn}, where a correct prescription of the shift of the co-precessing QNM frequencies was  developed, we compute in the \seobfivephm~model the co-precessing frame QNM frequencies from the QNM frequencies in the $\bm{J}_{\rm f}$-frame as,
\begin{equation}
  \omega^{\rm QNM, P}_{l m } = \omega^{\rm QNM, J}_{l m}-m (1- |\cos \beta(t_{\rm match})|) \omega_{\rm prec}.
\label{eq:QuasinormalModes}
\end{equation}

Another essential aspect in the construction of the merger-rindown waveforms is the mapping from binary component masses and spins to the final mass and spin, required to evaluate the QNM frequencies of the remnant. Several groups have developed fitting formulas based on large sets of NR simulations (see Ref.~\cite{Varma:2018aht} for a brief overview of the literature). To ensure agreement in the non-precessing limit with \seobfivehm~\cite{Pompiliv5}, we employ the fits for the final mass from Ref.~\cite{Jimenez-Forteza:2016oae}, and the fits from Ref.~\cite{Hofmann:2016yih} for the final spin.

The application of the fitting formulae for the final mass and spin requires choosing a time during the inspiral at which to evaluate the spins, as for precessing binaries the individual components of the spins vary with time. In the \seobfivephm~model, we choose to evaluate the spins at a time corresponding to an orbital separation $r=10M$. Similarly as in Ref.~\cite{Ossokine:2020kjp}, this choice is based on good agreement with NR configurations, and by the restriction that the smallest initial orbital separation must be $r>10.5M$ to ensure small initial eccentricities \cite{Babak:2016tgq}. Additionally, this choice guarantees that a given physical configuration always produces the same waveform regardless of the initial starting frequency, as all configurations will pass through an orbital separation $r=10M$.

Finally, the inspiral-merger-ringdown GW modes in the inertial frame $h^I_{lm}$  are obtained by rotating the inspiral-merger-ringdown modes $h_{lm}^P$ from the co-precessing frame to the ${\bm J }_{\rm f}$-frame, and then from  ${\bm J }_{\rm f}$-frame to the inertial observer's frame using the expressions for the rotations in Appendix A of Ref.~\cite{Babak:2016tgq}. The inertial frame GW polarizations at a time $t$, and location in the sky of the observer $(\varphi_0,\iota)$ can be expressed in terms of the $-2$-spin-weighted spherical harmonics, as follows
\begin{equation}
h^{\rm I}_{+}(t; \bm{\lambda},\varphi_0,\iota) - i h^{\rm I}_{\times}(t; \bm{\lambda},\varphi_0,\iota) = \sum_{\ell, m} {}_{-2} Y_{\ell m}(\varphi_0,\iota)\,h^{\rm I}_{\ell m}(t; \bm{\lambda}) \,,
\end{equation}
where $\bm{\lambda}$ represents the set of intrinsic parameters (masses and spins), and $\{\varphi_0,\iota\}$ the coalescence phase and the inclination angle of the signal.

\subsection{Efficient calculation of the GW polarizations}\label{sec:polarizations}

For applications in which only the GW polarizations are required, as for most of the current parameter-estimation codes, we introduce an alternative and computationally more efficient method to obtain the polarizations directly in terms of the co-precessing $-2$-spin-weighted spherical harmonic modes.  This involves  rotating the spin-weighted spherical harmonic basis, instead of computing the full set of spin-weighted spherical harmonic modes in the inertial frame.

The inertial-frame (\textit{I-frame}) modes are related to the co-precessing-frame (\textit{P-frame}) modes by a time-dependent rotation from the co-precessing frame to the frame where the $z$-axis is aligned with the final angular momentum of the system (\textit{$\bm{J}_{\rm f}$-frame}~\footnote{The $\bm{J}_{\rm f}$-frame is the frame where the approximation of the Euler angles in Eqs. \eqref{eq:alphaRD},~\eqref{eq:betaRD} and~\eqref{eq:gammaRD} is applied.}), and a time-independent rotation from the  $\bm{J}_{\rm f}$-frame to the final inertial frame
\begin{equation}
h^{\rm I}_{\ell m}(t)=\sum_{m', m''} \Big(\textbf{R}^{\rm J \rightarrow I}\Big)_{m,m'}\Big(\textbf{R}^{\rm P \rightarrow J}\Big)_{m',m''}h^{\rm P}_{\ell m''}(t),
\end{equation}
where $\textbf{R}^{\rm X \rightarrow Y}$ indicates the rotation operator from the frame $X$ to the frame $Y$, and the indices $m', m''$ denote summation indices over the modes available in the co-precessing frame.

Factoring out the source orientation information from the spin-weighted spherical harmonic basis as a rotation of the basis
\begin{equation}
{}_{-2} Y_{\ell m}(\varphi_0,\iota)=\sum_{m'}\Big(\textbf{R}^{\varphi_0,\iota}\Big)_{m,m'}{}_{-2} Y_{\ell m}(0,0),
\end{equation}
the complete rotation of the basis functions from the co-precessing frame to the final inertial frame can be constructed composing the individual rotations as
\begin{equation}
\textbf{R}^{\rm P \rightarrow I}=\textbf{R}^{\varphi_0,\iota}   \textbf{R}^{\rm J \rightarrow I}   \textbf{R}^{\rm P \rightarrow J},
\end{equation}
with associated Euler angles $\{\alpha_{\rm P \rightarrow I}, \beta_{\rm P \rightarrow I}, \gamma_{\rm P \rightarrow I}\}$. Applying this rotation operator, the spin-weighted spherical harmonic basis can be written as
\begin{equation}
\sum_{m'}\Big(\textbf{R}^{\rm P \rightarrow I}\Big)_{m,m'}{}_{-2} Y_{\ell m}(0,0)=e^{2i\alpha_{\rm P \rightarrow I}}{}_{-2} Y_{\ell m}(\gamma_{\rm P \rightarrow I},\beta_{\rm P \rightarrow I}),
\end{equation}
and the GW polarizations in the inertial frame can therefore be expressed as
\begin{equation}
\label{eq:direct_hpc}
h^{\rm I}_{+}(\varphi_0,\iota;t) - i h^{\rm I}_{\times}(\varphi_0,\iota;t) =e^{2i\alpha_{\rm P \rightarrow I}}\sum_{\ell, m} {}_{-2} Y_{\ell m}(\gamma_{\rm P \rightarrow I},\beta_{\rm P \rightarrow I})\,h^{\rm P}_{\ell m}(t).
\end{equation}
Eq. \eqref{eq:direct_hpc} is only summed over the set of 7 co-precessing modes\footnote{The negative m-modes in the co-precessing frame are obtained by the symmetry relation  $h^P_{l,-m}= (-1)^l h^{P *}_{l,m}$.}, and the computation of the complete rotation and its application to the basis functions\footnote{In this method we have 14 basis functions corresponding to the positive and negative \textit{m}-modes.} is more efficient than the corresponding (double) rotation of the GW modes, which requires  the rotation of 33 GW modes.

\section{Performance of the multipolar precessing-spin effective-one-body waveform model}\label{sec:WaveformValidation}

In this section we assess the accuracy of the multipolar
precessing-spin \seobfivephm~model by comparing it, as well as other
models, to NR simulations of quasi-circular precessing-spin
BBHs. Particularly, we consider state-of-the-art precessing-spin
models within the EOB framework, such as
\seobfourphm~\cite{Ossokine:2020kjp} and the public version of
\teob\footnote{In this paper we employ the \teob~model from the public
  bitbucket repository \url{https://bitbucket.org/eob_ihes/teobresums}
  with the git hash \texttt{fc4595df72b2eff4b36e563f607eab5374e695fe},
  which is the latest release at the time of this publication.}
\cite{Gamba:2021ydi}, and within the phenomenological approach, the
frequency-domain \xphm~\cite{Pratten:2020ceb} (and in Appendix \ref{app:PhenomModels} the
time-domain \tphm~\cite{Estelles:2021gvs} model). All the previous
models, including \seobfivephm, are not calibrated to precessing-spin
NR waveforms. We also investigate the validity and systematics of
models by comparing them against the surrogate
\nrsur~\cite{Varma:2019csw} model. Finally, we assess the
computational efficiency of the \seobfivephm~model to be used for data analysis.

\subsection{Brief overview of the faithfulness function}
\label{sec:NRunfaithfulness}
The GW signal emitted by a quasi-circular precessing-spin BBH system depends on 15 parameters: the component masses, $m_{1,2}$ (or equivalently mass ratio $q$ and total mass $M$), the dimensionless spin vectors  $\bm{\chi}_{1,2}(t)$, the direction of the observer from the source can be described by the angles $(\varphi_0, \iota)$, the luminosity distance $d_L$, polarization angle $\psi$, the location in the sky of the detector $(\theta, \phi)$, and the time of arrival $t_c$. The strain in the detector caused by a passing GW can be  expressed as
\begin{align}
\label{eq:det_strain}
h(t) \equiv & F_+(\theta,\phi,\psi) \ h_+(t;\iota,\varphi_0, d_L, \bm{\lambda},t_{\mathrm{c}}) \nonumber \\
&+ F_\times(\theta,\phi,\psi)\ h_\times(t;\iota,\varphi_0, d_L, \bm{\lambda},t_\mathrm{c})\,,
\end{align}
where $\boldsymbol{\lambda}=\{q,M,\bm{\chi}_{1,2}(t)\}$ is introduced to simplify the notation, and $F_{+,\times}$ are the antenna pattern functions \cite{Sathyaprakash:1991mt,Finn:1992xs}. The strain in Eq. \eqref{eq:det_strain} can be expressed in terms of an effective polarization angle $\kappa(\theta,\phi,\psi)$ as
\begin{align}
\label{eq:strainKappa}
h(t) =\mathcal{A}(\theta, \phi) (h_+ \cos \kappa + h_\times \sin \kappa),
\end{align}
where the dependences of $\kappa$, $h_+$ and $h_\times$ have been removed to ease notation, and the definition of the coefficient $\mathcal{A}(\theta,\phi)$ can be found in Refs.~\cite{Cotesta:2018fcv, Ossokine:2020kjp}. As discussed, the GW polarizations can be decomposed in the basis of $-2$-spin weighted spherical harmonics as
\begin{equation}
h_+-i h_\times = \sum_{l=2}^\infty \sum_{m=-l}^{m=+l} {}_{-2}Y_{lm}(\varphi_0, \iota)h_{lm}(t; \bm{\lambda}),
\label{eq:polarizations}
\end{equation}
 where $h_{lm}(t; \bm{\lambda})$ are the GW multipolar modes.

We introduce the inner product between two waveforms, $h_1$ and $h_2$ \cite{Sathyaprakash:1991mt,Finn:1992xs}
\begin{equation}
\langle h_1, h_2 \rangle \equiv 4\ \textrm{Re}\int_{f_{\rm in}}^{f_{\rm max}} df\,\frac{\tilde{h}_1(f) \ \tilde{h}_2^*(f)}{S_n(f)},
\end{equation}
 where a tilde indicates Fourier transform, a star complex conjugation and $S_n(f)$   the power spectral density (PSD) of the detector noise. In this work, we employ for the PSD the LIGO’s “zero-detuned high-power” design sensitivity curve~\cite{Barsotti:2018}. When both waveforms are in band we use $f_{\rm in} = 10 {\rm Hz}$  and $f_{\rm max} = 2048 {\rm Hz}$. For cases where this is
not the case (e.g., the NR waveforms are used), we employ $f_{\rm in}=1.35 f_{\rm peak}$, where $f_{\rm peak}$ corresponds to the peak amplitude of the frequency-domain strain of the signal, and the factor $1.35$ accounts for possible artifacts coming from the Fourier transform of the time domain waveforms\footnote{The factor $1.35$ has been chosen after experimenting with several values, and finding that this value provides more stable mismatch results and less Fourier transform artifacts. Note that this choice is the same as in Ref. \cite{Hinder:2017sxy}. We also tested applying the same procedures to set $f_{\rm in}$ as in Refs. \cite{Ossokine:2020kjp,Pratten:2020ceb}, and obtained qualitatively similar results.}.

To assess the agreement between two waveforms --- for instance, the signal, $h_s$, and the template, $h_t$, observed by a detector, we define the faithfulness function \cite{Cotesta:2018fcv,Ossokine:2020kjp},
\begin{equation}
\label{eq:eq16}
\mathcal{F}(M_{\textrm{s}},\iota_{\textrm{s}},{\varphi_0}_{\textrm{s}},\kappa_{\textrm{s}}) =  \max_{t_c, {\varphi_0}_{t}, \kappa_{t}} \left[\left . \frac{ \langle h_s|h_t \rangle}{\sqrt{  \langle h_s|h_s \rangle  \langle h_t|h_t \rangle}}\right \vert_{\substack{\iota_{\mathrm{s}} = \iota_{t} \\\boldsymbol{\lambda}_\mathrm{s}(t_{\mathrm{s}} = t_{0_\mathrm{s}}) = \boldsymbol{\lambda}_{t}(t_t = t_{0_\mathrm{t}})}} \right ].
\end{equation}
When comparing waveforms with higher-order multipoles ~\cite{Cotesta:2018fcv,Ossokine:2020kjp,Garcia-Quiros:2020qpx} a typical choice in Eq. \eqref{eq:eq16} is to set the inclination angle of the template and the signal to be the same, while the coalescence time, azimuthal and effective polarization angles of the template, $(t_{0_t},\varphi_{0_t}, \kappa_t)$, are adjusted to maximize the faithfulness of the template.  The maximizations over the coalescence time $t_c$ and coalescence phase ${\varphi_0}_{t}$ are performed numerically, while the optimization over the effective polarization angle $\kappa_{t}$ is done analytically as described in Ref.~\cite{Capano:2013raa}.

In Eq. \eqref{eq:eq16} the condition $\boldsymbol{\lambda}_{\rm s} (t_{\rm s} = t_{0_{\rm s }}) = \boldsymbol{\lambda}_{\rm t}(t_{\rm t} = t_{0_{\rm t}}) $ enforces that the intrinsic properties (mass ratio $q$, total mass $M$, and spins $\bm{\chi}_{1,2}$) of the template waveform at $t=t_{0}$ (typically the start of the waveform) are the same as those of the signal waveform at its $t_0$. However, such identification of the same $t_0$ is not trivially satisfied between different waveforms, including NR and waveform models. As a consequence, several approaches can be applied to mitigate such a choice. For instance, in Ref.~\cite{Ossokine:2020kjp} $t_{0_{\rm t}}$ is chosen such that the time elapsed from $t_{0_{\rm s}}$ and $t_{0_{\rm t}}$ to the peak of the frame-invariant amplitude $\sum_{l,m} |h_{lm}|^2$ occurs at the same time for NR and \seobfourphm, while in Refs.~\cite{Khan:2019kot,Ossokine:2020kjp} numerical optimizations over the reference frequency of the waveform were performed for waveforms of the \texttt{IMRPhenom} family. Here, we instead optimize numerically over a rigid rotation $\delta\in[0,2\pi]$ of the in-plane spin components of the template $\{\chi^{\rm t}_{i,x},\chi^{\rm t}_{i,y}\}$ with $i=1,2$, at the reference frequency \cite{Pratten:2020ceb,Gamba:2021gap}, such that
\begin{equation}
\begin{split}
\chi^{\rm t}_{i,x} &= \chi^{\rm s}_{i,x}\cos(\delta) - \chi^{\rm s}_{i,y} \sin(\delta),\\
\chi^{\rm t}_{i,y} &= \chi^{\rm s}_{i,x}\sin(\delta) + \chi^{\rm s}_{i,y} \cos(\delta), \quad i=1,2,\\
\end{split}
\label{eq:rigidRot}
\end{equation}
where $\{\chi^{\rm s}_{i,x},\chi^{\rm s}_{i,y}\}$ denote the in-plane spin components of the signal.
This method, contrary to the procedure of optimizing over the reference frequency of the template, has unambiguous bounds for the parameters involved.

It is convenient to introduce the \textit{sky-and-polarization averaged faithfulness} to reduce the dimensionality of the faithfulness function and
express it in a more compact form \cite{Cotesta:2018fcv,Ossokine:2020kjp},
\begin{equation}
\label{eq:eq17}
\overline{ \mathcal{F}}(M_{\textrm{s}}) =  \frac{1}{8 \pi^2}\int^{1}_{-1} d{(\cos \iota_{\textrm{s}})} \int^{2 \pi}_0 d{\varphi_0}_{\textrm{s}} \int^{2 \pi}_0 d \kappa_{\textrm{s}} \mathcal{F}(M_{\textrm{s}},\iota_{\textrm{s}},{\varphi_0}_{\textrm{s}},\kappa_{\textrm{s}}).
\end{equation}
Another useful metric to assess the closeness between waveforms is the
signal-to-noise (SNR)-weighted faithfulness \cite{Ossokine:2020kjp}
 \begin{widetext}
\begin{equation}
\overline{\mathcal{F}}_{\mathrm{SNR}}(M_\mathrm{s}) = \sqrt[3]{\frac{\int^{1}_{-1} d{(\cos \iota_{\textrm{s}})} \int_{0}^{2\pi} d\kappa_ {\mathrm{s}} \int_{0}^{2\pi} d{\varphi_0}_{\mathrm{s}} \ \mathcal{F}^{3}(M_{\textrm{s}},\iota_{\textrm{s}},{\varphi_0}_{\textrm{s}},\kappa_{\textrm{s}}) \ \mathrm{SNR}^3(\iota_{\textrm{s}},{\varphi_0}_{\textrm{s}},\kappa_{\textrm{s}})}{\int^{1}_{-1} d{(\cos \iota_{\textrm{s}})} \int_{0}^{2\pi} d\kappa_{\mathrm{s}} \int_{0}^{2\pi} d{\varphi_0}_{\mathrm{s}} \ \mathrm{SNR}^3(\iota_{\textrm{s}},{\varphi_0}_{\textrm{s}},\kappa_{\textrm{s}})}},
\label{eq:eq18}
\end{equation}
\end{widetext}
where the SNR is defined as
\begin{equation}
\mathrm{SNR}(\iota_{\textrm{s}},{\varphi_0}_{\textrm{s}},\theta_\textrm{s}, \phi_\textrm{s}, \kappa_{\textrm{s}},{d_{\mathrm{L}}}_{\mathrm{s}},\boldsymbol{\lambda}_\mathrm{s},{t_c}_\mathrm{s}) \equiv \sqrt{\langle h_{\mathrm{s}},h_{\mathrm{s}}\rangle}.
\label{eq:eq19}
\end{equation}
In Eq. \eqref{eq:eq18} the weighting by the SNR takes into account the
dependence on the phase and effective polarization of the signal at a fixed
distance. Finally, we introduce the unfaithfulness or mismatch as
\begin{equation}
\overline{\mathcal{M}}_{\rm SNR}=1-\overline{\mathcal{F}}_{\rm SNR}.
\label{eq:eq20}
\end{equation}

\subsection{Assessment in modeling spin effects in EOB Hamiltonian}

In Secs. \ref{sec:EOBexpressions} and \ref{sec:EOBwaveforms} we have described the construction of the \seobfivephm~model, here we assess the impact of several approximations in the description of the precessing-spin dynamics as well as in the waveform multipoles. Differently from the \seobfourphm~model, in \seobfivephm~the full precessing-spin Hamiltonian and spin equations are not evolved. By contrast, we build on recent waveform models, \tphm~\cite{Estelles:2021gvs} and \teob~\cite{Gamba:2021ydi}, which couple a purely aligned-spin dynamics (only $\bm{a}_\pm \cdot \LNhat$) with PN-expanded equations for the spins, angular-momentum and frequency. However, in the new \seobfivephm~model there are significant differences with respect to previous approaches:
\begin{itemize}
\item The spin, velocity and angular momentum equations in \seobfivephm~are fully PN-expanded in the velocity parameter $v$, and include SO and SS couplings through NNLO, thus differ from the ones employed in Refs. \cite{SpinTaylorNotes,Estelles:2021gvs,Gamba:2021ydi}.
\item The SO contributions to the angular momentum equations in \seobfivephm~are consistent with the fully generic canonical Hamiltonian $H^{\rm prec}_{\rm EOB}$ \cite{Khalilv5} (i.e., they use the same spin-supplementary condition, and thus differ from the ones in Refs. \cite{SpinTaylorNotes,Estelles:2021gvs,Gamba:2021ydi}).
\item In \seobfivephm, the orbital equations of motion are evolved using a partial precessing-spin EOB Hamiltonian, $H^{\rm pprec}_{\rm EOB}(\Lhat^2,\bm{a}_\pm\cdot \LNhat,\bm{a}_\pm\cdot \Lhat, \bm{a}_+\cdot \bm{a}_-)$, which has all spin components (also orbit-averaged in-plane spin components instead of only $\bm{a}_\pm \cdot \LNhat$).
\end{itemize}

\begin{figure*}
	\includegraphics[width=\columnwidth]{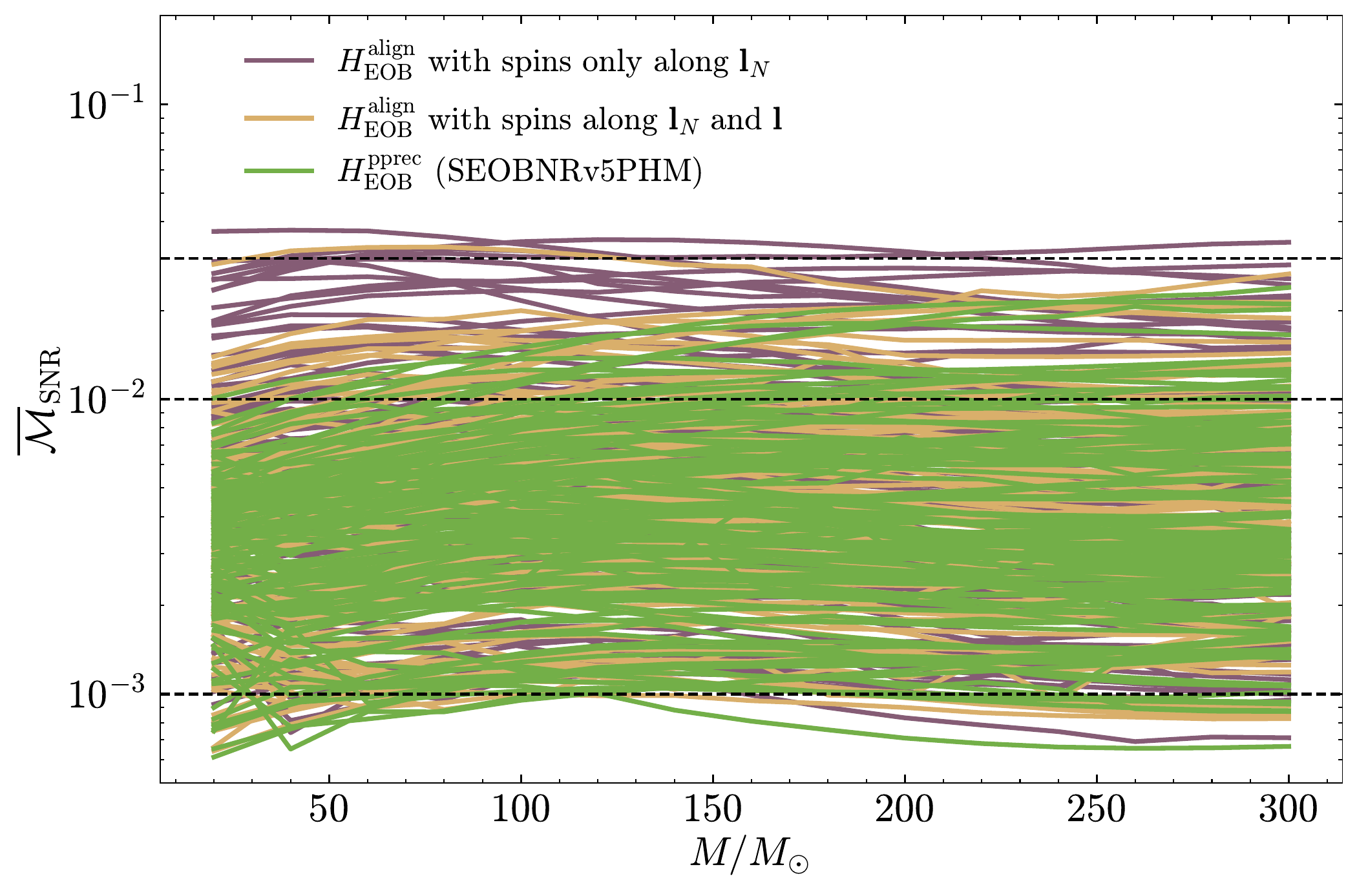}
	\includegraphics[width=\columnwidth]{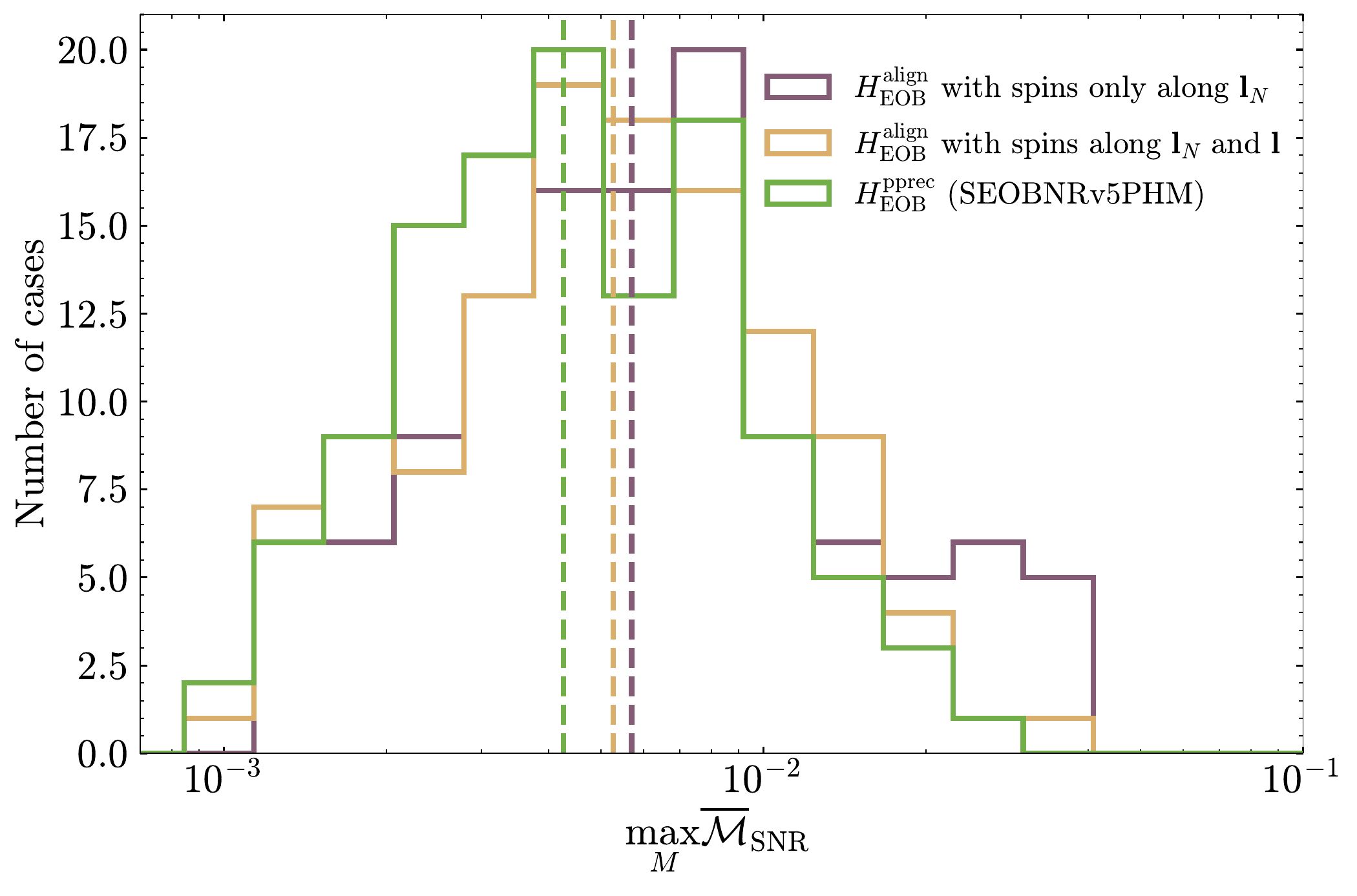}
	\caption{\textit{Left panel:} Sky-and-polarization averaged, SNR-weighted unfaithfulness in the total mass range between $[20-300] M_\odot$ for an inclination $\iota=\pi/3$, of \seobfivephm~with different prescriptions for the dynamics against the 118 highly precessing-spin NR simulations from Ref.~\cite{Ossokine:2020kjp}. The different prescriptions for the dynamics correspond to using the {\tt SEOBNRv5HM} Hamiltonian, $H^{\rm align}_{\rm EOB}$, with the spins projected onto $\LNhat$ (purple),  using $H^{\rm align}_{\rm EOB}$  with the spins projected onto $\LNhat$ and $\Lhat$ (yellow), and using the partially precessing Hamiltonian $H^{\rm pprec}_{\rm EOB}$ of \seobfivephm~(green), with the spins projected onto $\LNhat$ and $\Lhat$, see the main text for details. The dashed horizontal vertical lines correspond to the $10^{-3}$, $0.01$ and $0.03$ unfaithfulness values. \textit{Right panel:} Distribution of the maximum unfaithfulness over the total mass range for each NR simulation considered in the left plot. The vertical dashed lines indicate the median values of the distribution.}
\label{fig:unf_testHams}
\end{figure*}

In Figure \ref{fig:unf_testHams} we assess the impact of these
improvements in the treatment of the precessing-spin dynamics by
computing the unfaithfulness of \seobfivephm~with different
prescriptions for the conservative dynamics against the set of 118
highly precessing\footnote{Note that highly precessing configurations typically are quoted in the literature as binaries with high mass ratios and/or high values of the in-plane spin components. Here, we refer to these simulations as highly precessing in comparison to the ones available in the SXS catalog, as the set of 118 simulations are longer, more accurate and have larger in-plane spin than most of the simulations in the catalog.} BBH simulations from Ref.~\cite{Ossokine:2020kjp}.

The different prescriptions for \seobfivephm~correspond to 1) using the
aligned-spin Hamiltonian $H^{\rm align}_{\rm EOB}$ of \texttt{SEOBNRv5HM}~\cite{Khalilv5,Pompiliv5} with the
spins only projected onto $(\bm{a}_\pm \cdot \LNhat)$, such that the spin variables are computed like $a^2_\pm =(\bm{a}_\pm \cdot \LNhat)^2$ (i.e., a purely aligned-spin dynamics as in the \teob~\cite{Gamba:2021ydi} and \tphm~\cite{Estelles:2021gvs} models),  2) employing $H^{\rm align}_{\rm EOB}$ with a spin treatment consisting in using the full spin components for the scalar products (i.e. $a^2_\pm = (\bm{a}_\pm\cdot \bm{a}_\pm)^2$), as well as the spins projected onto $\Lhat$ in the spin-orbit sector, and onto $\LNhat$ in the rest of the spin sector, and 3) using the partially precessing Hamiltonian $H^{\rm pprec}_{\rm EOB}$ of \seobfivephm~with the latter treatment of the spins projections (see Appendix~\ref{app:Ham}).
 In the left panel of
Fig. \ref{fig:unf_testHams} we show the unfaithfulness as a function
of the total mass of the binary, while in the right panel the
distributions of the maximum unfaithfulness in the total-mass range
are displayed. The results show that using the aligned-spin
Hamiltonian with the projections of the spins onto $\LNhat$ (i.e., a
purely aligned-spin dynamics as in \texttt{TEOBResumS} and \texttt{IMRPhenomT}),
leads to $95.8 \%$ $(75.4\%)$ of cases with a maximum unfaithfulness over the total mass range considered of
$[20,300]M_\odot$, lower than $3\%$ ($1\%$),
while considering projections onto $\LNhat$, $\Lhat$ and the full spin-components entering
the aligned-spin Hamiltonian improves the previous numbers to $99.2
\%$ $(80.5\%)$, and it reduces significantly the tail of cases with
unfaithfulness larger than $3\%$. Finally, keeping the latter
treatment of the spin projections and using the partially precessing
Hamiltonian, $H^{\rm pprec}_{\rm EOB}$, which includes in-plane spin
effects in an orbit-average approximation for quasi-circular orbits
(see Appendix \ref{app:Ham} for details), leads to a further increase
in accuracy with $100 \%$ $(86.4\%)$ of cases with a maximum
unfaithfulness below $3\%$ ($1\%$). As a consequence, the latter
Hamiltonian and treatment of spin effects is the one that we adopt in the \seobfivephm~model.


\subsection{Comparison against numerical-relativity waveforms}
\label{sec:NRquasicircular}

The accuracy of the \seobfivephm~model is assessed by comparing it
to the publicly available simulations of the SXS catalogue~\cite{Boyle:2019kee}, as well as
the 118 highly precessing-spin simulations produced in
Ref.~\cite{Ossokine:2020kjp}. We also perform such a comparison for
other state-of-the-art precessing-spin EOB waveform models,
\seobfourphm~and \teob, as well as the phenomenological
frequency-domain \xphm~model. (To ease the
comparisons we compare against phenomenological \tphm~ model
in Appendix \ref{app:PhenomModels}). In
Fig. \ref{fig:paramsSpaceNR} we provide an overview of the NR
simulations employed to assess the accuracy of the different
models. The precessing-spin simulations considered here\footnote{In
  the extra material, we provide the SXS IDs of the precessing-spin NR
  simulations employed in this section.} were produced with the
\texttt{SpEC} code \cite{SpECwebsite}, and they correspond to the 118
SXS runs from Ref.~\cite{Ossokine:2020kjp}, and 1425 simulations
available in the public SXS catalog~\cite{Boyle:2019kee}.

We start by comparing the unfaithfulness\footnote{We always refer to the sky-and-polarization averaged, SNR-weighted unfaithfulness, $\overline{\mathcal{M}}_{\rm SNR}$, as unfaithfulness to ease the notation.} of the precessing-spin models  against the set of 118 highly precessing-spin simulations including all the modes up to $l=5$ in the NR waveforms.
The waveform modes included in the co-precessing frame for the different models is done consistently with Ref.~\cite{Pompiliv5} for the non-spinning approximants, and they are specifically
$(\ell, |m|)=\{(2,2),(2,1),(3,3),(4,4),(5,5)\}$ for \seobfourphm,
$(\ell, |m|)=\{(2,2),(2,1),(3,3),(3,2),(4,4),(4,3),(5,5)\}$ for \seobfivephm,
$(\ell, |m|)=\{(2,2),(2,1),(3,3),(3,2),(4,4)\}$ for \xphm~and
$(\ell, |m|)=\{(2,2),(2,1)(3,3)$ $(3,2),(3,1)$,$(4,4),(4,3),(4,2)\}$ for \teob\footnote{We note that  \teob~\cite{Gamba:2021ydi} models contains also the $(5,5)$-mode in the co-precessing frame, but in order to be consistent with Ref.~\cite{Pompiliv5} (see the reasons for its exclusion in Sec. V therein) we do not include such multipole. Additionally, we have tested that the unfaithfulness results for \teob~against NR when including and excluding the $(5,5)$-mode  are very similar.}.

\begin{figure}[!] 
	\includegraphics[scale=0.22]{ 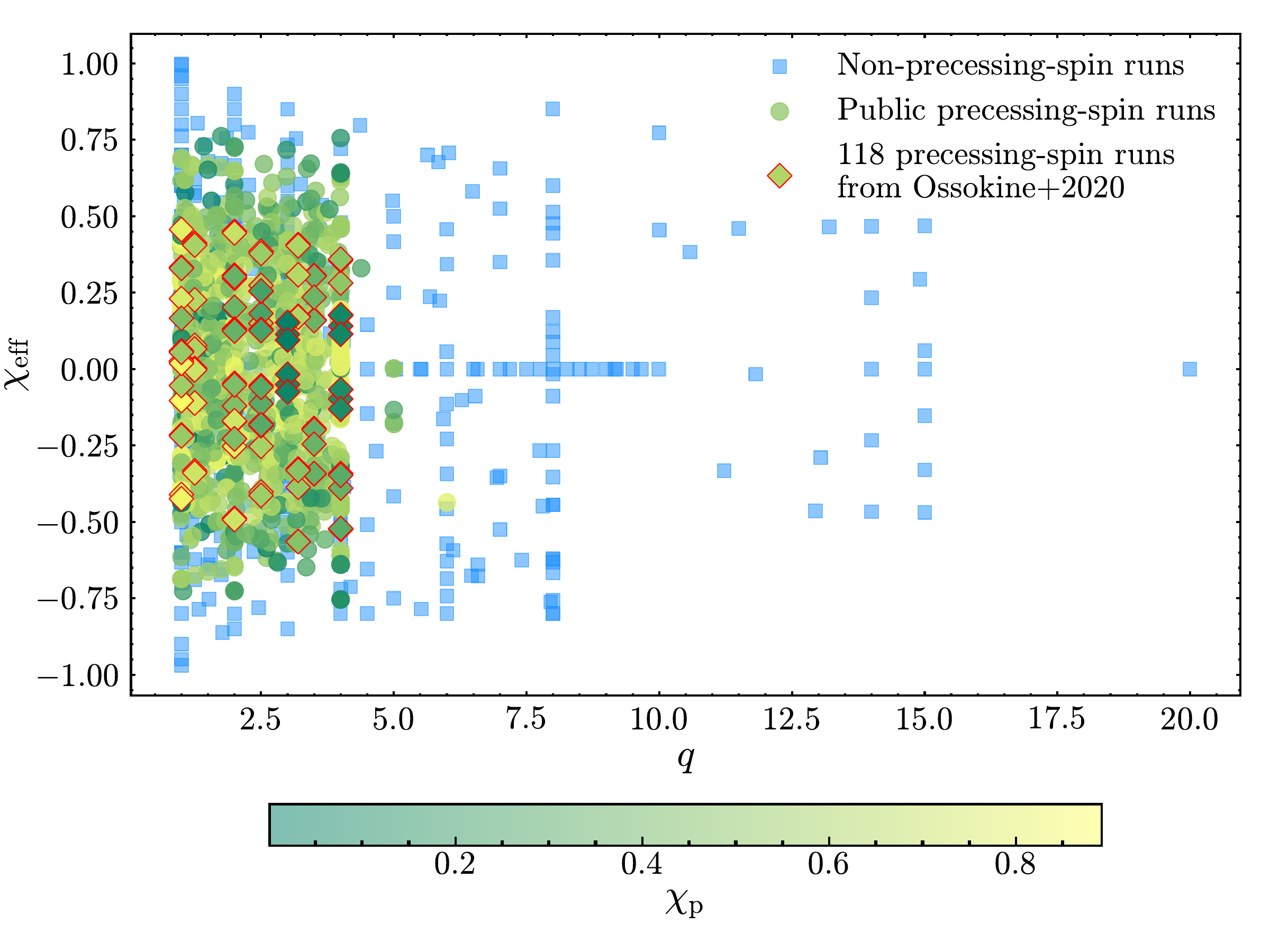}
	\caption{Parameter space coverage in $q-\chi_{\rm eff}-\chi_{\rm p}$ space for NR simulations used to build and validate the \texttt{SEOBNRv5} models. For the non-precessing runs used in the construction of the \seobfivehm~model the color is fixed to blue (see Ref.~\cite{Pompiliv5} for details about these simulations). The precessing-spin runs follow a color map depending on the value of the effective spin parameter $\chi_{\rm p}$ at the reference time of the simulation. The set of precessing NR waveforms is built upon simulations from the public SXS catalog \cite{Boyle:2019kee}, as well as the 118 simulations from Ref.~\cite{Ossokine:2020kjp},  which are highlighted with red diamonds to ease their visualization.}
\label{fig:paramsSpaceNR}
\end{figure}

\begin{figure*}[!]
	\includegraphics[width=\columnwidth]{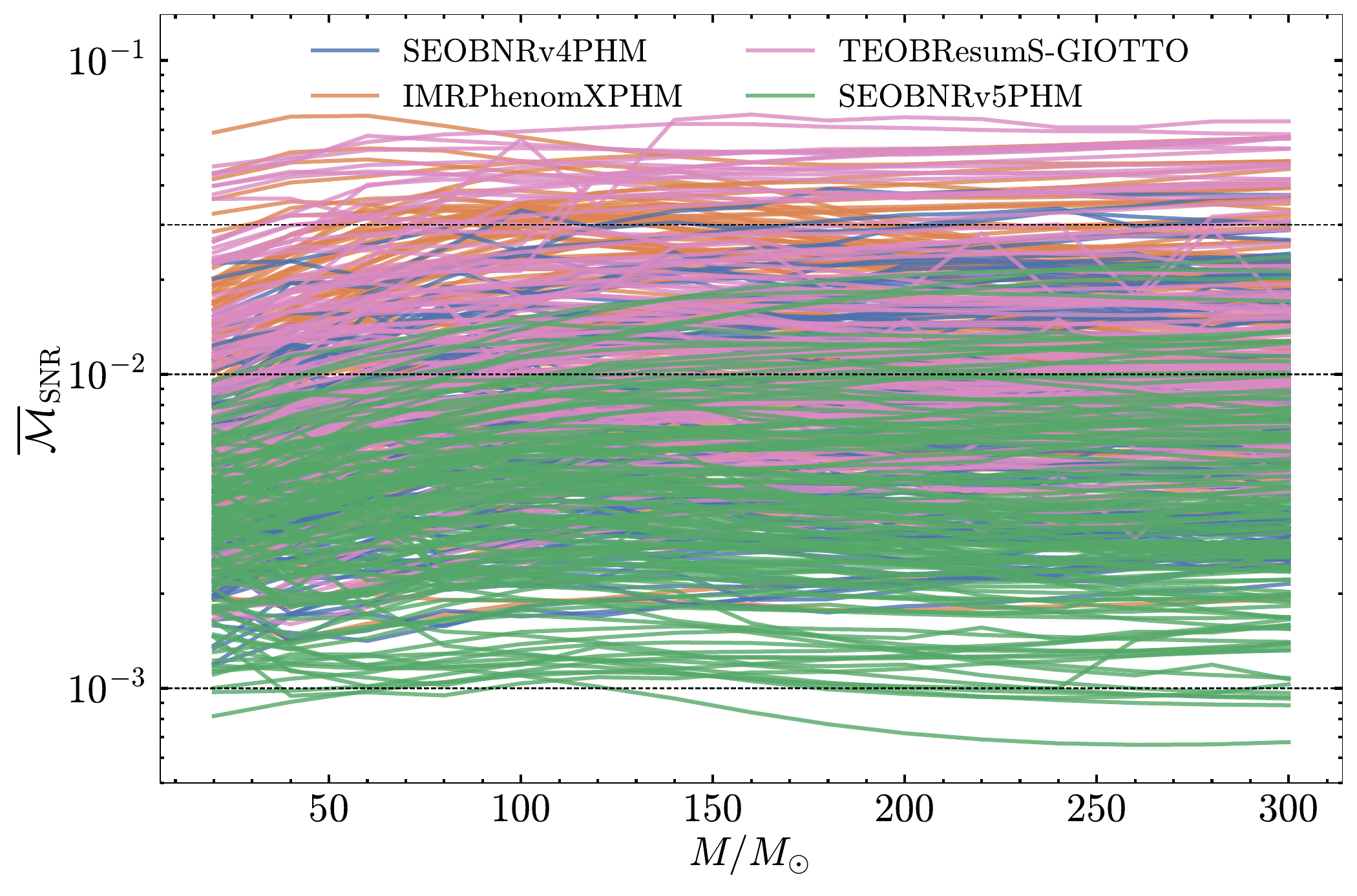}
	\includegraphics[width=\columnwidth]{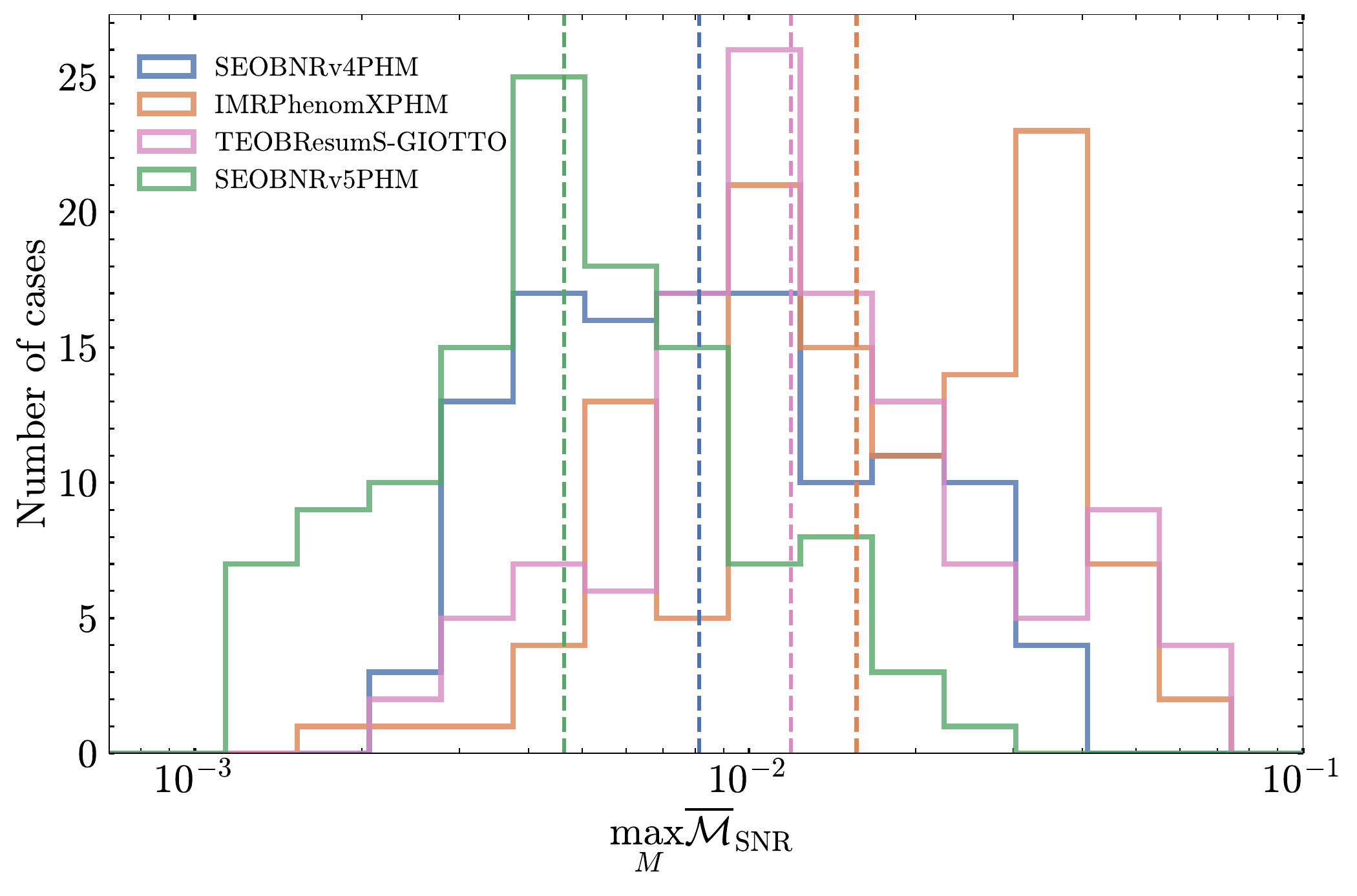}
	\caption{\textit{Left panel:} Sky-and-polarization averaged, SNR-weighted unfaithfulness in the total mass range between $[20-300] M_\odot$ for an inclination $\iota=\pi/3$, between \seobfourphm~(blue), \xphm~(orange), \teob~(pink) and \seobfivephm~(green) against NR for the 118 highly precessing-spin BBH simulations from Ref.~\cite{Ossokine:2020kjp}. The dashed horizontal vertical lines correspond to the $10^{-3}$, $0.01$ and $0.03$ unfaithfulness values. \textit{Right panel:} Distribution of the maximum unfaithfulness over the total mass range for each NR simulation considered in the left plot. The vertical dashed lines indicate the median values of the distribution.}
\label{fig:histogramMax_public118}
\end{figure*}

In the left panel of Fig. \ref{fig:histogramMax_public118} the unfaithfulness is shown as a function of total mass, $[20-300] M_\odot$, for each NR simulation, while in the right panel the distribution of the maximum unfaithfulness over the total mass range is displayed. The two panels of Fig. \ref{fig:histogramMax_public118} show that the phenomenological model, \xphm, and the EOB model \teob, have a tail of large unfaithfulness reaching $\sim 7 \%$.
Precisely, they have $78.3 \%$ $(38.3\%)$ and $83.3 \%$ $(44.9 \%)$ of cases with a maximum unfaithfulness, in  the total mass range considered, below $3\%$ $(1\%)$, respectively. This tail of large unfaithfulness is not present in the \seobfourphm~and \seobfivephm~models, and it is consistent with the fact that both models include effects due to the evolution of the in-plane spin components in the co-precessing frame dynamics. More specifically, the \seobfourphm~model has $96.6\%$ $(57.6 \%)$ of cases with maximum unfaithfulness, in the total mass range considered, below $3\%$ $(1\%)$, while these numbers increase to $100\%$ $(85.6\%)$ for the \seobfivephm~model, which has lower unfaithfulness (higher accuracy) than \seobfourphm. We suspect this is due to the more accurate underlying aligned-spin model, \seobfivehm~\cite{Pompiliv5}, as well as the new improvements included in  \seobfivephm, such as the shift in the co-precessing QNM frequencies, described in Secs. \ref{sec:EOBexpressions} and \ref{sec:EOBwaveforms}.

\begin{figure*}[!] 
	\includegraphics[width=\linewidth]{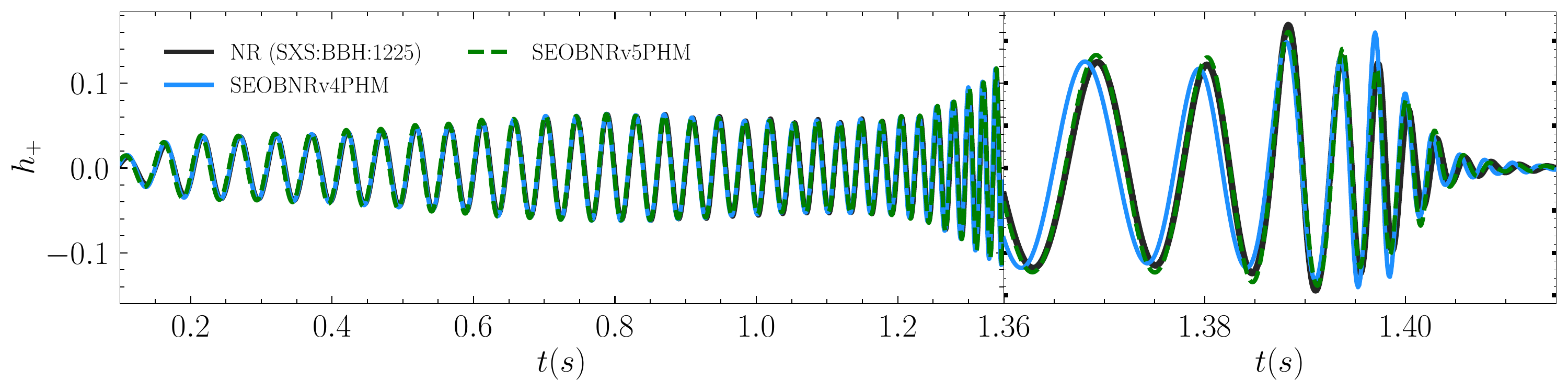}
	\caption{Time-domain comparison of the \seobfivephm~and \seobfourphm~models to the NR waveform \texttt{PrecBBH000001} from Ref.~\cite{Ossokine:2020kjp} with mass ratio 1.25, black-hole spin magnitudes 0.8 and total mass $M=60 M_\odot$. The source parameters are $\iota_s = \pi/3$, $\phi_s = \pi$, $\kappa_s = 0$. The NR waveform includes all the multipoles with $l\leq 5$. Both waveform models resemble accurately the features of the NR waveform at the inspiral, merger and ringdown, with a more faithful agreement of \seobfivephm~which translates into an unfaithfulness  of $0.69 \%$, while for \seobfourphm~it increases to $1.1\%$.}
	 \label{fig:waveform_plot}
\end{figure*}

In Fig. \ref{fig:waveform_plot} we show the polarizations of \seobfivephm~and \seobfourphm~for the precessing NR simulation \texttt{PrecBBH000001} with mass ratio 1.25, spin magnitudes $\chi_i \equiv |\bm{\chi}_i| = 0.8$, total mass $60 M_\odot$ and all the modes $l\leq 5$.  Specifically, we plot the plus polarization, $h_+$, leaving out the overall constant amplitude.  We note that  \seobfivephm~reproduces more accurately the features of the NR waveform at merger and ringdown, which translates into an unfaithfulness of $0.69 \%$ against the NR waveform, while for \seobfourphm~the unfaithfulness is  $1.1\%$.

\setlength{\extrarowheight}{8pt}
\begin{table*}[!]
    \centering
    \begin{tabular}{ c  c  c  c  c  c   }
 \hline
 \hline
 Approximant & \seobfourphm & \seobfivephm~&  \xphm &  \teob \\
 \hline
 \centering
median $\max_M \overline{\mathcal{M}}_{\rm SNR}$ & $7.49 \cdot 10^{-3}$ & $4.75 \cdot 10^{-3}$ &  $14.35 \cdot 10^{-3}$ & $11.47 \cdot 10^{-3}$ \\
$\%$ cases with $\max_M \overline{\mathcal{M}}_{\rm SNR}<1\%$ & $60.8\%$ & $84.4\%$ & $38.3\%$ &  $44.9\%$\\
$\%$ cases with $\max_M \overline{\mathcal{M}}_{\rm SNR}<3\%$ & $95.3\%$ & $99.8\%$ & $78.3\%$ &  $83.3\%$\\
 \hline
 \hline
    \end{tabular}
 \caption{Summary of the sky-and-polarization averaged, SNR-weighted unfaithfulness in the total mass range between $[20-300] M_\odot$ for an inclination $\iota=\pi/3$, between different precessing-spin approximants and the 1543 SXS NR simulations from Refs.~\cite{Boyle:2019kee,Ossokine:2020kjp}. The table shows the median of the maximum unfaithfulness across total mass, and the percentage of cases with mismatches below $1\%$ and $3\%$.}
 \label{tab:NRmismatches}
\end{table*}

\begin{figure*}
	\includegraphics[width=\textwidth]{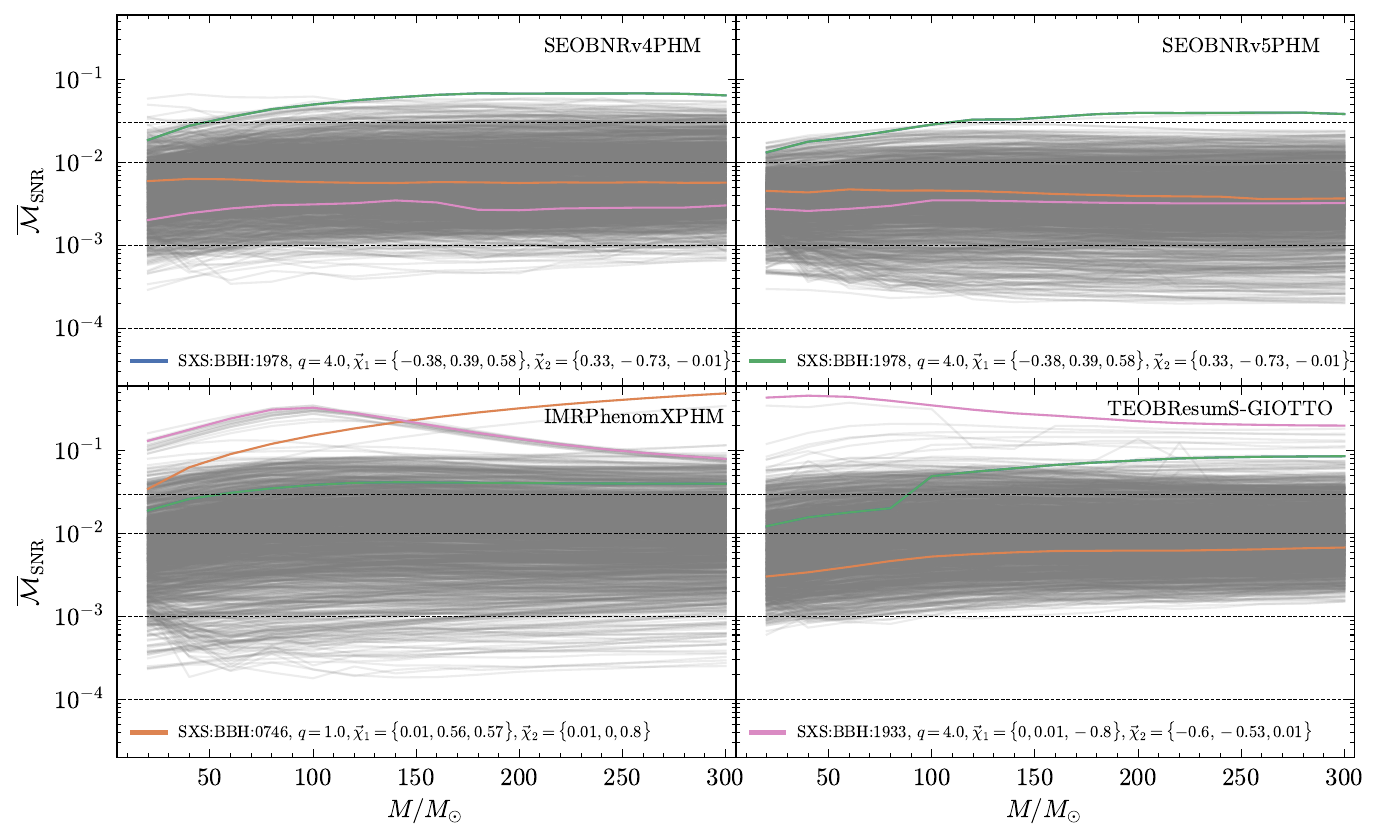}
	\caption{Sky-averaged SNR weighted unfaithfulness as a function of the total mass of the system $[20,300] M_\odot$, of the \seobfourphm~model (top left panel),  the \seobfivephm~model (top right panel), the \xphm~model (left bottom panel) and \teob~(right bottom panel),  against 1543 precessing-spin SXS simulations. The simulations with the highest unfaithfulness for each model are highlighted in each panel. For \seobfivephm~and \seobfourphm~the case with highest unfaithfulness coincides, and thus the highlighted curves overlap.} 
	 \label{fig:spaghetti_eob}
\end{figure*}

We now turn to exploring the broader parameter space by computing the unfaithfulness  against a set of 1543 precessing-spin NR waveforms (1425 public + 118 highly precessing configurations above). In Fig. \ref{fig:spaghetti_eob} we show the unfaithfulness as a function of the total mass of the system for each model against all the simulations. Additionally, we highlight the simulations with the largest unfaithfulness for each waveform model in each panel. The simulations with larger unfaithfulness differ depending on the waveform approximant considered. For the EOB models they correspond to high mass ratios $q=4$ and high in-plane spin components where the modeling approximations are expected to perform worse, while the phenomenological model presents the largest unfaithfulness for an equal-mass simulation with high-in plane component.

\begin{figure*}[!]
	\includegraphics[width=\linewidth]{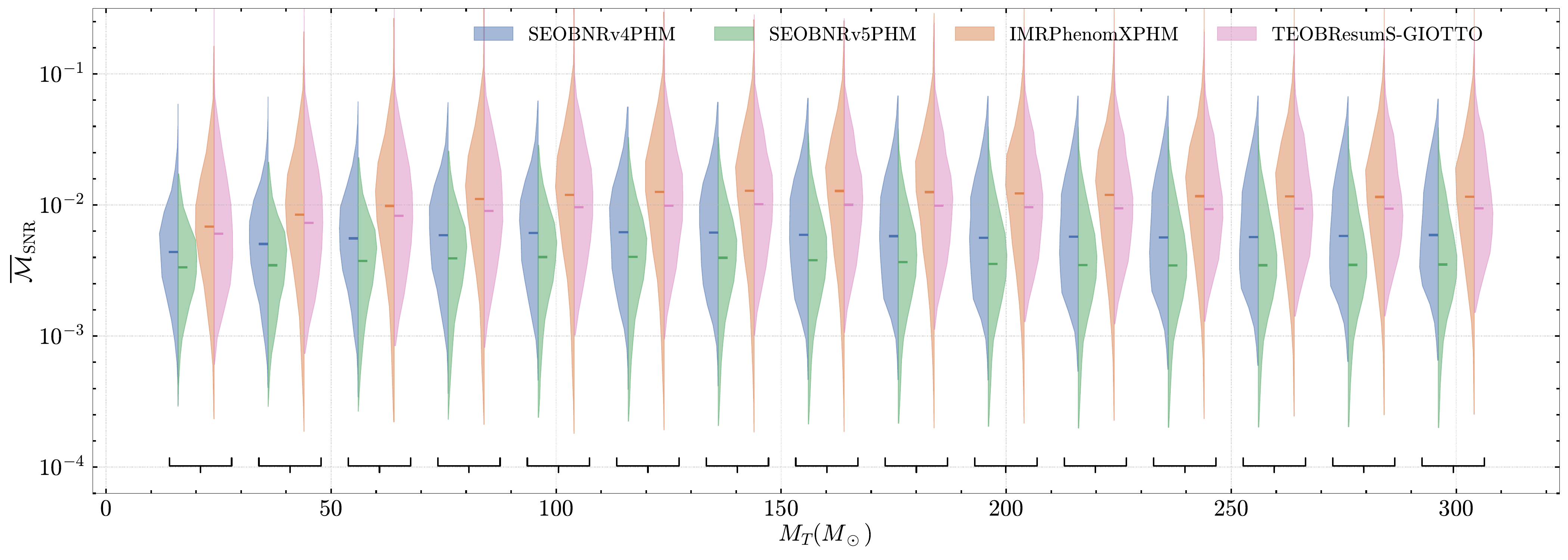}
	\caption{Distribution of the sky-and-polarization averaged, SNR-weighted           unfaithfulness as a function of binary's total mass for inclination $\iota=\pi/3$, between \seobfourphm~(blue), \seobfivephm~(green), \xphm~(orange) and \teob~(pink) against NR for 1543 quasi-circular precessing-spin BBH simulations. For each total mass considered the distributions of \seobfourphm~and \seobfivephm, and \xphm~and \teob~have been shifted in the x-axis by $4 M_\odot$  from the total mass value used to compute the unfaithfulness  to ease the visualization of the  results. The bracket indicates the total mass value used for the unfaithfulness calculation. In each distribution the median value is highlighted with thicker lines. }
	 \label{fig:violin_precessing}
\end{figure*}

The results from Fig. \ref{fig:spaghetti_eob} indicate that the \seobfivephm~model has lower values of unfaithfulness with respect to the rest of the models. The information in Fig. \ref{fig:spaghetti_eob} is more quantitatively represented in  Fig. \ref{fig:violin_precessing} as a violin plot of the distribution of unfaithfulness of the different models against NR for each total mass considered between $[20-300] M_\odot$. We note that the trend in the unfaithfulness is similar to the one for the 118 highly precessing-spin simulations. The \xphm~model has the largest tails of unfaithfulness reaching $10 \%$, followed by the \teob~model, which generally has lower unfaithfulness than \xphm~as shown in Ref.~\cite{Gamba:2021ydi}. The \seobfourphm~model gives an even lower unfaithfulness, while the distributions of the \seobfivephm~model have less support at high unfaithfulness than the rest of the models and lower median values for all the total masses considered with respect to the next more accurate model, \seobfourphm. A more quantitative analysis of the unfaithfulness against NR can be found in Table \ref{tab:NRmismatches}, which reveals that \seobfivephm~has $99.8 \%$ ($84.4 \%$) cases with a maximum unfaithulness, in the total mass range considered, below $3 \%$  ($1\%$). These numbers reduce to $95.3 \%$ ($60.8 \%$) for  \seobfourphm, to $83.3 \%$ $(44.9 \%)$ for  \teob~and to $78.3\%$ $(38.3\%)$ for \xphm.  Regarding the improvements of \seobfivephm~at high total masses with respect to \seobfourphm, this is a combination of all the new additions in the merger and ringdown part explained in Sec. \ref{sec:MergerWaveforms}. For instance, the co-precessing frame modes based on \seobfivehm~produce an overall more accurate merger and ringdown, but the application of a consistent calculation of the quasi-normal modes in the $\bm{J}_{\rm f}$-frame from Ref. \cite{Hamilton:2023znn} also reduces the number of cases with mismatch larger than $3\%$. A detailed study of the approximations at merger and ringdown, their limitations and possible improvements by including NR information is ongoing, and we leave for future work their application to the next generation of  precessing-spin \texttt{SEOBNR} models.

\begin{figure*}[!]
	\includegraphics[width=\linewidth]{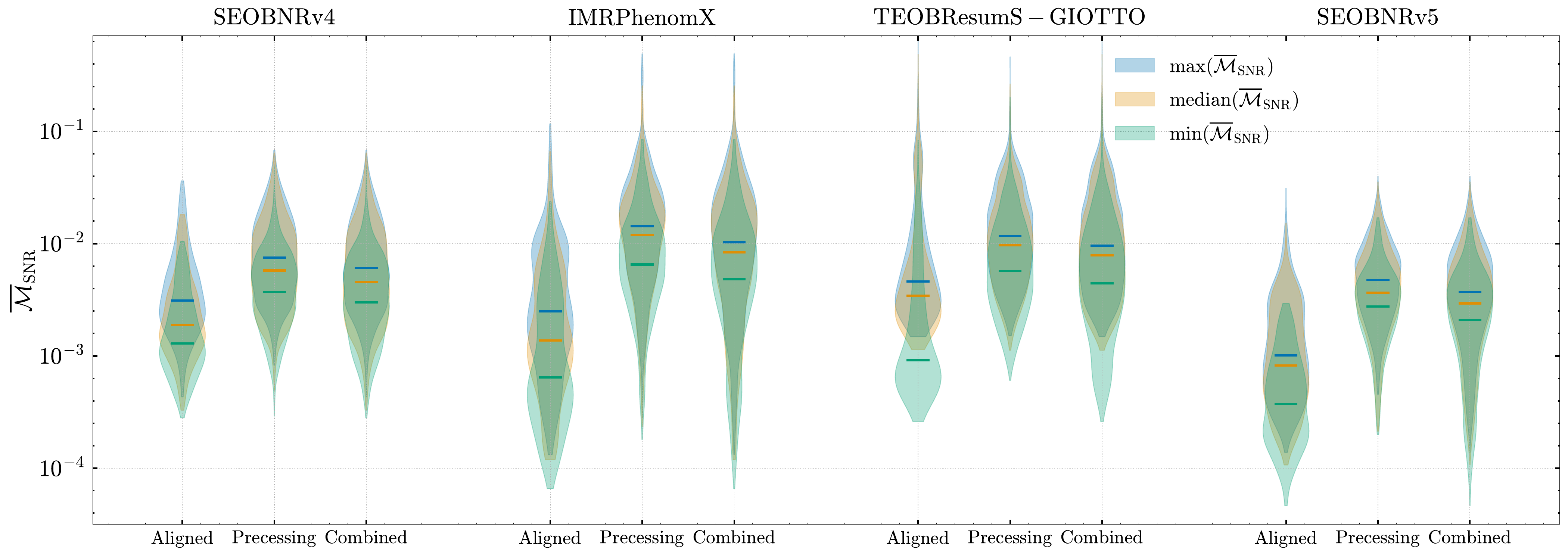}
	\caption{Distribution of maximum (blue), median (orange) and minimum (green) sky-and-polarization averaged, SNR-weighted
          unfaithfulness over the binary's total mass range $[20-300] M_\odot$ for inclination
          $\iota=\pi/3$, between the different waveform families ({\tt SEOBNRv4}, {\tt IMRPhenomX}, \teob~and {\tt SEOBNRv5}) against NR for aligned spins (\textit{Aligned}), precessing spins (\textit{Precessing}) and combining the two previous distributions (\textit{Combined}). The non-precessing NR simulations correspond to the 441 cases presented in Ref.~\cite{Pompiliv5}, while the precessing NR simulations correspond to the 1543 cases used in Fig. \ref{fig:spaghetti_eob}. In the violin plots the median values of the distributions are highlighted with thicker lines.}
	 \label{fig:violin_ASPrec}
\end{figure*}

Finally, we provide a more complete picture of the accuracy of the different models against NR in the quasi-circular limit by incorporating to our precessing results the unfaithfulness corresponding to 441 non-precessing SXS NR waveforms computed in Ref.~\cite{Pompiliv5}. Fig. \ref{fig:violin_ASPrec} shows violin plots of the maximum, median and minimum unfaithfulness distributions of the different waveform models considered in the aligned-spin, precessing-spin case and with the combined distributions. A thorough discussion of the accuracy of the different models in the non-precessing case can be found in  \cite{Pompiliv5}, but we remark that the new aligned-spin \seobfivehm~model presents the lowest unfaithfulness distribution when compared to the other models.  As discussed above, in the precessing case the \seobfivephm~model leads to the lowest unfaithfulness values followed closely by the \seobfourphm~model.  We also observe that the lack of calibration to precessing-spin NR waveforms causes a shift in the unfaithfulness of the precessing-spin models (with respect to the nonprecessing models) towards larger values. This points out that in order to increase further the accuracy of the models in the precessing-spin case, calibration to NR precessing waveforms is required, which we leave to the future.

\subsection{Comparison against other precessing-spin waveform models} \label{sec:ModelsComparison}

We now study the performance of the \seobfivephm~model in a larger parameter space. First we compute the unfaithfulness of \seobfivephm~against the NR surrogate model \nrsur~\cite{Varma:2019csw}, which includes all $l\leq 4$ waveform multipoles,  in the region in which it was built, that is mass ratios $q\in[1,4]$, spin magnitudes up to 0.8 and total masses larger than $~60 M_\odot$. Specifically, we generate a set of 5000 cases uniformly distributed in mass ratios $q\in[1,4]$ and effective precessing-spin parameter\footnote{We do not sample uniformly in spin magnitudes and orientations to avoid having most of the cases clustering at low values of  $\chi_p$, where precession effects are less significant.} $\chi_p$~\cite{Schmidt:2014iyl},  with spin magnitudes up to $0.8$ and initial geometric frequency of $M \omega = 0.023$, large enough such that all the configurations have a length compatible with the one of the surrogate waveforms. We also compute the unfaithfulness of the state-of-the-art precessing-spin models, \seobfourphm, \xphm~and \teob, against the \nrsur~model.

\begin{figure*} 
	\includegraphics[width=\linewidth]{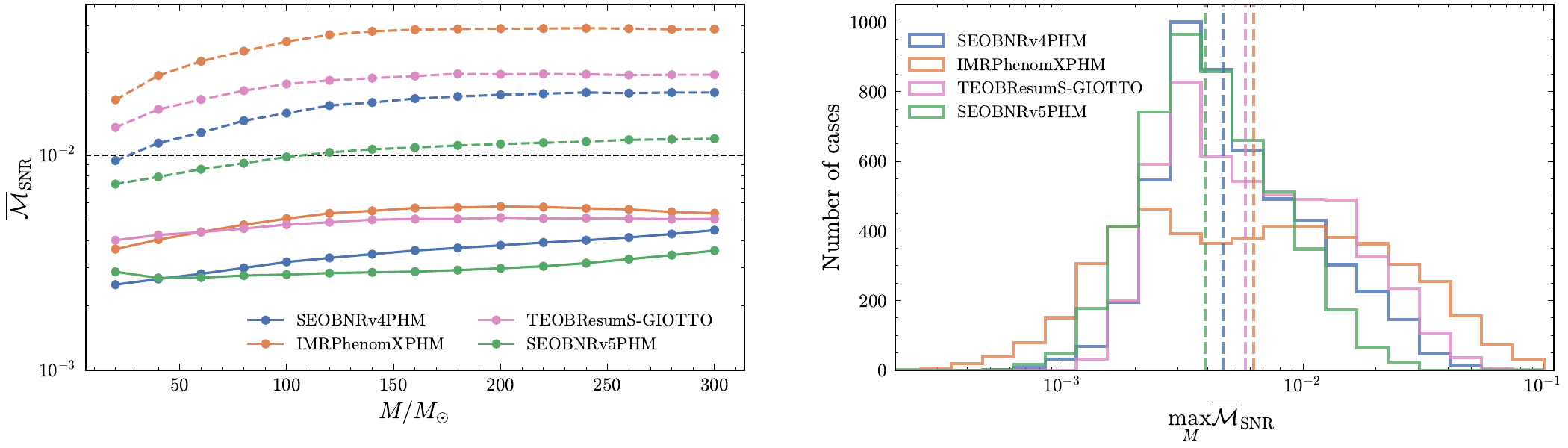}
	\caption{Sky-and-polarization-averaged, SNR-weighted unfaithfulness as a function of the total mass of the binary for inclination $\iota_s = \pi/3$, among the \nrsur~model and the \seobfourphm~(blue), \xphm~(orange), \teob~(pink) and \seobfivephm~(green) models for 5000 randomly distributed precessing-spin configurations. \textit{Left:} The solid (dashed) lines show the median (95th percentile) as a function of the total mass. \textit{Right:} Distribution of maximum unfaithfulness over all the total masses considered. The vertical dashed lines indicated the median values of the distributions.}
	 \label{fig:unf_nrsur}
\end{figure*}

The results of such study are summarized in Fig. \ref{fig:unf_nrsur}, where in the left panel the median and the 95th percentile of the unfaithfulness, as a function of the total mass of the binary are shown, while in the right plot the distributions of the maximum unfaithfulness, over the total mass range $[20-300] M_\odot$, are displayed. We find that the behavior of the unfaithfulness resembles those of the comparisons against the NR waveforms in Figs. \ref{fig:histogramMax_public118} and \ref{fig:violin_precessing}. All the models have median unfaithfulness below $1\%$ with the \seobfivephm~model showing the lowest median~\footnote{The median unfaithfulness for the \seobfourphm~model is $0.46\%$, $0.62\%$ for \xphm~and $0.69 \%$ for \teob.} of $0.39 \%$ unfaithfulness values. We note that the median of unfaithfulness of \seobfivephm~is followed very closely by the other models, with the \seobfourphm~model being the closest one.
The difference between the \texttt{SEOBNR} models and the \xphm~and \teob~models is likely a consequence of neglecting the in-plane spin effects in the orbital dynamics in the co-precessing frame. As described in Sec. \ref{sec:EOBexpressions}, these effects are introduced in \seobfivephm~through the partially precessing Hamiltonian, $H^{\rm pprec}_{\rm EOB}$.
Furthermore, the increase in accuracy of \seobfivephm~with respect to \seobfourphm~can be understood due to the more accurate underlying co-precessing waveform model (\seobfivehm), as well as the improvements discussed in Sec. \ref{sec:EOBwaveforms}. More quantitatively, we find that for \seobfivephm~$100\%$ $(90.1\%)$ of cases have a maximum unfaithfulness, in the total mass range considered, against the \nrsur~model below $3\%$ $(1\%)$, while these numbers reduce to $98.7\%$ $(79.5\%)$ for \seobfourphm, $89.4\%$ $(62.8\%)$ for \xphm~
and $96.1\%$ $(66\%)$ for \teob. For all the models the cases with high unfaithfulness correspond to configurations with mass ratios $q \sim 4$ and $\chi_p \sim 0.8$, which is the boundary region of calibration of the \nrsur~model, and where the effects of spin precession are stronger in the waveform,  as already seen in previous comparisons to the NR surrogate in Refs.~\cite{Ossokine:2020kjp,  Gamba:2021ydi}.

\begin{figure*}
	\includegraphics[width=\linewidth]{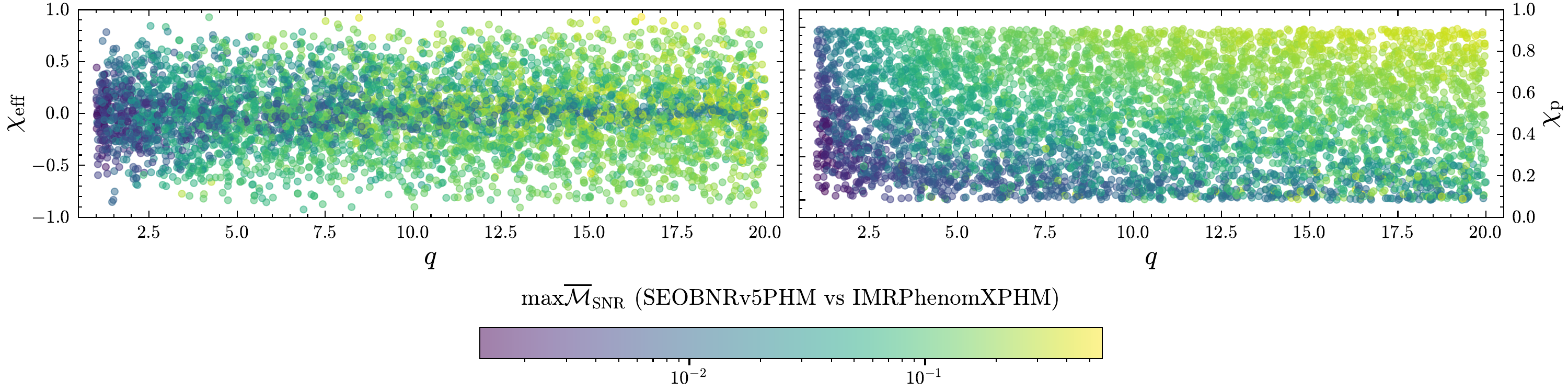}
	\caption{Maximum sky-and-polarization-averaged unfaithfulness weighted by the SNR over the total mass range $[20-300]M_\odot$ between \seobfivephm~and  \xphm~for 5000 random configurations with inclination $\iota_s = \pi/3$. The unfaithfulness grows with increasing mass ratio and spin magnitude values, and it can reach very large values for mass ratios $q\sim 20$ and $\chi_\mathrm{p} \sim 1$.}
	 \label{fig:unf_models}
\end{figure*}

Finally, we also examine the behavior of the precessing models in a wider parameter space outside the region of calibration of the underlying aligned-spin models, and where there are no precessing-spin NR simulations available. For this purpose we consider 5000 configurations randomly distributed in mass ratios  $q \in [1,20]$ and uniformly distributed in the effective precessing-spin $\chi_p$ parameter up to 0.99,  for inclination $\iota_s = \pi/3$, with an initial starting geometric frequency of $M \omega = 0.022$, and compute the unfaithfulness, $\overline{\mathcal{M}}_{\rm SNR}$, using the \xphm\footnote{We do not include the \teob~model in these comparisons as we have found some unphysical growth of the amplitude at merger of the $l=2$ inertial frame modes for large spins and mass ratios, which is likely due to the behavior of the NQC coefficients of the (2,1)-mode as already described in Ref.~\cite{Gamba:2021ydi}. We show the comparison against the \tphm~model in Appendix \ref{app:PhenomModels}.} as a signal, and the \seobfivephm~model as the template waveform.
Figure \ref{fig:unf_models} shows the unfaithfulness as a function of mass ratio ($q$), effective spin parameter ($\chi_\mathrm{eff}$), and effective precessing-spin parameter ($\chi_{\mathrm{p}}$). We find that for mass ratios $q<5$,  $96.84 \%$ ($41.3 \%$) of cases have a maximum unfaithfulness, in the total mass range $[20,300]M_\odot$, below $10 \%$ $(1\%)$.
The unfaithfulness increases significantly with mass ratio and spins, with the highest unfaithfulness values at the largest mass ratios $q \sim 20$, and effective spin precessing parameter $\chi_{\mathrm{p}} \sim 0.99$. In particular, when considering $q\leq 20$ we find that 
 $59.19\%$ $(13.45\%)$ cases with maximum unfaithfulness, in the total mass range considered, below $10\%$ $(1\%)$.
These unfaithfulness comparisons and the large differences between models point out the necessity to populate this challenging region of high mass ratio and high spins with NR simulations, which can be used to validate distinct waveform models, as well as to improve their accuracy by incorporating this NR information into them.

\subsection{Computational performance} \label{sec:benchmarks}

In previous sections we have demonstrated the accuracy of the \seobfivephm~model with respect to NR waveforms and predictions of other state-of-the-art waveforms models. Another key aspect to test is the computational efficiency of the model, as parameter-estimation runs with standard stochastic samplers require of the order of $10^7-10^{8}$ or more waveform evaluations (see e.g. Refs.~\cite{Ashton:2021anp,Williams:2021qyt, Williams:2023ppp}). Therefore, computational efficiency is a key feature for the model to be useful for the analysis of GW signals or Bayesian inference studies.

The  \seobfivephm~model is part of the fifth generation of \texttt{SEOBNR} models implemented in a high-performance Python package \texttt{pySEOBNR} \cite{Mihaylovv5}. As described in  Ref.~\cite{Mihaylovv5}, the \texttt{pySEOBNR} infrastructure offers a simple and modular procedure to develop highly accurate and computationally efficient waveform models. This new Python infrastructure moves the development of the \texttt{SEOBNR} family from the highly efficient, but more rigid C-99 \texttt{LALSuite} \cite{lalsuite} libraries to a more flexible and modular Python framework.

\begin{figure}
	\includegraphics[width=\columnwidth]{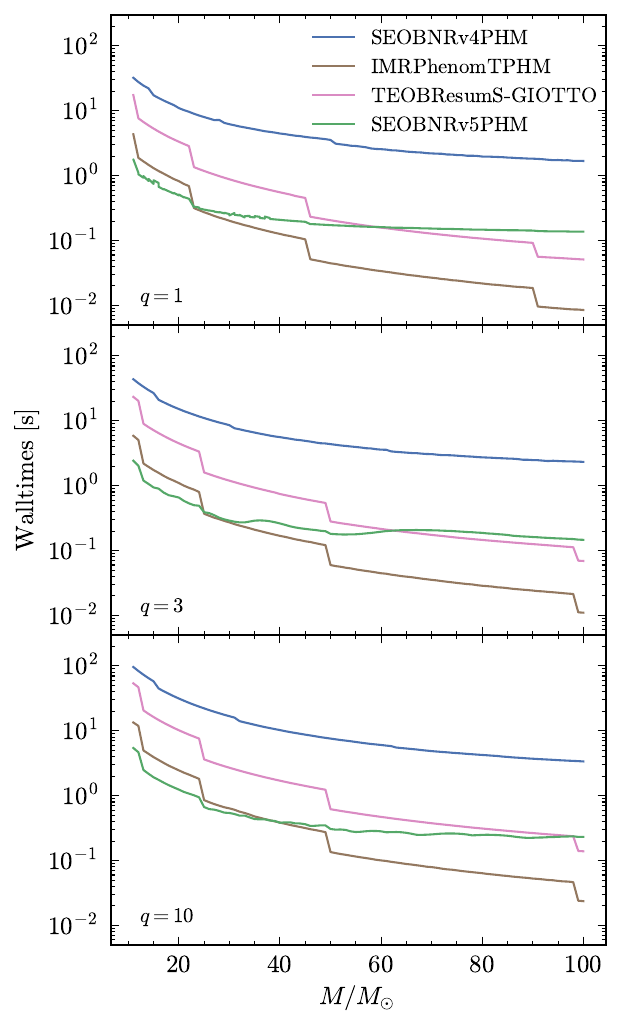}
	\caption{Walltimes of the \seobfourphm, \tphm, \teob~and \seobfivephm~models for a configuration with dimensionless spins $\bm{\chi}_1=[0.5,0,0.8]$, $\bm{\chi}_2=[0,0.5,0.3]$, total mass range $M \in[10,100]M_\odot$, starting frequency $f_{\rm start}=10$Hz and three different mass ratios 1 (top panel), 3 (mid panel) and 10 (bottom panel).}
\label{fig:benchmarks}
\end{figure}

In this section we asses the computational efficiency of the \seobfivephm~model implemented in {\tt pySEOBNR}, by timing the waveform generation and comparing it to other state-of-the-art time-domain multipolar precessing-spin models (\seobfourphm, \tphm~and \teob). We consider binary's configurations with mass ratios $q=1,3,10$, dimensionless spins $\bm{\chi}_1=[0.5,0,0.8]$, $\bm{\chi}_2=[0,0.5,0.3]$, total mass range $M \in[10,100]M_\odot$ at a starting frequency $f_{\rm start}=10$Hz. The results of the walltimes to generate the waveforms are shown in Fig. \ref{fig:benchmarks}, where we are including all the modes up to $l=4$, and a maximum frequency consistent with the Nyquist criterion satisfied for all the multipoles considered~\footnote{The benchmarks of the waveform generation timing were performed on a computing node (dual-socket, 32-cores per socket, SMT-enabled AMD EPYC (Milan) 7513 (2.60 GHz), with 8 GB RAM per core) of the \texttt{Hypatia} cluster at the Max Planck Institute for Gravitational Physics in Potsdam. We keep all default settings for every model}.
The outcome of the benchmark demonstrates the significant increase in speed of the \seobfivephm~model with respect to the previous generation \seobfourphm. For the arbitrary configurations considered for the benchmarks, we observe more than an order of magnitude improvement in speed.
The substantial increase in speed for \seobfivephm~is a consequence, not only of the fast and efficient implementation in the \texttt{pySEOBNR} infrastructure, but also to the use of the PN-expanded spin and angular-momentum evolution equations, Eqs. \eqref{eq:SLeqns}, which allow the use of the PA approximation~\cite{Nagar:2018gnk,Mihaylov:2021bpf} in the \seobfivephm~model. The PA approximation reduces the computational cost of evaluating the inspiral waveform as it replaces solving numerically the ordinary differential equations at every timestep of the EOB inspiral by an iterative procedure over a coarser radial grid (see Appendix \ref{app:PA} for details of the implementation in \seobfivephm). Besides the PA approximation, the \seobfivephm~model also implements an efficient calculation of the polarizations as described in  Se.~\ref{sec:polarizations},  which translates into a further increase in speed at lower total masses, where the computational cost of generating the waveform comes from the interpolation of the waveform multipoles into a constant time grid~\footnote{The interpolation of the waveform modes onto a time grid with constant timestep is needed to perform an efficient Fourier transform of the waveform for data-analysis studies.}. This can be seen in Fig.
~\ref{fig:benchmarks}, where the \seobfivephm~model outperforms the \teob~and \tphm~models at low total masses, while at high total masses where the interpolation of the modes is a subdominant operation in terms of computational cost, the \teob~and \tphm~perform faster. \tphm~is substantially faster at high total masses than the rest of the models, due to the fact that it is only integrating the evolution equations for the spins (i.e., no integration of the orbital dynamics as in the \seobfourphm, \seobfivephm~and \teob~models), and the waveform is evaluated using analytical closed expressions.  In summary, the \seobfivephm~model has a comparable speed to the current state-of-the-art precessing-spin time-domain models,  and it is in general between $8-20$ times faster than the \seobfourphm~model, and thus it can be used as a standard tool for data analysis as demonstrated in Sec.~\ref{sec:PE}.

\section{Bayesian analysis with multipolar precessing waveform models}\label{sec:PE}
The main application of the \seobfivephm~waveform model is the Bayesian inference of source parameters of GWs emitted by BBHs. Thus, we now assess how the accuracy of \seobfivephm~quantified in Sec. \ref{sec:WaveformValidation} through the unfaithfulness metric affects parameter-estimation studies. We perform first a synthetic NR signal injection into detector noise, in particular in zero-noise, which is equivalent to averaging over many noise realizations, to assess possible biases coming from waveform inaccuracies and avoid any biases introduced by a random noise realization. Then, we perform a re-analysis of 6 real GW events detected by the LVK collaboration: GW150914, GW190412, GW190521, GW190814, GW191109 and GW200129, and we compare with results from the literature.

\subsection{NR-injection recovery}
In this section we assess the accuracy of the \seobfivephm~model in parameter estimation by injecting a synthetic NR signal corresponding to the NR waveform {\tt SXS:BBH:0165} from the public SXS catalog, with mass ratio $q=6$, source-frame total mass $M=95 M_\odot$ and BH's dimensionless spin vectors defined at 20Hz of $\bm{\chi}_1 = [-0.06,0.78.-0.4]$ and $\bm{\chi}_2 = [0.08,-0.17,-0.23]$. This BBH system is strongly precessing, and it is one of the worst cases in terms of unfaithfulness for \seobfivephm, reaching $2 \%$ for the injected total mass.

For this injection we choose the inclination with respect to the line of sight of the BBH to be $\iota=0.69$ rad, to emphasize the effect of higher order modes. The injected coalescence and polarization phases are $\phi=0.6$ rad and $\psi=0.33$ rad, respectively. The sky-position is defined by its right ascension of 3.81 rad and declination of 0.63 rad at a geocentric time of 1126259600 s. The luminosity distance to the source is chosen to be 650 Mpc, which produces a three-detector (LIGO Hanford, LIGO Livingston and Virgo) network-SNR of $19.4$ when using the LIGO and Virgo PSD at design sensitivity \cite{Barsotti:2018}.

For the parameter estimation study we employ {\tt parallel Bilby} \cite{Smith:2019ucc}, a highly parallelized version of the Bayesian inference Python package {\tt Bilby} \cite{Ashton:2018jfp,Romero-Shaw:2020owr}, using the recommended LVK's setting for the number of auto-correlation times $\mathrm{nact}=50$, number of live points $\mathrm{nact}=2048$, and setting the remaining sampling parameters to their default values. We choose a uniform prior in inverse mass ratio and chirp mass, with ranges $1/q \in [0.05,1]$ and $\mathcal{M} \in [15,45] M_\odot$. The priors on the dimensionless spin vectors are uniform in magnitude  $a_i \in [0,0.99]$, and isotropically distributed in the unit sphere for the spin directions.
The luminosity distance prior is uniform in distance $\propto d_L$ as we are interested in the intrinsic ability of the models in recovering the parameters, since a prior uniform in the comoving-frame of the source $\propto d_L^2$ requires selecting a specific cosmology to compute the redshift \cite{Callister:2021gxf}, which may introduce an effect on the estimated posterior.
The rest of the priors are set according to Appendix C of Ref.~\cite{LIGOScientific:2018mvr}. We perform the injection-recovery with \seobfivephm~and \xphm~in order to compare the performance of both models with a highly precessing signal. We note that \xphm~has an unfaithfulness of
$\sim 12\%$ against the SXS NR-injected waveform, thus we expect some biases in the recovered parameters.

\begin{figure*}
	\includegraphics[width=\columnwidth]{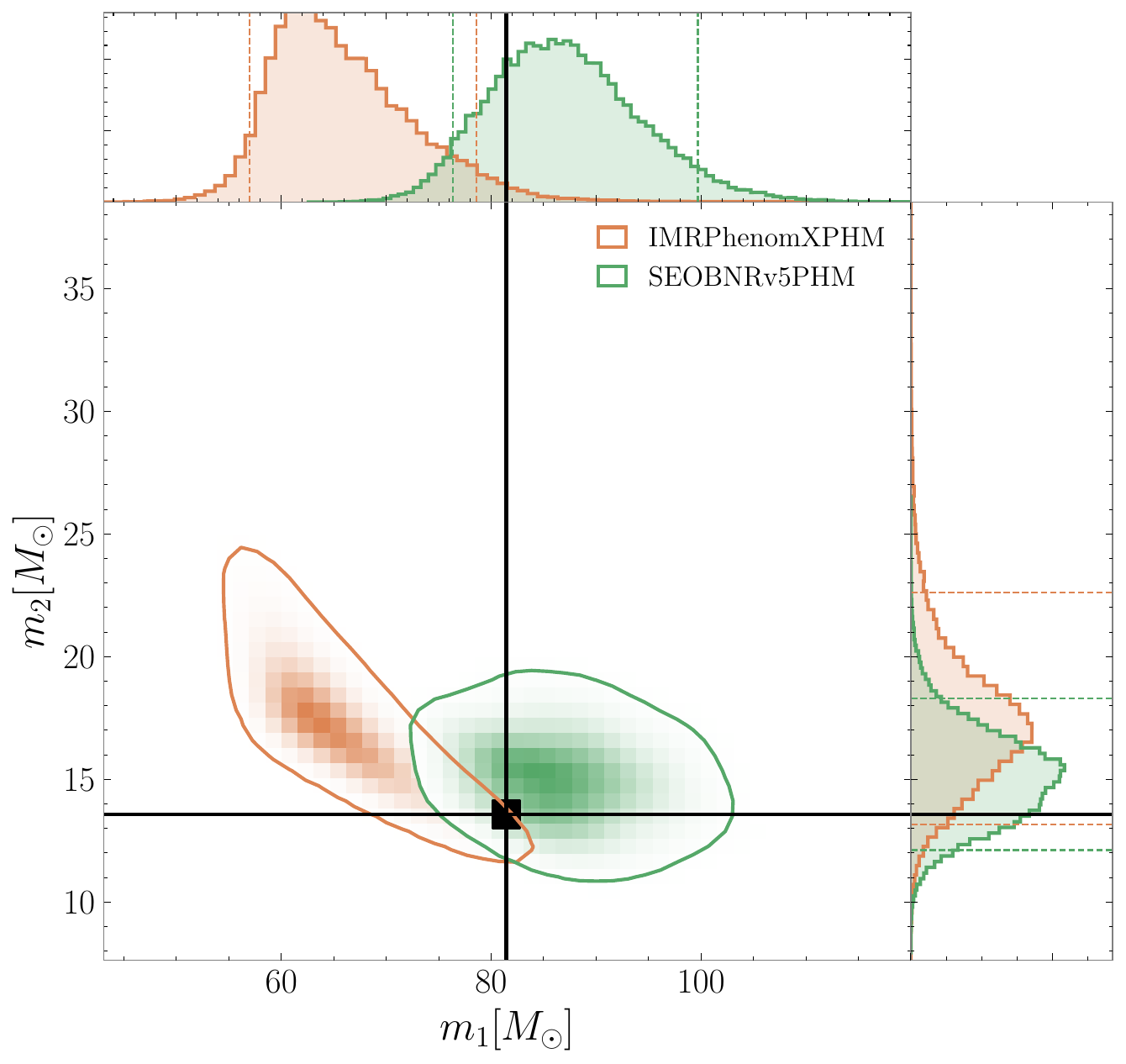}
	\includegraphics[width=\columnwidth]{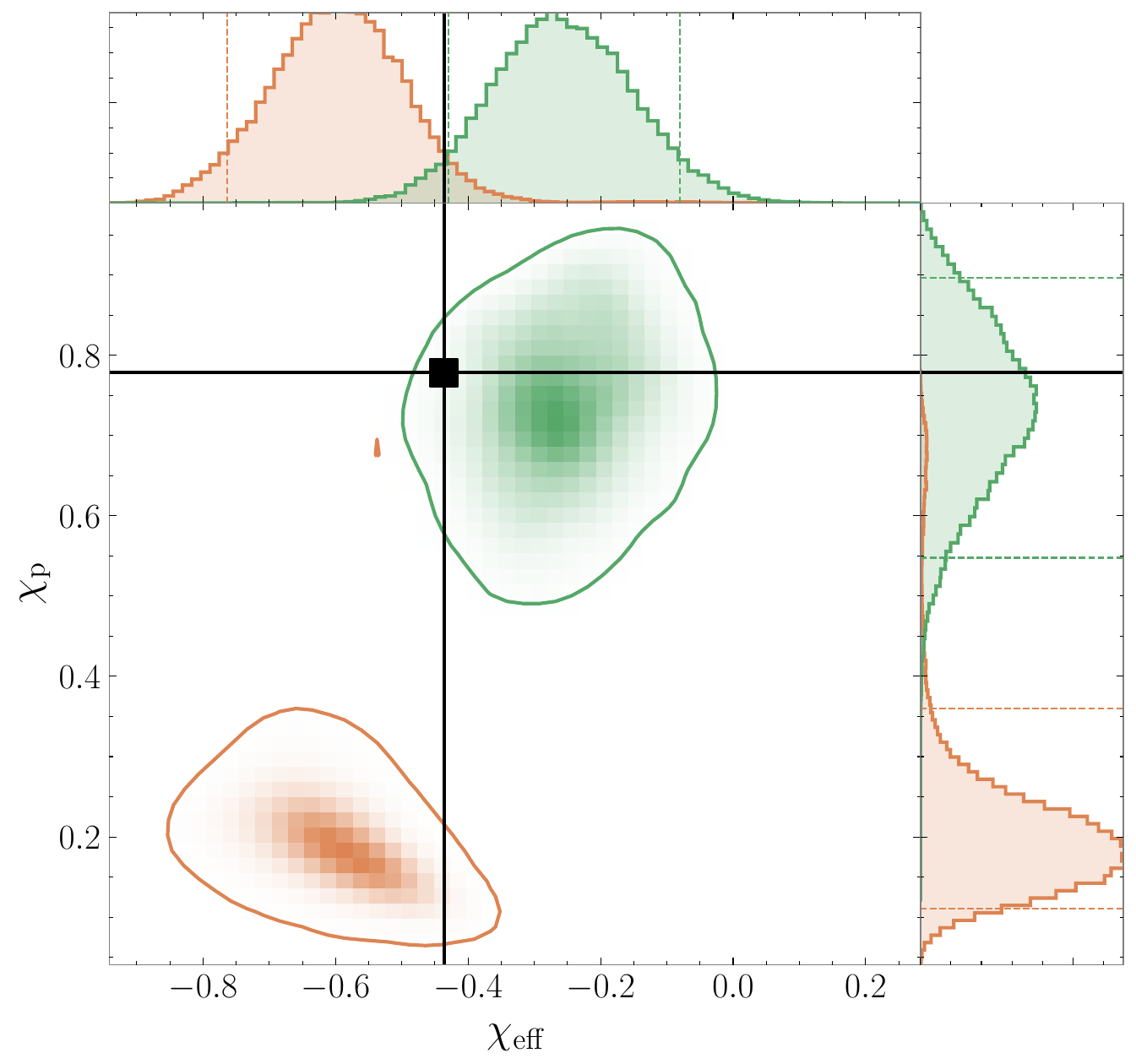}
	\caption{2D and 1D posterior distributions for some relevant parameters measured from the synthetic BBH signal with mass ratio  $q=6$, total source-frame mass of $95 M_\odot$, dimensionless spins of the BHs $\bm{\chi}_1 = [-0.06,0.78.-0.4]$ and $\bm{\chi}_2 = [0.08,-0.17,-0.23]$ defined at 20Hz. The inclination with respect to the line of sight of the binary is  $\iota=0.69$ rad. The other parameters are specified in the text and in Table \ref{tab:injection_settings}. The signal waveform is a NR waveform from the public SXS catalog {\tt SXS:BBH:0165}. In the 2D posteriors the solid contours represent the $95\%$ credible intervals and black dots show the values of the parameters of the injected signal. In the 1D posteriors they are represented by dashed and solid vertical lines, respectively. The parameter estimation is performed with the \seobfivephm~model (green) and the \xphm~model (orange). \textit{Left:} Component masses in the detector frame. \textit{Right:} Effective spin parameters, $\chi_{\mathrm{eff}}$ and $\chi_{\mathrm{p}}$.}
	 \label{fig:pe_nrInj}
\end{figure*}

In Fig. \ref{fig:pe_nrInj} we summarize the parameter-estimation results of the injection. We report the marginalized 1D and 2D posteriors for the detector-frame component masses $m_1$ and $m_2$, and the effective spin parameters, $\chi_{\mathrm{eff}}$ and $\chi_{\mathrm{p}}$. In Table \ref{tab:injection_settings} we provide the values of the injected parameters and the median of the inferred posterior distribution with the $90\%$ confidence intervals for both models. The results show that \seobfivephm~is able to recover the component masses within the $90\%$ confidence intervals, while \xphm~presents a significant bias in the primary mass, and the injected values are at the boundary of the 2D $95\%$ credible interval. For the effective spin parameters, both models present a biased result for the effective spin parameter $\chi_{\mathrm{eff}}$, but the precessing effective spin parameter $\chi_\mathrm{p}$ is highly biased in \xphm~towards lower values, while \seobfivephm~recovers an almost unbiased result. Moreover, the injected point is inside the 2D $95\%$ credible interval for \seobfivephm, while \xphm~predicts a region with lower precessing spins and highly anti-aligned spins. From Table \ref{tab:injection_settings} we observe that the spin tilt angles, $\theta_{1,2}$, are recovered within the $90\%$ confidence interval by \seobfivephm, but the phenomenological model \xphm~presents biases for both parameters. In terms of recovered matched filter SNR, \seobfivephm~recovers higher values in the three detectors with respect to \xphm, which is consistent with the higher Bayes factor obtained by \seobfivephm. This example shows the ability of \seobfivephm~to model more accurately precessing signals in comparison to \xphm, likely due to the inclusion of in-plane spin information in the conservative dynamics of the model.  It should be noted that there are some parameters for which \seobfivephm~presents small biases, such as the  effective-spin parameter $\chi_{\mathrm{eff}}$ and the tilt angle of the orbital plane $\theta_{\text{JN}}$, which might be expected since this simulation provides one of the highest unfaithfulness for the model of $\sim 2 \%$, while for the \xphm~model the unfaithfulness increases to $\sim 12 \%$, which explains the larger biases in more parameters than \seobfivephm. However, more studies will be needed in a larger region of the binary's parameter space to assess the efficiency of \seobfivephm~ in capturing spin precession.

\setlength{\extrarowheight}{8pt}
\begin{table}[h!]
    \centering
    \begin{tabular}{ c  c  c  c   }
 \hline
 \hline
 Parameter & \makecell[cc]{Injected \\ value} & \makecell[cc]{\xphm} & \makecell[cc]{\seobfivephm}  \\
 \hline
 \centering
$M/M_\odot$ &   $95.02$ & $82.51^{+9.6}_{-5.27}$ &  $101.59^{+12.96}_{-9.56}$ \\
 $\mathcal{M}/M_\odot$ &   $21.85$ & $27.76^{+3.34}_{-1.88}$ &  $29.3^{+3.74}_{-3.14}$   \\
 $1/q$ & $0.167 $ & $0.27^{+0.12}_{-0.1}$ & $0.17^{+0.05}_{-0.04}$  \\
 $\chi_{\text{eff}}$ & $-0.437$ & $-0.6^{+0.16}_{-0.17}$ & $-0.26^{+0.18}_{-0.17}$ \\
 $\chi_{\text{p}}$ & $0.779$ & $0.19^{+0.17}_{-0.08}$ & $0.74^{+0.16}_{-0.19}$ \\
 $\theta_{1}$ & $2.11$ & $3.01^{+0.09}_{-0.21}$ & $1.97^{+0.25}_{-0.25}$ \\
 $\theta_{2}$ & $2.46$ & $1.5^{+0.57}_{-0.56}$ & $1.4^{+1.11}_{-0.93}$  \\
$\theta_{\text{JN}}$ & $1.28$ & $0.81^{+0.43}_{-0.28}$ & $0.46^{+0.21}_{-0.22}$  \\
 $d_{L}$ & $1200$ & $1444^{+223}_{-237}$ &  $1374^{+325}_{-248}$ \\
$\phi_{\text{ref}}$ & $1.2$ & $3.66^{+1.19}_{-1.13}$ &  $3.01^{+2.86}_{-2.61}$ \\
 $\psi$ & $0.7$ & $2.4^{+0.52}_{-1.79}$ &  $0.89^{+0.84}_{-0.57}$ \\
 $\rho^{\mathrm{H1}}_{\mathrm{mf}}$  & $13.92$ & $13.55^{+0.1}_{-0.19}$ & $13.68^{+0.09}_{-0.16}$  \\
 $\rho^{\mathrm{L1}}_{\mathrm{mf}}$  & $16.03$ & $15.61^{+0.11}_{-0.2}$ & $15.75^{+0.1}_{-0.17}$  \\
 $\rho^{\mathrm{V1}}_{\mathrm{mf}}$  & $6.66$ & $6.47^{+0.09}_{-0.28}$ & $6.52^{+0.06}_{-0.23}$  \\ $\log {\mathcal{BF}}$  &  & $194.33 \pm 0.19$ & $205.65 \pm 0.18$  \\[0.1cm]
 \hline
 \hline
    \end{tabular}
 \caption{Injected and median values of the posterior distributions for the synthetic NR injection, corresponding to the NR simulation {\tt SXS:BBH:0165} of the public SXS catalog, recovered  with \xphm~and \seobfivephm. The binary parameters correspond to the total mass $M$, chirp mass $\mathcal{M}$, mass ratio $q$, effective spin parameter  $\chi_{\text{eff}}$, effective precessing-spin parameter  $\chi_{\text{p}}$, tilt angles  $\theta_{1,2}$, angle between the total angular momentum and the line of sight $\theta_{\text{JN}}$, luminosity distance $d_{L}$, coalescence phase $\phi_{\text{ref}}$,  polarization angle $\psi$,  matched-filtered SNR for LIGO-Hanford/Livingston and Virgo detectors  $\rho^{\mathrm{H1},\mathrm{L1},\mathrm{V1}}_{\mathrm{mf}}$, and signal-versus-noise log Bayes factor $\log {\mathcal{BF}}$.}
 \label{tab:injection_settings}
\end{table}

\subsection{Real events}

In this section we re-analyze 6 GW events recorded by the LIGO and Virgo detectors \cite{LIGOScientific:2018mvr,LIGOScientific:2021usb,LIGOScientific:2021djp}: GW150914, GW190412, GW190521, GW190814, GW191109 and GW200129. We employ strain data from the Gravitational Wave Open Source Catalog (GWOSC) \cite{LIGOScientific:2019lzm} and the released PSD and calibration envelopes included in the Gravitational Wave Transient Catalogs GWTC-2.1 \cite{LIGOScientific:2021usb} and GWTC-3 \cite{LIGOScientific:2021djp},  and their respective parameter-estimation samples releases.

We perform the analysis using the parameter-estimation code \texttt{Bilby}\footnote{In this paper we employ the \texttt{Bilby} code from the public repository \url{https://git.ligo.org/lscsoft/bilby} with the git hash \texttt{507d93c8950e7f62cd5ff5792aab6cdf2d76d21f}, which correspond to the version \texttt{2.0.1}.} \cite{Ashton:2018jfp}, and the nested sampler \texttt{dynesty} \cite{Speagle:2019ivv} using the \texttt{acceptance-walk} method, which is well-suited for executing on a multicore single-computing node\footnote{See \url{https://lscsoft.docs.ligo.org/bilby/dynesty-guide.html} for details on the  \texttt{acceptance-walk} method.}, and we perform the run for GW190521 with the parameter-estimation code \texttt{parallel Bilby}\footnote{In this paper we employ the \texttt{parallel Bilby} code from the public repository \url{https://git.ligo.org/lscsoft/parallel_bilby} with the git hash \texttt{97df49f75ef5f240164e5fc44b6074c33e694a35}, which correspond to the version \texttt{1.1.0}.} \cite{Smith:2019ucc} as the nested sampler settings for this event are more expensive and the parallelization of this code ensures results in a short timescale. Both \texttt{Bilby} and  \texttt{parallel Bilby} employ the nested sampler \texttt{dynestey} \cite{Speagle:2019ivv}. The list of parameter-estimation runs and the main settings are specified in Table \ref{tab:peruntime}, together with the runtime and the number of cores employed. We find that results can be obtained using \texttt{Bilby} on just one computing node within days.

\begin{table}[h!]
\centering
\begin{tabular}{l   c c c c c c }
 \hline
 \hline
  \makecell[cc]{GW event \\ sampler} & \multicolumn{2}{c}{\makecell[cc]{Data \\ settings}} &\multicolumn{2}{c}{\makecell[cc]{Sampler \\ settings}} & \makecell[cc]{Computing \\ resources} & Runtime   \\
 \hline
 &  \makecell[cc]{srate  \\ (Hz)} & \makecell[cc]{seglen\\(s)} & \makecell[cc]{naccept/\\nact} & nlive & cores$\times$nodes &   \\
 \hline
 \makecell[cc]{GW150914 \\ \texttt{Bilby}} & 2048  & 8 & 60 & 1000 & $64\times 1$ & 1d 17h\\
 \makecell[cc]{GW190412 \\ \texttt{Bilby}} &  4096 & 8 & 60 & 1000 & $64\times 1$ & 4d 3h\\
 \makecell[cc]{GW190521 \\ \texttt{Bilby}} &  2048 & 8 & 60 & 1000 & $64\times 1$ & 1d 17h \\
 \makecell[cc]{GW190521 \\ \texttt{parallel Bilby}} &  2048 & 8 & 30 & 8192 & $64\times 8$ & 3d 4h \\
 \makecell[cc]{GW190814 \\ \texttt{Bilby}} &  4096 & 32 & 60 & 1000 & $64\times 1$& 5d 23h\\
 \makecell[cc]{GW191109 \\ \texttt{Bilby}} &  1024 & 8 & 60 & 1000 & $64\times 1$& 2d 1h\\
 \makecell[cc]{GW200129 \\ \texttt{Bilby}} & 2048 & 8 & 60 & 1000 & $64\times 1$& 2d 21h\\
 \hline
 \hline
\end{tabular}
\caption{Settings and evaluation time for the different parameter estimation runs on real GW events with the \seobfivephm~model. Sampling rate ($\text{srate}$) and data segment duration ($\text{seglen}$) are specified in the data settings, while the number of accepted MCMC-chains $\text{naccept}$ for bi and number of live points $\text{nlive}$ are specified in the sampler settings (for the GW190521 \texttt{parallel Bilby} run, the number quoted is the number of auto-correlation times). The time reported is walltime, while the total computational cost in CPU hours can be obtained multiplying this time by the reported number of CPU cores employed.}
\label{tab:peruntime}
\end{table}

\begin{figure*}[htpb!]
\includegraphics[width=0.65\columnwidth]{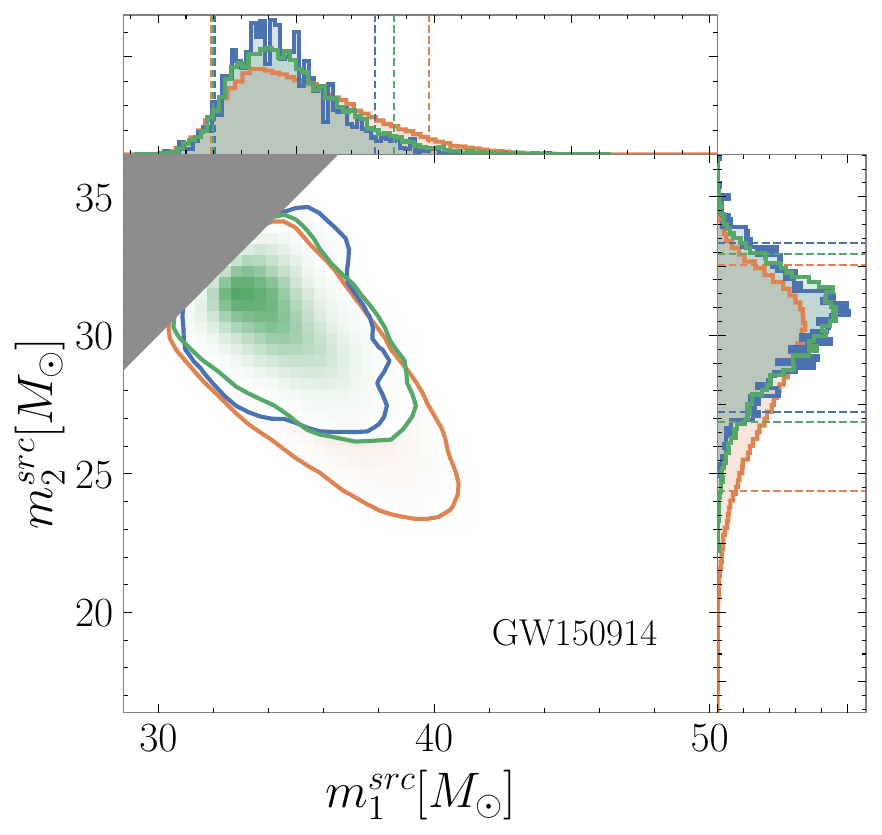}
\includegraphics[width=0.65\columnwidth]{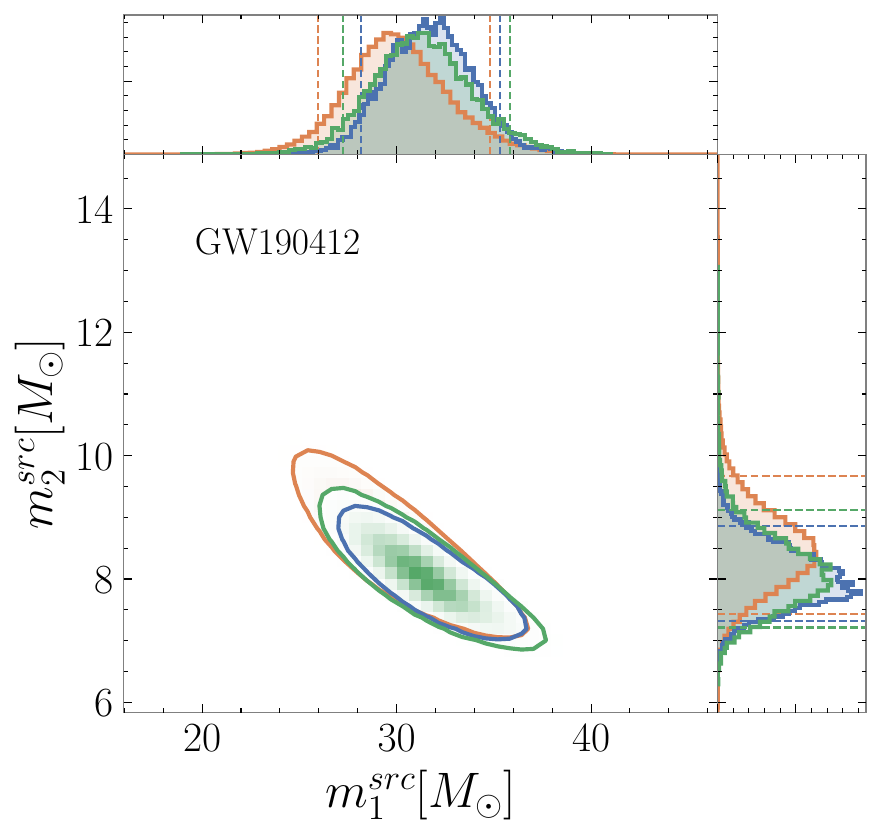}
\includegraphics[width=0.65\columnwidth]{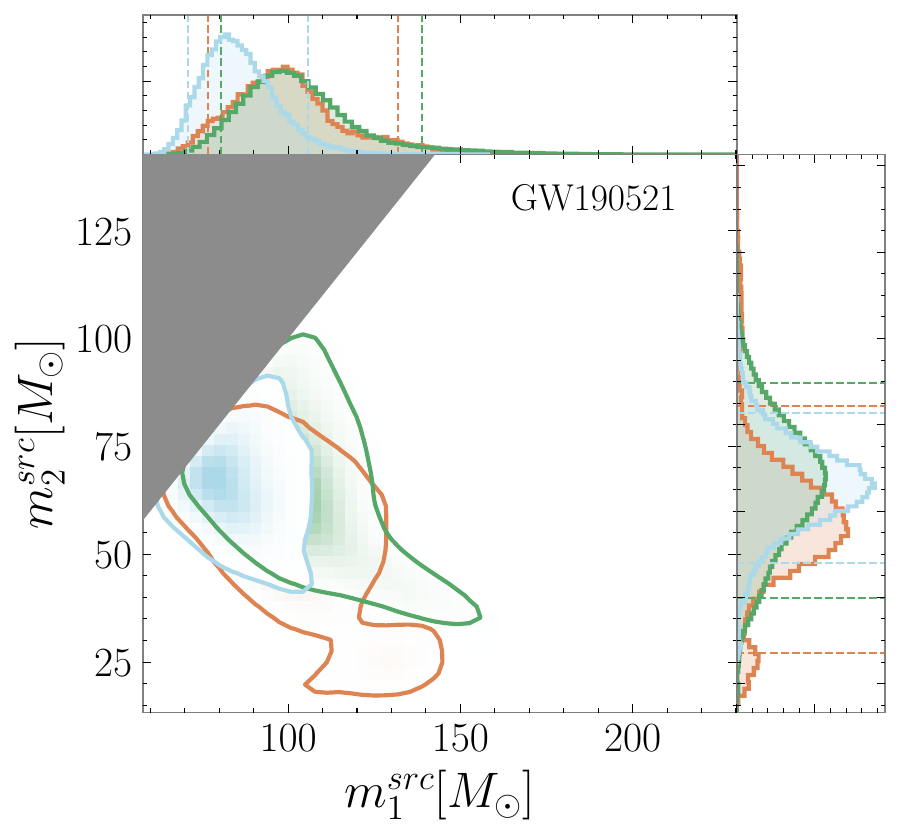}\\
\includegraphics[width=0.65\columnwidth]{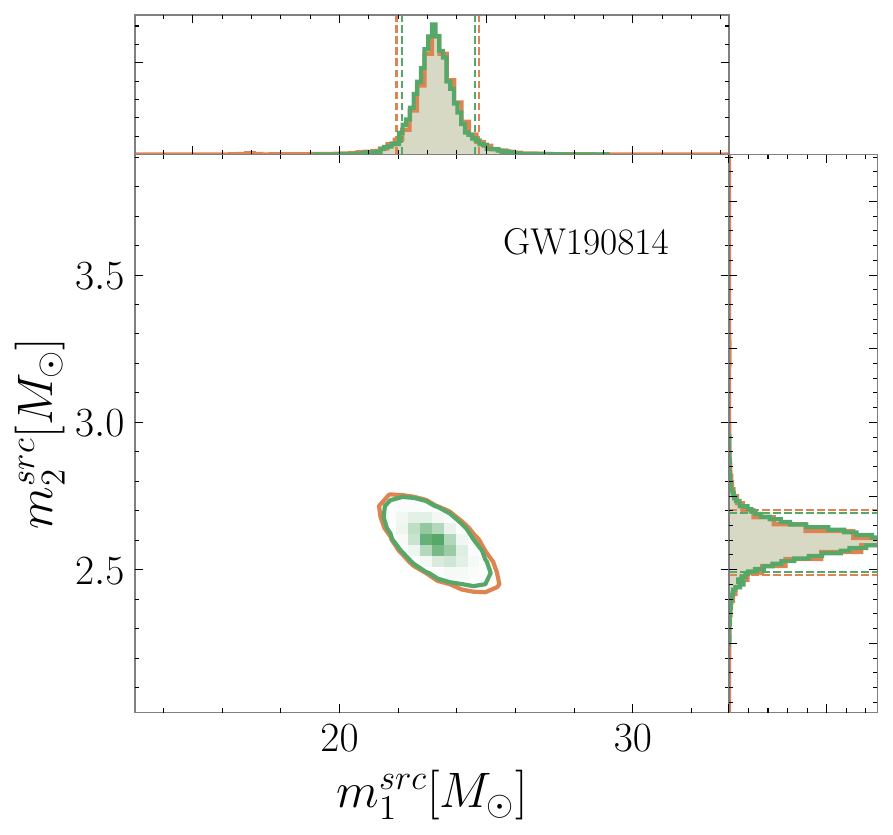}
\includegraphics[width=0.65\columnwidth]{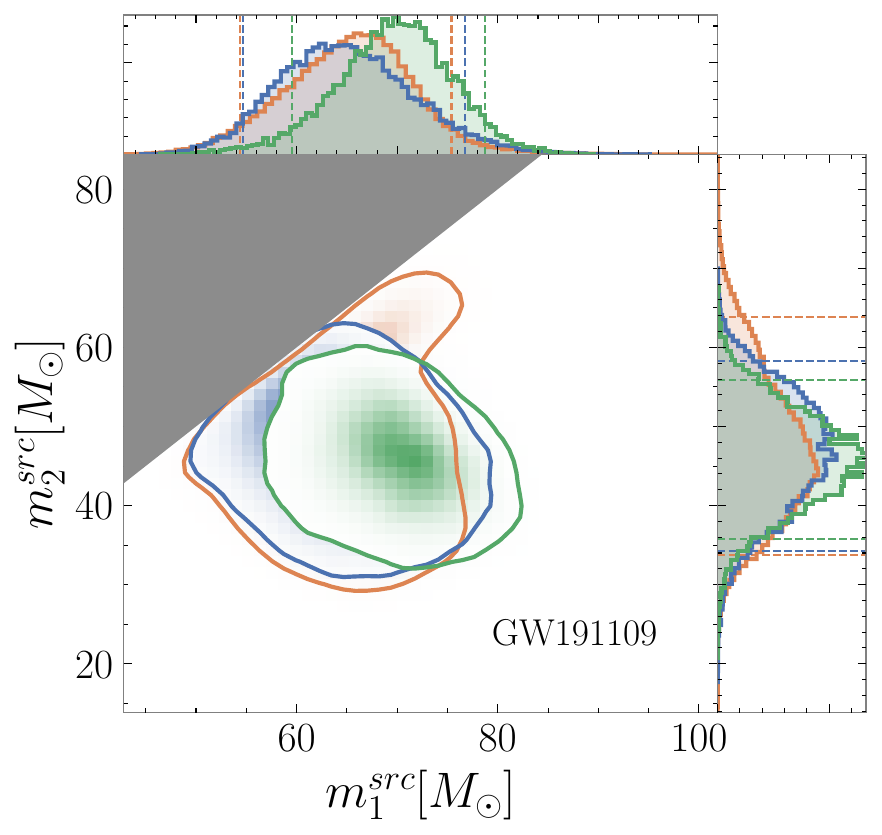}
\includegraphics[width=0.65\columnwidth]{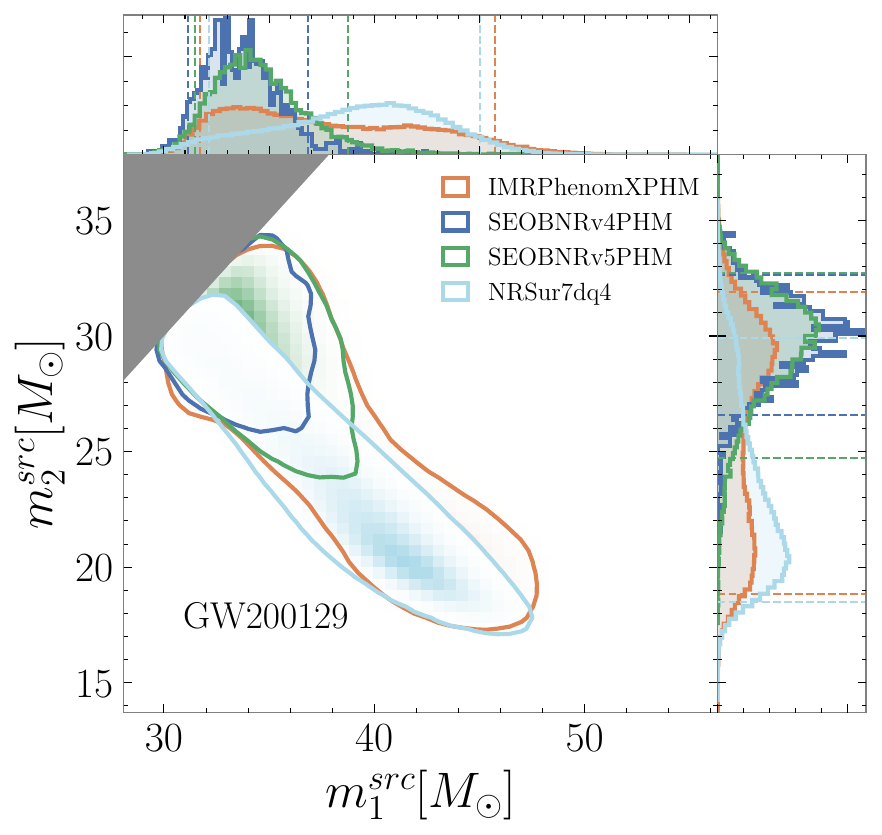}
 \caption{Component masses in the source-frame inferred for the real GW events re-analysed with \seobfivephm. Comparisons are presented with \seobfourphm~(when available) and \xphm~from GWTC-2.1 \cite{LIGOScientific:2021usb} and GWTC-3 \cite{LIGOScientific:2021djp} catalogs, except for GW190412 for which we present the \seobfourphm~from the discovery paper \cite{LIGOScientific:2020stg}, since the convergence of the posteriors is larger than in the GWTC-2.1 catalog. For GW190521 and GW200129 we include the posterior samples of the \texttt{NRSur7dq4} model from \cite{LIGOScientific:2020iuh} and \cite{Hannam:2021pit}, respectively. }
\label{fig:PEsourcemass}
\end{figure*}
\begin{figure*}[htpb!]
\includegraphics[width=0.65\columnwidth]{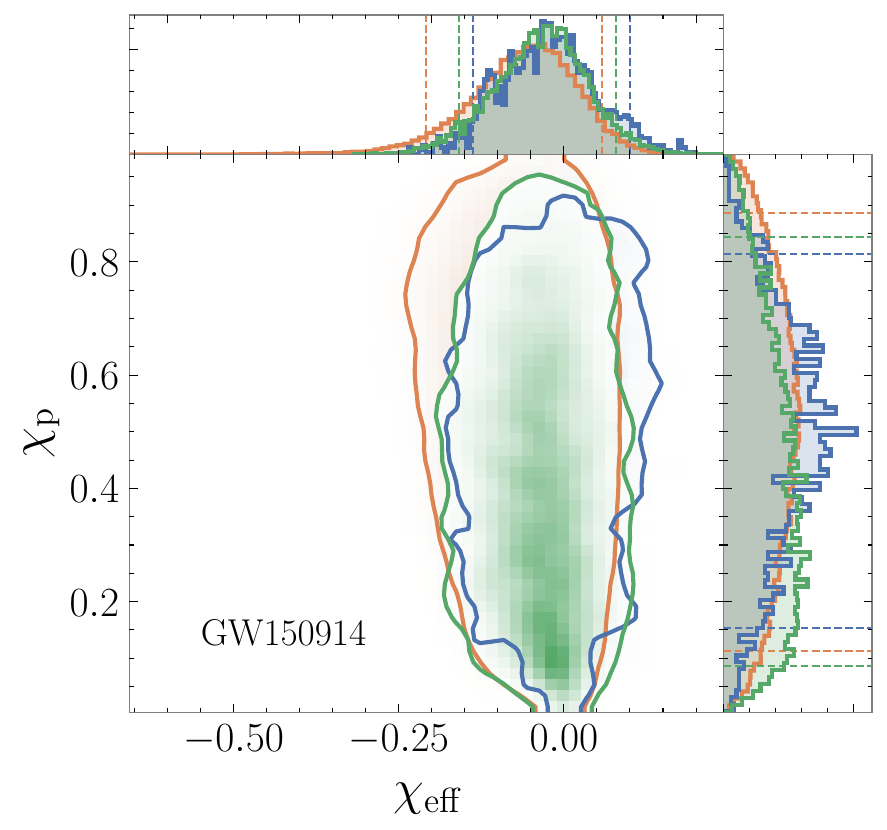}
\includegraphics[width=0.65\columnwidth]{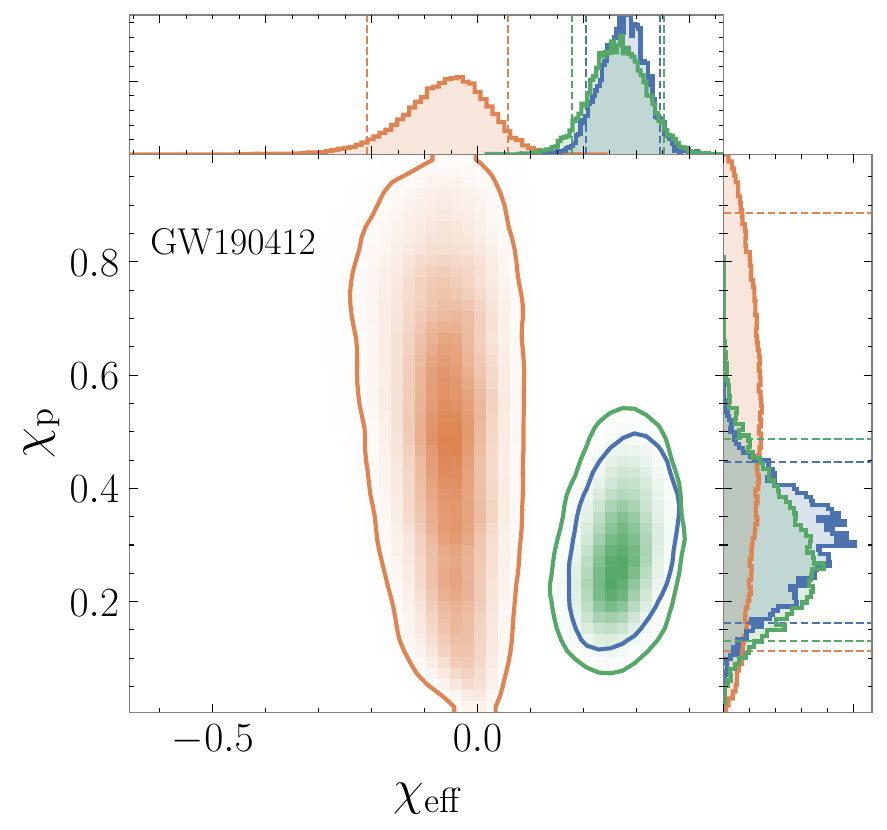}
\includegraphics[width=0.65\columnwidth]{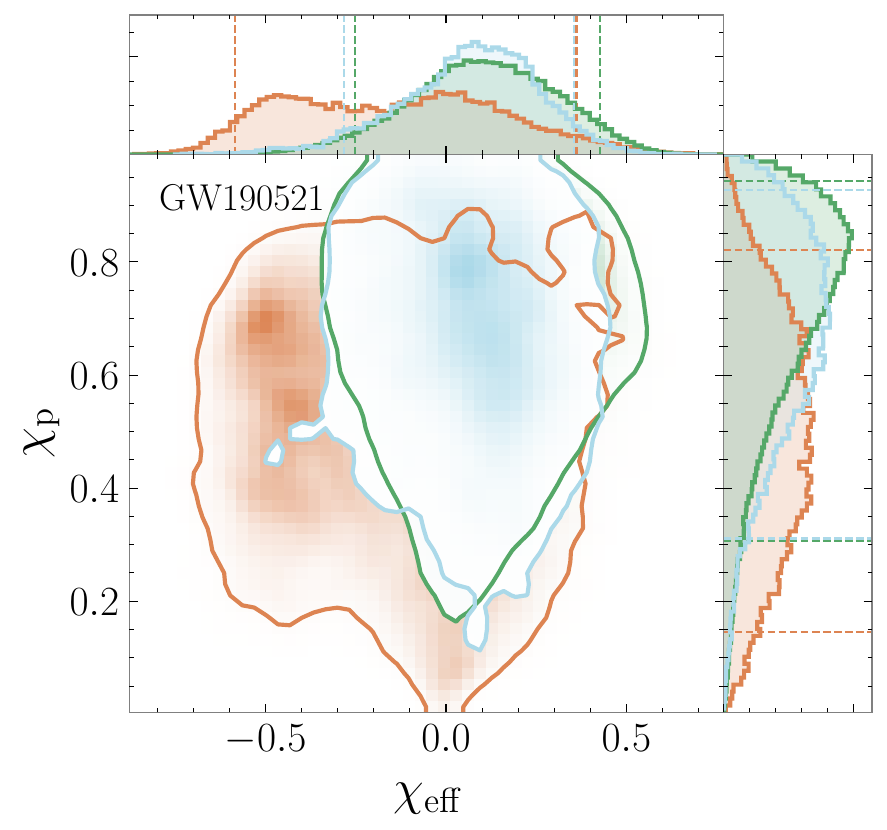}\\
\includegraphics[width=0.65\columnwidth]{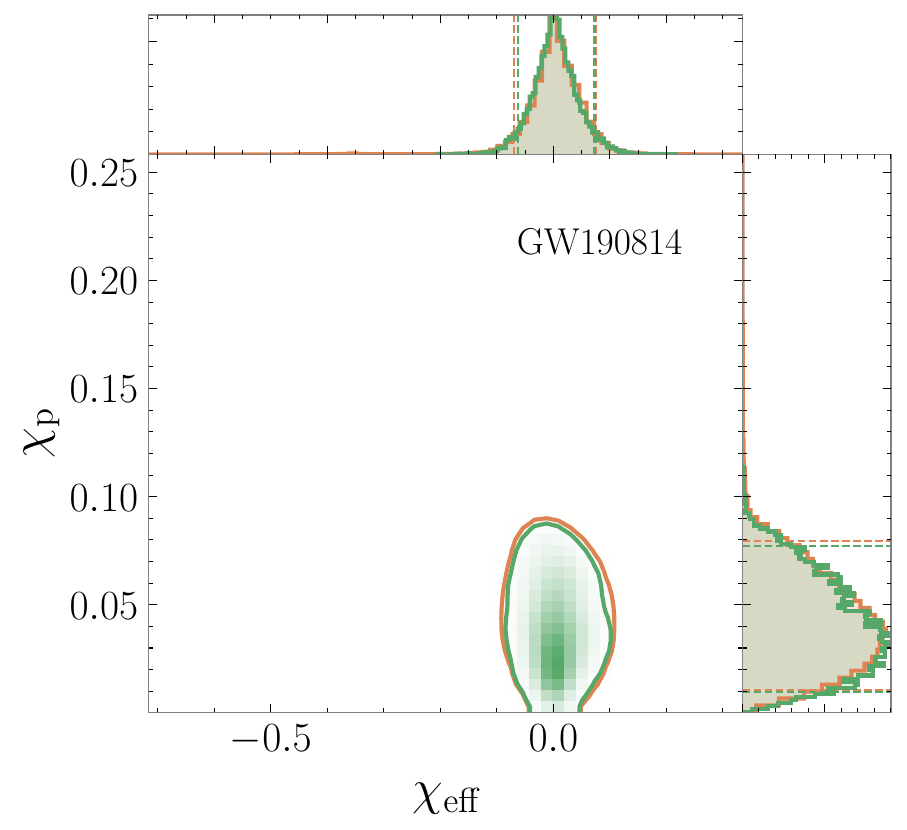}
\includegraphics[width=0.65\columnwidth]{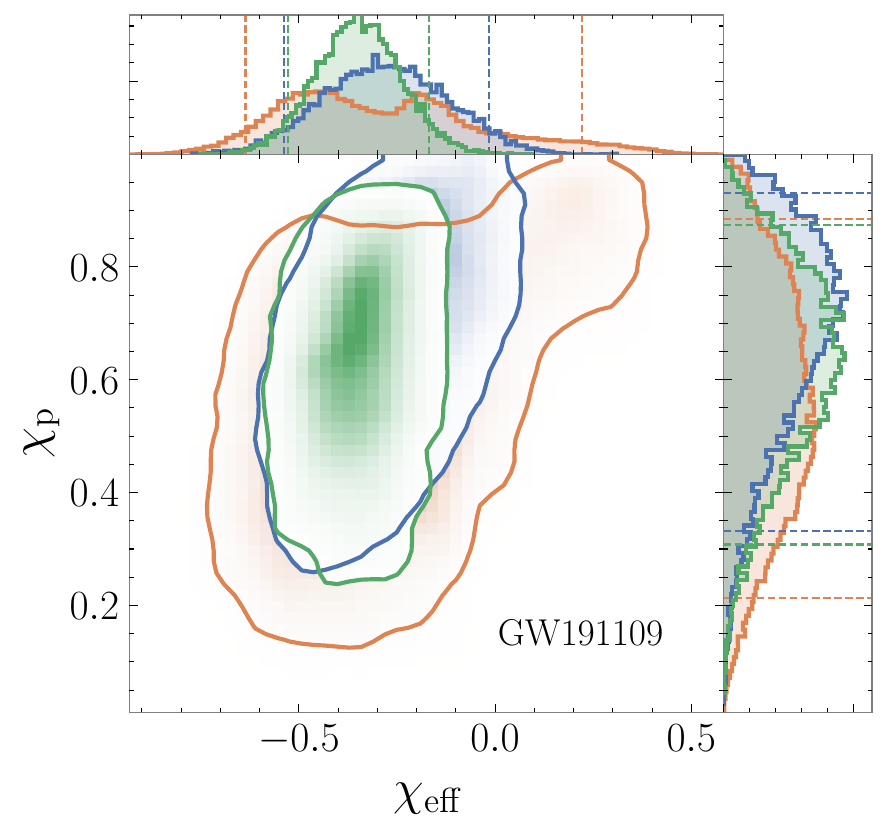}
\includegraphics[width=0.65\columnwidth]{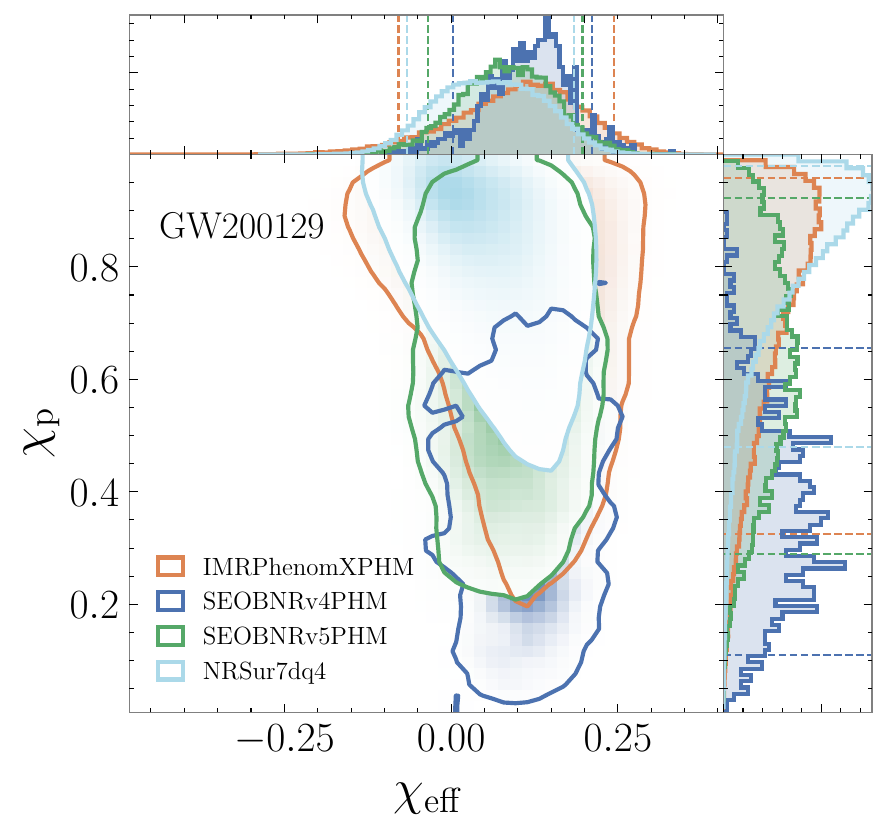}
 \caption{Effective-spin parameters $\chi_{\text{eff}}$ and $\chi_{\text{p}}$ inferred for the GW events re-analysed with \seobfivephm. Comparisons are presented with \seobfourphm~(when available) and \xphm~from GWTC-2.1 \cite{LIGOScientific:2021usb} and GWTC-3 \cite{LIGOScientific:2021djp} catalogs, except for GW190412 for which we present the \seobfourphm~from the discovery paper \cite{LIGOScientific:2020stg} as in Fig. \ref{fig:PEsourcemass}. For GW190521 and GW200129 we include the posterior samples of the \texttt{NRSur7dq4} model from \cite{LIGOScientific:2020iuh} and \cite{Hannam:2021pit}, respectively.}
\label{fig:PEspins}
\end{figure*}

In Figure \ref{fig:PEsourcemass} we summarize the results for the source component masses for the 6 re-analyzed events with \seobfivephm~and we compare with results from the \xphm~model released in GWTC-2.1 and the previous generation {\tt SEOBNR} model \seobfourphm~(when available) also from  GWTC-2.1 (obtained with the parameter-estimation code \texttt{RIFT} \cite{Pankow:2015cra,Lange:2018pyp}), except for the event GW190412 in which we show the \seobfourphm~results from the discovery paper \cite{LIGOScientific:2020stg} (obtained with \texttt{parallel Bilby}) due to a better convergence of the posteriors than in the GWTC-2.1 catalog \cite{LIGOScientific:2021usb}. Similarly, in Figure \ref{fig:PEspins} we summarize the results for the effective spin parameters $\chi_{\text{eff}}$ and $\chi_{\text{p}}$. In general, we observe broad consistency between our results and the GWTC results, but differences are stronger in some of the events, with \xphm~being, in general, more in tension with our results than \seobfourphm.

For GW150914 we observe good consistency between the \seobfivephm~and \seobfourphm~models, however the source mass posteriors are less constrained for \xphm.

For GW190412, the first confident mass-asymmetric event reported by the LIGO-Virgo collaboration \cite{LIGOScientific:2020stg}, we observe a better agreement between the time-domain models \seobfivephm~and \seobfourphm, which also are consistent with results from the phenomenological time-domain model \tphm~from Ref.~\cite{Colleoni:2020tgc}. For this event, the higher-mode content is important, and the more accurate precessing dynamics provides a more reliable multipolar structure of the waveforms, therefore the tension with \xphm in the recovery of $\chi_p$ (see Fig. \ref{fig:PEspins}) can be explained by the fact that the precessing description contains more approximations in this model.

\begin{figure}[htpb!]
\includegraphics[width=\columnwidth]{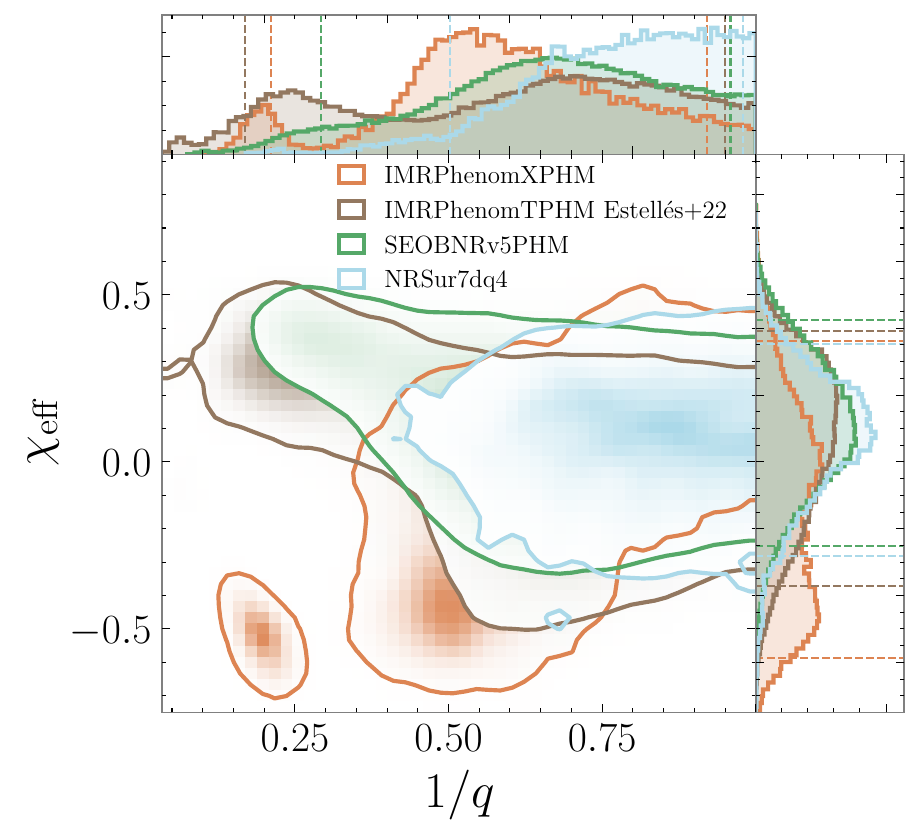}
\caption{Comparison of mass-ratio and effective spin parameter inferred for GW190521 between \seobfivephm, the phenomenological models \tphm~from Ref.~\cite{Estelles:2021jnz} and \xphm~from GWTC-2.1 \cite{LIGOScientific:2021usb} and the samples of the \texttt{NRSur7dq4} model from \cite{LIGOScientific:2020iuh}.}
\label{fig:gw190521}
\end{figure}

The GW190521 signal is particularly interesting, with only 4 cycles in band in the detectors,  thus being consistent with a merger-ringdown dominated signal. It  has been attributed to a variety of physical systems, from eccentric binaries \cite{Gayathri:2020coq,Romero-Shaw:2020thy}, non-spinning hyperbolic capture  \cite{Gamba:2021gap} and head-on collision of exotic compact objects \cite{CalderonBustillo:2020fyi}. Under the conservative assumption of a quasi-circular binary system, we observe differences with respect to the \xphm~results from GWTC-2.1. We have compared our results with the re-analysis of Ref.~\cite{Estelles:2021jnz} in which the phenomenological time-domain model \tphm~was employed using \texttt{LALInference MCMC} \cite{Veitch:2014wba}, and in Fig.~\ref{fig:gw190521} we present the 2D distribution of mass-ratio and effective spin $\chi_\text{eff}$. 
We also include the posterior distribution for the \texttt{NRSur7dq4} model from \cite{LIGOScientific:2020iuh}. 
We observe a better agreement of the \seobfivephm~distributions with the time-domain models \texttt{NRSur7dq4} and \tphm, in particular the mass asymmetric support for the posterior is correlated with positive effective spin, instead of negative effective spin as the results from {\xphm} suggest. The reason for the tension with \xphm~can be explained by the fact that this Fourier-domain model lacks a description of the effective precessing motion of the ringdown signal, which is present in the \texttt{NRSur7dq4} model and in an approximate way in \seobfivephm~and \tphm. We also note that the measurement of the effective precessing-spin parameter $\chi_p$ by \seobfivephm~is more consistent with the result obtained by the \texttt{NRSur7dq4} model, which is calibrated to precessing-spin NR simulations, while the lack of a secondary in the inverse mass ratio posterior in the \texttt{NRSur7dq4} model can be explained by the mass ratio prior used $1/q\in [0.17,1]$, which is restricted to region where the \texttt{NRSur7dq4} model can be generated, while for \seobfivephm~a wider prior $1/q \in [0.05,1]$ is used consistent with other analysis in the literature~\cite{Estelles:2021jnz}.


The next event we re-analyze is GW190814, a computationally challenging signal due to its low chirp mass and high-mass asymmetry, compatible with a heavy neutron star black-hole  system. For this event we find very good agreement between the \xphm~results from GWTC-2.1 and our results, in essentially all the parameters. The good agreement can be explained by the fact that this signal is consistent with a non-spinning configuration, and in the small spin-magnitude region the systematics between models is less severe, due to the underlying calibration of the non-precessing baselines. It is worth noting that the result for this event can be obtained within days with \seobfivephm~employing \texttt{Bilby} (see Table \ref{tab:peruntime} for details).

We also re-analyze GW191109, an interesting signal with support for \emph{negative} effective spin and non-negligible in-plane spin. For this event, we observe a slightly better consistency for the source component masses between \seobfourphm~and \xphm, although the spin distribution is more consistent between \seobfourphm~and \seobfivephm. Note that that the \xphm~results present multimodality in some parameters, like the effective spin parameter $\chi_{\rm eff}$, while this feature is not present both in the \seobfourphm~and \seobfivephm~results, therefore the more accurate modeling of the precessing dynamics could help in solving this degeneracy. Another interesting feature is that \seobfivephm~seems to  produce more constrained parameters than the other two models.

The last event we re-analyze is GW200129, which has been-claimed to be the first confident precessing-spin detection \cite{Hannam:2021pit} (although there are some concerns with data quality issues and glitch substraction that were discussed in Ref.~\cite{Payne:2022spz}). Our results do not recover a high support for high precessing spin values as the \texttt{NRSur7dq4} results from \cite{Hannam:2021pit} show (see Fig. \ref{fig:PEspins}), although the support is greater in \seobfivephm~than in \seobfourphm~results. Also the \seobfivephm~results for the source masses in Fig. \ref{fig:PEsourcemass}
prefer a region of low probability of the \texttt{NRSur7dq4} posteriors, while the \xphm~results show a bimodality in the posteriors. These differences in the posteriors can be explained by possible waveform systematics between the different state-of-the-art precessing-spin models for this particular event, as well as possible systematics in the glitch subtraction methods, which may impact differently the measurement of the binary parameters depending on the waveform model employed as pointed in Ref. \cite{Payne:2022spz}. 

\begin{figure*}[htpb!]
\includegraphics[scale=0.7]{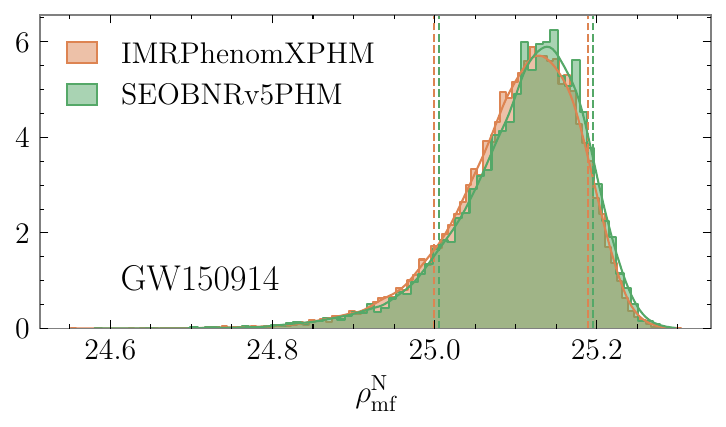}
\includegraphics[scale=0.7]{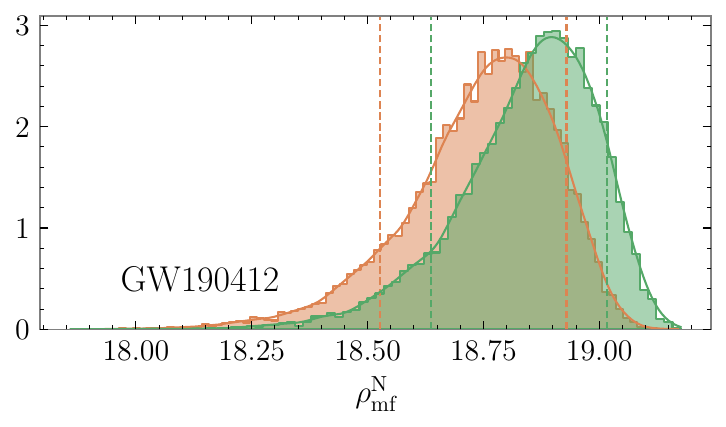}\\
\includegraphics[scale=0.7]{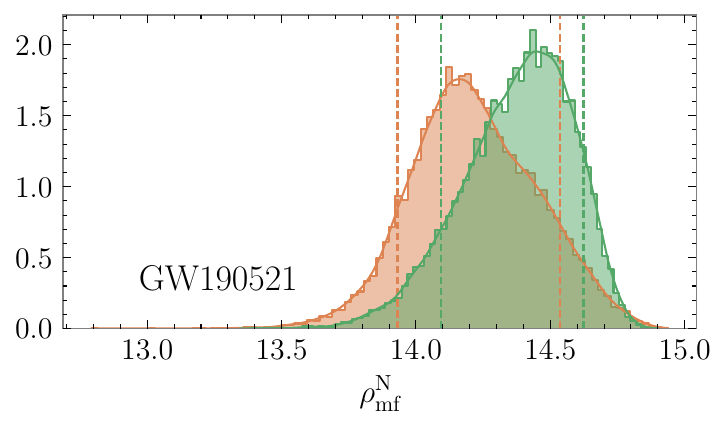}
\includegraphics[scale=0.7]{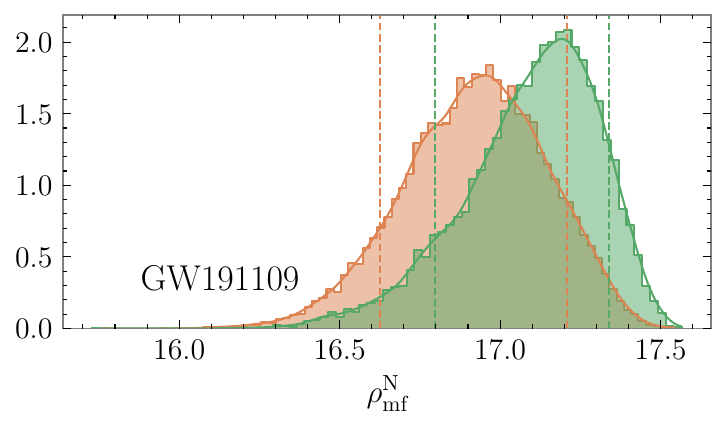}
 \caption{Distribution of network matched-filter SNR inferred for some of the GW events re-analysed with \seobfivephm. Comparisons are presented with \xphm~results obtained with the same settings and data as the \seobfivephm~results.}.
\label{fig:PEsnr}
\end{figure*}

Finally, in Fig.~\ref{fig:PEsnr}, we present the posterior distribution of the network matched-filter SNR $\rho^{\text{N}}_{\text{mf}}$ for some of the events, computed from the results of \seobfivephm, as well as \xphm~that we obtain running this model with the same settings as \seobfivephm. We can observe that in general greater SNR values are recovered with \seobfivephm, in particular for the events that show higher support for precession. This is likely due to the better description of the precessing dynamics included in \seobfivephm, as well as the modeling of the precessing ringdown, which is absent in the Fourier-domain model \xphm. This, together with the differences we have observed in the parameter posteriors,  emphasizes  the importance of using several accurate models such as \seobfivephm~for production analysis of GW events.

\section{Conclusions}\label{sec:conclusions}
In this paper we have developed and validated the multipolar precessing-spin \seobfivephm~model, of the fifth generation of \texttt{SEOBNR} models. This work is the culmination of a series of papers developing the \texttt{SEOBNRv5} models ahead of the fourth observing of the LVK Collaboration.

The \texttt{SEOBNRv5} models are built upon the most recent analytical PN results and improved resummations for the Hamiltonian \cite{Balmelli:2015lva,Khalil:2020mmr,Khalilv5}, the RR force and waveform modes \cite{Henry:2022ccf,Henry:2022dzx}, including information from second-order gravitational self-force \cite{Pound:2019lzj,VandeMeentv5} in the modes/RR force. The new analytical information and improvements in the conservative dynamics are derived in Ref.~\cite{Khalilv5}, while the inclusion of second order self-force results in the RR force and modes of \texttt{SEOBNRv5} is obtained in  Ref.~\cite{VandeMeentv5}. All these new analytical improvements are combined with input from NR waveforms to improve the calibration of the non-precessing \seobfivehm~model in Ref.~\cite{Pompiliv5}. The NR calibration in the aligned-spin sector is extended to 442 NR waveforms, in addition to 13 Teukolsky waveforms. The multipolar \seobfivehm~model includes the $(2,2),(2,1),(3,3),(4,4),(5,5)$ plus the $(3,2),(4,3)$ modes for which the mode-mixing during ringdown is modelled, and it improves substantially the accuracy of the \texttt{SEOBNR} family against non-precessing NR waveforms \cite{Pompiliv5}.

This modeling effort is developed within a new Python infrastructure \texttt{pySEOBNR} \cite{Mihaylovv5}, which offers more flexibility in including new analytical information, it is highly modular and it produces faster and more efficient \texttt{SEOBNR} models than the current ones in \texttt{LALSuite} \cite{lalsuite}.

More specifically, regarding the \seobfivephm~model developed here, following previous precessing \texttt{SEOBNR} models \cite{Taracchini:2013rva,Ossokine:2020kjp}, we have built such a model twisting up the non-precessing waveforms of \seobfivehm~\cite{Pompiliv5} from the co-precessing frame~\cite{Buonanno:2002fy,Schmidt:2010it,Boyle:2011gg,OShaughnessy:2011pmr,Schmidt:2012rh} to the inertial frame. With respect to the previous \texttt{SEOBNR} model, \seobfourphm~\cite{Ossokine:2020kjp}, which has been used in LVK data analysis \cite{LIGOScientific:2021usb,LIGOScientific:2021djp}, the new model: 1) does not evolve the EOB equations for the spins, but building on previous works \cite{Estelles:2020osj,Akcay:2020qrj, Gamba:2021ydi} decouples the spin evolution equations from the evolution of the orbital dynamics allowing for the specification of a reference frequency distinct from the starting frequency of the evolution, 2) employs PN-expanded EOB spin evolution equations derived from the generic {\tt SEOBNRv5} Hamiltonian in an orbit-average approximation \cite{Khalilv5},  3) evolves the conservative dynamics using a partially precessing Hamiltonian, $H^{\rm pprec}_{\rm EOB}$, which includes in-plane spin terms in an orbit average and reduces to the \seobfivehm~Hamiltonian in the aligned-spin limit, 4) employs a more accurate aligned-spin two-body dynamics, since in the non-precessing limit it reduces to \seobfivehm, 5) includes in the co-precessing frame two new modes $(3,\pm 2)$ and $(4,\pm 3)$, instead of only the $(2,\pm 2),(2,\pm 1),(3,\pm 3), (4,\pm 4), (5, \pm 5)$, 6) applies the PA scheme~\cite{Nagar:2018gnk} to the EOB orbital evolution, which increases the efficiency of the model, 7) implements an efficient calculation of the polarizations based on the rotation of the basis of -2 spin-weighted spherical harmonics, which further accelerates the evaluation of the model, and 8) incorporates latest insights from NR waveforms by properly rotating the quasi-normal mode frequencies \cite{Hamilton:2023znn}.

The improvement in accuracy between \seobfivephm~and \seobfourphm~is evident from Figure \ref{fig:histogramMax_public118}, where we have compared these models, as well as other state-of-the-art precessing-spin models (\xphm, \tphm~and \teob) to the public SXS catalog of 1425 precessing-spin NR waveforms, and the 118 SXS NR waveforms from Ref.~\cite{Ossokine:2020kjp}. When comparing to the highly precessing 118 simulations from Ref.~\cite{Ossokine:2020kjp}, the \seobfivephm~provides the highest accuracy with respect to NR waveforms (see Fig. \ref{fig:spaghetti_eob}), as it includes effects of in-plane spin components in its dynamics, unlike \xphm, \tphm~and \teob, while having a more accurate description of the co-precessing waveforms through the non-precessing \seobfivehm~model than \seobfourphm. When turning to a broader comparison and including all the 1543 SXS precessing-spin NR waveforms available, we have found that for the \seobfivephm~model,   $99.8 \%$ ($84.4 \%$) of cases have a maximum unfaithfulness value, in the total mass range $[20,300]M_\odot$, below $3 \%$  ($1\%$). These numbers reduce to $95.3 \%$ ($60.8 \%$) for  \seobfourphm, to $83.3 \%$ $(44.9 \%)$ for  \teob, to $91.6\%$ $(62.4\%)$ for \tphm~and to $78.3\%$ $(38.3\%)$ for \xphm. We have also investigated the accuracy of the previous models, which are not calibrated to precessing-spin NR waveforms, against the NR surrogate \nrsur~model by computing the unfaithfulness for 5000 configurations in the parameter space of the calibration of the surrogate model ($q\in[1-4]$, and $a_{1,2}\in[0,0.8]$). The configurations have been uniformly distributed in the effective precessing-spin parameter, $\chi_p$, to increase the number of configurations with highly precessional effects. We have found in Fig. \ref{fig:unf_nrsur} that \seobfivephm~provides the lowest unfaithfulness against the surrogate model, with $100\%$ $(90.1\%)$ cases with maximum unfaithfulness, over the total mass range considered, below $3\%$ $(1\%)$, while these numbers reduce to $98.7\%$ $(79.5\%)$ for \seobfourphm, $89.4\%$ $(81.4\%)$ for \xphm~and $96.1\%$ $(66\%)$ for \teob.
The largest values of unfaithfulness against the surrogate model occur at high mass ratios and high values of the in-plane spin components, where the in-plane spin effects and mode asymmetries play an important role in the description of the waveforms.
We have also compared \seobfivephm~against \xphm~in a larger region of parameter space $q\in [1,20]$ and $\chi_p\in[0,0.99]$ outside the region of calibration of the underlying aligned-spin models. We have found that the largest differences occur at mass ratios larger than 4 and spin magnitudes larger than 0.8 (see Fig. \ref{fig:unf_models}). These results are consistent with the differences found in the comparisons of non-precessing models in Ref.~\cite{Pompiliv5}, and highlight the need to improve the parameter-space coverage of the NR waveforms combined with improved analytical information in the spinning sector, such as gravitational self-force, so that the accuracy of the models can be further improved  in these challenging regions of the parameter space.

The improvement in accuracy of the \seobfivephm~model is also accompanied by an improvement in the speed of the model with respect to \seobfourphm. The acceleration in waveform evaluation of the  model is a consequence of several factors: 1) its implementation in the high-performance \texttt{pySEOBNR} Python package \cite{Mihaylovv5}, which allows to incorporate new analytical information combined with NR calibration in a flexible, modular and efficient way, 2) the PA routine, which accelerates the evaluation of the two-body dynamics (see Appendix \ref{app:PA}), and 3) an efficient procedure to compute the polarizations as described in Sec. \ref{sec:polarizations}. As a result, we find that \seobfivephm~is overall $\sim 8-20$ times faster than \seobfourphm, and comparable in speed to other state-of-the-art time-domain precessing-spin models (\teob~and \tphm).

Given the high accuracy and computational efficiency of \seobfivephm, we have performed a Bayesian inference study on mock signals and real GW events detected by the
LVK Collaboration. We have first investigated how the modeling inaccuracy impacts the inference of parameters by injecting a synthetic NR signal into a network of LIGO-Virgo detectors at design sensitivity. We have injected in zero noise a precessing-spin NR waveform  (\texttt{SXS:BBH:0165}) with mass ratio 6,  total mass 95 $M_\odot$, SNR 19.4, inclination $0.69$ with respect to the line of sight, and recovered it with \seobfivephm~and \xphm. The unfaithfulness values of these models against the synthetic signal is $2\%$ for \seobfivephm~and $12\%$ for \xphm. The results are summarized in Fig. \ref{fig:pe_nrInj} and Table \ref{tab:injection_settings}. We have found that the recovery of the parameters with \seobfivephm~does not produce significant biases, except for the effective spin parameter, for which the injected value lies at the boundary of the $90\%$ credible intervals, while the rest of the binary parameters are accurately recovered. While in the case of the \xphm~model a $12\%$ value of unfaithfulness translates into larger biases in several parameters, like the component masses or the effective precessing-spin parameter.  A more comprehensive Bayesian inference study will be needed to quantify the modeling inaccuracies and systematics, and how they translate into biases in the inference of binary parameters. Here, new methods of Bayesian inference through machine learning techniques, like \texttt{DINGO}  \cite{Green:2020dnx,Dax:2021tsq,Dax:2022pxd}, may offer an alternative and efficient method to perform large-scale injection campaigns and assess waveform systematics with a significant reduction of its computational cost. We leave such waveform systematics studies using Bayesian inference methods for future work.

Besides injection studies, we have demonstrated that \seobfivephm~can be used as a standard tool in Bayesian inference studies of real GW events. We have reanalyzed several GW events (GW150914, GW190412, GW190521, GW190814, GW191109 and GW200129) detected by the LVK Collaboration in the first and third observing runs, with two different parameter estimation codes \texttt{serial Bilby} \cite{Ashton:2018jfp} and \texttt{parallel Bilby} \cite{Smith:2019ucc}. We have found that the parameters inferred by \seobfivephm~are consistent with the ones obtained in the literature for most of the events. For instance, in the case of the massive GW190521, we find consistency in the recovery of the mass ratio and effective spin parameter with other time-domain precessing models in the literature, while for GW200129, consistently with \seobfourphm, we do not find support for high precession as claimed in Ref.~\cite{Hannam:2021pit} using the \nrsur~model. Furthermore, we find that for all the events considered in this paper \seobfivephm~recovers systematically more SNR than the \xphm~model (see Fig. \ref{fig:PEsnr}). The \seobfivephm~model results have been obtained in a few days when using \texttt{parallel Bilby}, and on the order of a week when using \texttt{serial Bilby} (see Table \ref{tab:peruntime}). This makes \seobfivephm~a standard tool that can be used with a variety of stochastic samplers, and we plan in the future to extend the Bayesian inference study  presented here,  including the machine-learning code {\tt DINGO}, to all the GW events detected during the third-observing run \cite{Estellesv5}.

Finally, the \seobfivephm~model is not calibrated to precessing-spin NR waveforms,  which  limits its  accuracy. To overcome this limitation, calibration to NR waveforms in the conservative dynamics, as well in the waveform modes with the inclusion of mode asymmetries\footnote{Similarly as done in Refs. \cite{Varma:2019csw}.} will be developed in the future. In this context, the \texttt{pySEOBNR} infrastructure provides an ideal framework to incorporate such improvements,  as well as other physical effects, such as eccentricity and tidal effects, which have been already incorporated in \texttt{SEOBNRv4} models \cite{Ramos-Buades:2021adz,Hinderer:2016eia,Steinhoff:2016rfi,Lackey:2018zvw,Steinhoff:2021dsn,Matas:2020wab}, and that we are in the process of implementing in the \texttt{SEOBNRv5} models. Further improvements for the near future concern with the adoption of the \texttt{SEOBNRv5} models to perform theory agnostic tests of GR \cite{Ghosh:2021mrv,Mehta:2022pcn,Maggio:2022hre}, as well as developing \texttt{SEOBNR} waveforms in specific
beyond-GR theories and calibrating/comparing them to beyond-GR NR waveforms of BBHs \cite{Okounkova:2019zjf,Okounkova:2019dfo,Silva:2020omi,Okounkova:2022grv,Corman:2022xqg}.

\section*{Acknowledgments}\label{sec:acknowledgements}

It is our pleasure to thank Geraint Pratten, Stanislav Babak, Alice Bonino, Eleanor Hamilton, N.V. Krishnendu, Sylvain Marsat, Piero Rettegno and Riccardo Sturani for perfoming the LVK Collaboration internal review of the implementation of the \texttt{SEOBNRv5} models. We would like to thank Lucy Thomas for comments and suggestions to improve the manuscript.

Part of M.K.'s work on this paper is supported by Perimeter Institute for Theoretical Physics. Research at Perimeter Institute is supported in part by the Government of Canada through the Department of Innovation, Science and Economic Development and by the Province of Ontario through the Ministry of Colleges and Universities.
M.S. acknowledges  the Fulbright U.S. Student Program Study/Research Fellow and the Max Planck Institute for Gravitational Physics in Potsdam. The computational work for this manuscript was carried out on the computer clusters \texttt{Minerva} and \texttt{Hypatia} at the Max Planck Institute for Gravitational Physics in Potsdam.

\texttt{SEOBNRv5PHM} is publicly available through the python package \texttt{pySEOBNR} \href{https://git.ligo.org/waveforms/software/pyseobnr}{\texttt{git.ligo.org/waveforms/software/pyseobnr}}. Stable versions of \texttt{pySEOBNR} are published through the Python Package Index (PyPI), and can be installed via ~\texttt{pip install pyseobnr}.

This research has made use of data or software obtained from the Gravitational Wave Open Science Center (gwosc.org), a service of LIGO Laboratory, the LIGO Scientific Collaboration, the Virgo Collaboration, and KAGRA. LIGO Laboratory and Advanced LIGO are funded by the United States National Science Foundation (NSF) as well as the Science and Technology Facilities Council (STFC) of the United Kingdom, the Max-Planck-Society (MPS), and the State of Niedersachsen/Germany for support of the construction of Advanced LIGO and construction and operation of the GEO600 detector. Additional support for Advanced LIGO was provided by the Australian Research Council. Virgo is funded, through the European Gravitational Observatory (EGO), by the French Centre National de Recherche Scientifique (CNRS), the Italian Istituto Nazionale di Fisica Nucleare (INFN) and the Dutch Nikhef, with contributions by institutions from Belgium, Germany, Greece, Hungary, Ireland, Japan, Monaco, Poland, Portugal, Spain. KAGRA is supported by Ministry of Education, Culture, Sports, Science and Technology (MEXT), Japan Society for the Promotion of Science (JSPS) in Japan; National Research Foundation (NRF) and Ministry of Science and ICT (MSIT) in Korea; Academia Sinica (AS) and National Science and Technology Council (NSTC) in Taiwan.


\appendix

\section{Precessing-spin effective Hamiltonian}
\label{app:Ham}

In this Appendix, we provide the \textit{partial-precession} Hamiltonian derived in Ref.~\cite{Khalilv5}, which reduces to the Hamiltonian of \seobfivehm~\cite{Pompiliv5} in the aligned-spin limit and includes orbit-average in-plane spin components for quasi-circular orbits.
The effective Hamiltonian is given by
\begin{align}
\label{HeffAnzSimp}
H_\text{eff}^\text{pprec} &= \frac{M p_{\phi} \Lhat \cdot (g_{a_+} \bm{a}_+ + g_{a_-} \delta \bm{a}_-) + \text{SO}_\text{calib} + \left\langle G_{a^3}^\text{pprec}\right\rangle}{r^3+a_+^2 (r+2M)}  \nonumber\\
&\quad+
\bigg[A^\text{pprec} \,\bigg(
\mu^2 + B_p^\text{pprec} \frac{p_\phi^2}{r^2} + \left(1 + B_{np}^\text{pprec}\right) (\bm{n}\cdot\bm{p})^2  \nonumber\\
&\qquad\qquad
+ B_{npa}^\text{Kerr\,eq} \frac{p^2_\phi (\Lhat \cdot\bm{a}_+)^2}{r^2}  + Q^\text{pprec} \bigg)\bigg]^{1/2},
\end{align}
where the gyro-gravitomagnetic factors are the same as in the aligned-spin case, which are given by Eq.~(28) of Ref.~\cite{Khalilv5}, and the SO calibration term is given by
\begin{equation}
\text{SO}_\text{calib} = \nu d_\text{SO} \frac{M^4}{r^3} p_\phi \Lhat\cdot\bm{a}_+.
\end{equation}
with the same value of $d_\text{SO}$ as in the aligned-spin model~\cite{Pompiliv5}.
The cubic-in-spin term $\left\langle G_{a^3}^\text{pprec}\right\rangle$ reads
\begin{align}
\left\langle G_{a^3}^\text{pprec}\right\rangle &=  p_\phi \delta (\Lhat \cdot \bm{a}_-) \Bigg\lbrace
\frac{M}{r^2}\Bigg[
\frac{a_+^2}{4} - \frac{5}{24} (\bm{l}_N\cdot\bm{a}_+)^2
\Bigg]\nonumber\\
&\qquad
-\frac{p_\phi^2 }{8\mu^2 r^3}\left( a^2_+-(\bm{l}_N\cdot\bm{a}_+)^2\right)
\Bigg\rbrace \nonumber\\
&\quad
+ p_\phi (\Lhat \cdot \bm{a}_+) \Bigg\lbrace
\frac{p_\phi^2}{\mu^2r^3} \left[
\frac{\delta}{4} \left[\bm{a}_+\cdot \bm{a}_- - (\LNhat \cdot\bm{a}_+) (\LNhat \cdot\bm{a}_-)\right]\right. \nonumber\\
&\quad\qquad \left. -\frac{1}{8}\left( a^2_+-(\bm{l}_N\cdot\bm{a}_+)^2\right)
\right]\nonumber\\
&\qquad
+\frac{M}{r^2} \Bigg[
-\frac{5a_+^2}{8} + \frac{3}{8} (\bm{l}_N\cdot\bm{a}_+)^2
-\delta\frac{5}{8} (\bm{a}_+\cdot\bm{a}_-) \nonumber\\
&\quad\qquad
+\delta\frac{5}{6}  (\LNhat \cdot\bm{a}_+) (\LNhat \cdot\bm{a}_-)
\Bigg]
\Bigg\rbrace.
\end{align}
The potential $B_{npa}^\text{Kerr\,eq}$ in Eq.~\eqref{HeffAnzSimp} is the same as in the Kerr Hamiltonian for equatorial orbits, and is given by
\begin{equation}
B_{npa}^\text{Kerr\,eq} = -\frac{1+2M/r}{r^2+a_+^2 (1+2M/r)}.
\end{equation}
\onecolumngrid
The other potentials $A^\text{pprec}$, $B_p^\text{pprec}$, $B_{np}^\text{pprec}$, and $Q^\text{pprec}$ include nonspinning and SS PN terms, and read:
\begin{equation}
\begin{aligned}
A^\text{pprec} &= \frac{a_+^2/r^2+A_\text{noS}+A_\text{SS}^\text{prec} +\left\langle \tilde{A}_\text{SS}^\text{in\,plane}\right\rangle}{1+ (1+2M/r)a_+^2/r^2}, \\
B_p^\text{pprec} &= 1 + \left\langle \tilde{B}_{p,\text{SS}}^\text{in\,plane}\right\rangle, \\
B_{np}^\text{pprec} &= -1 + a_+^2/r^2 + A_\text{noS} \bar{D}_\text{noS} + B_{np,\text{SS}}^\text{prec}, \\
Q^\text{pprec} &= Q_\text{noS} + Q_\text{SS}^\text{prec},
\end{aligned}
\end{equation}
where the nonspinning contributions $A_\text{noS}$, $\bar{D}_\text{noS}$ and $Q_\text{noS}$ are given by Eqs.~(21)--(25) of Ref.~\cite{Khalilv5}, while the SS corrections read
%
\begin{subequations}
\begin{align}
A_\text{SS}^\text{prec} &= \frac{M^2}{r^4}\left[
\frac{9 a_+^2}{8}-\frac{5}{4} \delta \bm{a}_- \cdot \bm{a}_+ +a_-^2 \left(\frac{\nu }{2}+\frac{1}{8}\right)\right]
+\frac{M^3}{r^5}\left[
a_+^2 \left(-\frac{175 \nu }{64}-\frac{225}{64}\right)
+\delta \bm{a}_- \cdot\bm{a}_+ \left(\!\frac{117}{32}-\frac{39 \nu }{16}\right)
+a_-^2 \left(\!\frac{21 \nu ^2}{16}-\frac{81 \nu }{64}-\frac{9}{64}\right)\right], \\
B_{np,\text{SS}}^\text{prec} &= \frac{M}{r^3} \left[a_+^2 \left(3 \nu +\frac{45}{16}\right)-\frac{21}{8}\delta \bm{a}_- \cdot\bm{a}_+ +a_-^2 \left(\frac{3 \nu }{4}-\frac{3}{16}\right)\right] \nonumber\\
&\quad
+ \frac{M^2}{r^4} \left[
a_+^2 \left(-\frac{1171 \nu }{64}-\frac{861}{64}\right)
+\delta \bm{a}_- \cdot\bm{a}_+ \left(\frac{13 \nu }{16}+\frac{449}{32}\right)
+a_-^2 \left(\frac{\nu ^2}{16}+\frac{115 \nu }{64}-\frac{37}{64}\right)
\right], \\
Q_\text{SS}^\text{prec} &= \frac{Mp_r^4}{\mu^2r^3} \left[
a_+^2 \left(-5 \nu ^2+\frac{165 \nu }{32}+\frac{25}{32}\right)
+\delta \bm{a}_- \cdot\bm{a}_+ \left(\frac{45 \nu }{8}-\frac{5}{16}\right)
+a_-^2 \left(-\frac{15 \nu ^2}{8}+\frac{75 \nu }{32}-\frac{15}{32}\right)
\right],\\
\left\langle \tilde{A}_\text{SS}^\text{in\,plane}\right\rangle &= \frac{M}{r^3} \left[a_+^2 - (\LNhat \cdot\bm{a}_+)^2\right]
+ \frac{M^2}{r^4} \left\lbrace\frac{33}{16} \delta \left[\bm{a}_+\cdot \bm{a}_- - (\LNhat \cdot\bm{a}_+) (\LNhat \cdot\bm{a}_-)\right]
+ \left(-\frac{\nu }{4}-\frac{3}{16}\right) \left[a_-^2 - (\LNhat \cdot\bm{a}_-)^2\right]
+\left(\frac{7 \nu }{8}-\frac{31}{8}\right) \left[a_+^2 - (\LNhat \cdot\bm{a}_+)^2\right]\right\rbrace \nonumber\\
&\quad
+ \frac{M^3}{r^5} \Bigg\lbrace
\delta  \left(\frac{17}{2} \nu -\frac{1}{8}\right)\left[\bm{a}_+\cdot \bm{a}_- - (\LNhat \cdot\bm{a}_+) (\LNhat \cdot\bm{a}_-)\right]
+\left(-\frac{41 \nu ^2}{16}+\frac{583 \nu }{64}-\frac{171}{128}\right) \left[a_-^2 - (\LNhat \cdot\bm{a}_-)^2\right] \nonumber\\
&\quad\qquad
+\left(-\frac{11 \nu ^2}{16}+\frac{1435 \nu }{192}+\frac{187}{128}\right) \left[a_+^2 - (\LNhat \cdot\bm{a}_+)^2\right]\Bigg\rbrace, \\
\left\langle \tilde{B}_{p,\text{SS}}^\text{in\,plane}\right\rangle &= -\frac{a_+^2 - (\LNhat \cdot\bm{a}_+)^2}{2r^2}
+ \frac{M}{r^3} \left\lbrace
\frac{3}{8} \delta  \left[\bm{a}_+\cdot \bm{a}_- - (\LNhat \cdot\bm{a}_+) (\LNhat \cdot\bm{a}_-)\right]
+\left(\frac{3}{32}-\frac{3 \nu }{8}\right) \left[a_-^2 - (\LNhat \cdot\bm{a}_-)^2\right]
+\left(-\frac{7 \nu }{8}-\frac{15}{32}\right) \left[a_+^2 - (\LNhat \cdot\bm{a}_+)^2\right]\right\rbrace \nonumber\\
&\quad
+ \frac{M^2}{r^4} \Bigg\lbrace\delta  \left(-\frac{49 \nu }{8}-\frac{43}{16}\right) \left[\bm{a}_+\cdot \bm{a}_- - (\LNhat \cdot\bm{a}_+) (\LNhat \cdot\bm{a}_-)\right]
+\left(\frac{19 \nu ^2}{16}-\frac{545 \nu }{64}+\frac{219}{128}\right) \left[a_-^2 - (\LNhat \cdot\bm{a}_-)^2\right] \nonumber\\
&\quad\qquad
+\left(\frac{11 \nu ^2}{16}-\frac{805 \nu }{192}+\frac{125}{128}\right) \left[a_+^2 - (\LNhat \cdot\bm{a}_+)^2\right]\Bigg\rbrace,
\end{align}
\end{subequations}
%

\twocolumngrid
where $\left\langle \tilde{A}_\text{SS}^\text{in\,plane}\right\rangle$ and $\left\langle \tilde{B}_{p,\text{SS}}^\text{in\,plane}\right\rangle$ only contain in-plane spin components that have been orbit-averaged using~\cite{Khalilv5}
\begin{equation}
\label{naAvg}
\begin{aligned}
\langle(\bm{n} \cdot \bm{a}_\pm)^2\rangle &= \frac{1}{2} \left[a_\pm^2 - (\LNhat \cdot\bm{a}_\pm)^2\right], \\
\langle(\bm{n} \cdot \bm{a}_+) (\bm{n} \cdot \bm{a}_-)\rangle &= \frac{1}{2} \left[\bm{a}_+\cdot \bm{a}_- - (\LNhat \cdot\bm{a}_+) (\LNhat \cdot\bm{a}_-)\right].
\end{aligned}
\end{equation}

\section{Post-adiabatic dynamics}\label{app:PA}

Since the EOB evolution equations in the \seobfivephm~model are of the same form as the aligned-spin ones in \seobfivehm~, we can apply
the iterative PA approach which was pioneered in  Ref.~\cite{Nagar:2018gnk} and used in
subsequent \texttt{TEOBResumS}~\cite{Nagar:2018plt,Nagar:2018zoe,Akcay:2020qrj,Gamba:2021ydi}~\footnote{For example, the latest
precessing-spin model \teob~\cite{Gamba:2021ydi} uses the PA approximation.} and also the \texttt{SEOBNRv4\_PA} model~\cite{Mihaylov:2021bpf}.

The crucial difference with the non-precessing case is the evolution of the spins, which enter the Hamiltonian and the flux at different points in the radial grid.
Following the procedure outlined in Ref.~\cite{Nagar:2018gnk}, we obtain the following
explicit equations for the corrections to the momenta:
\begin{align}
  p_{r_*}  &= \frac{\xi}{2\left(1+B_{np}^\text{pprec}\right)}\left[\mathcal{F}_\phi{\left(\frac{dp_{\phi}}{dr}\right)}^{-1}\frac{2H_{EOB}H_{\rm even}}{MA^\text{pprec}}-\xi\frac{\partial Q^\text{pprec}}{\partial p_{r_*}} \right], \\
  K_{0}p_{\phi}^{2}&+2H_{\rm even}\frac{\partial \bar{H}_{\rm odd}}{\partial r} p_{\phi}  + K_{1}\nonumber\\
  &\quad+\frac{2H_{\rm even}H_{EOB}}{M\xi}\left(\frac{dp_{r_*}}{dr}\frac{dr}{dt} - \frac{p_{r_*}}{p_{\phi}}\mathcal{F}_\phi\right) = 0,
  \label{eq:PApr}
\end{align}\\
where we split the effective Hamiltonian from Eq.~\eqref{HeffAnzSimp} into odd and even-in-spin parts, $H_\text{eff}^\text{pprec} \equiv H_\text{odd}+H_\text{even}$,
defined $\bar{H}_{\rm odd} \equiv H_{\rm odd}/p_{\phi}$, while the factors $K_0$ and $K_1$ are defined as,
\begin{widetext}
\begin{align}
\begin{aligned}
  K_{0} &\equiv \frac{dA}{dr}\left(\frac{B_p^\text{pprec}}{r^2}+\frac{(\bm{l}\cdot \bm{a}_{+})^2}{r^2} B_{npa}^\text{Kerr\,eq}\right)+A^\text{pprec}\left(-\frac{2}{r^3}\left(B_p^\text{pprec}+B_{npa}^\text{Kerr\,eq} (\bm{l}\cdot \bm{a}_{+})^2\right)+\frac{(\bm{l}\cdot \bm{a}_{+})^2}{r^2}\frac{d B_{npa}^\text{Kerr\,eq}}{dr}+ \frac{1}{r^2}\frac{d{B_p^\text{pprec}}}{dr}\right), \\
  K_{1} &\equiv \frac{dA}{dr}\left(\mu^2+\frac{p_{r_*}^{2}}{\xi^{2}}\left(1+B_{np}^\text{pprec}\right)+Q^\text{pprec}\right)
  + A\left(\frac{p_{r_*}^{2}}{\xi^{2}}\left[\frac{d B_{np}^\text{pprec}}{dr}-\frac{2}{\xi}\frac{d\xi}{dr}\left(1+B_{np}^\text{pprec}\right)\right] + \frac{\partial Q^\text{pprec}}{\partial r}\right),
\end{aligned}
\end{align}
\end{widetext}
where the different potentials are defined in Sec. \ref{app:Ham}. At each point, the radial Eqs.~\eqref{eq:PApr} are solved analytically for $p_\phi$ and $p_r$. In the \seobfivephm~model, we iteratively find the solution up to 8th post-adiabatic order.

\section{Comparison against the precessing-spin time-domain phenomenological model}\label{app:PhenomModels}

\begin{figure*}[!]
	\includegraphics[width=\columnwidth]{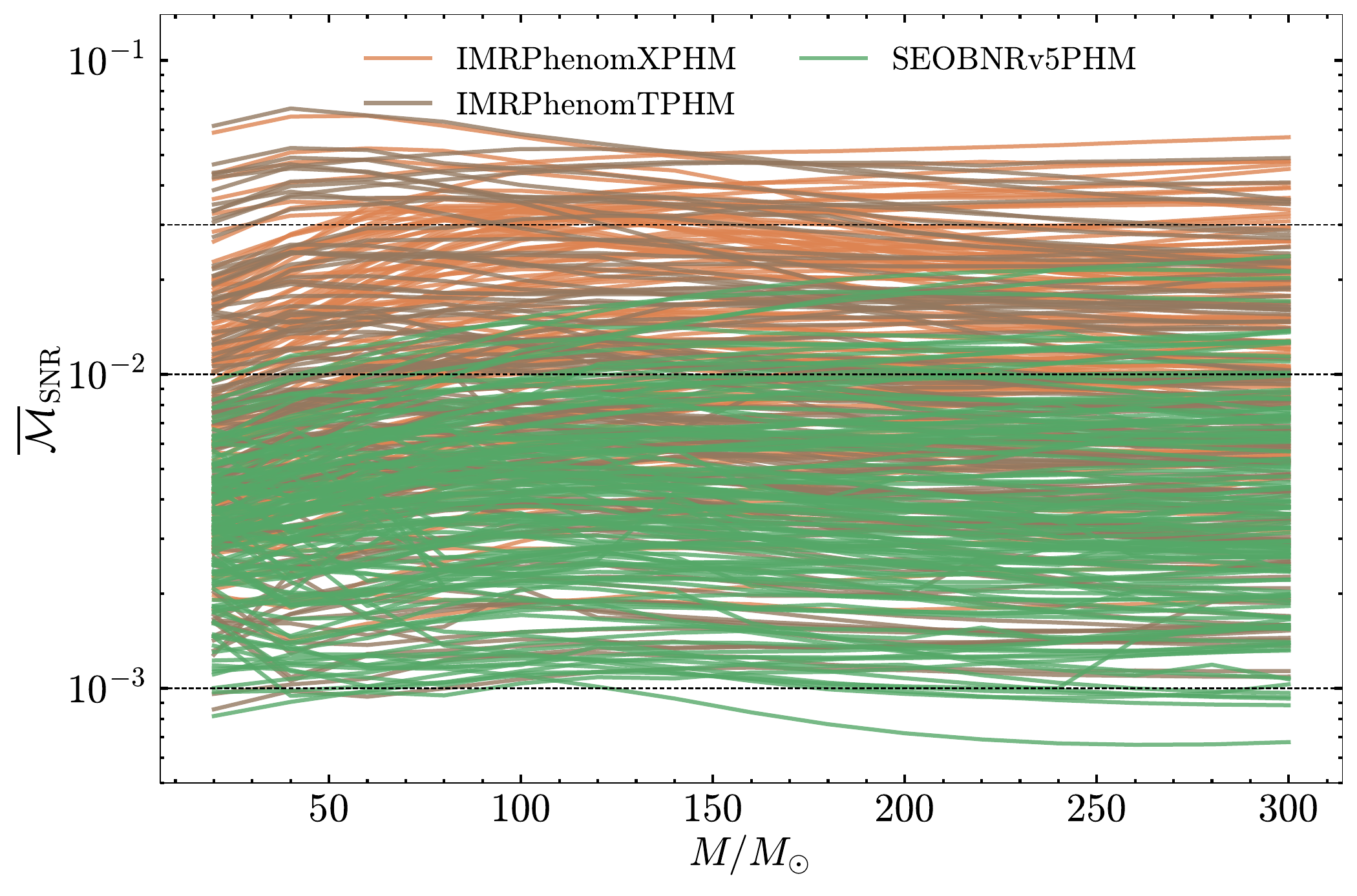}
	\includegraphics[width=\columnwidth]{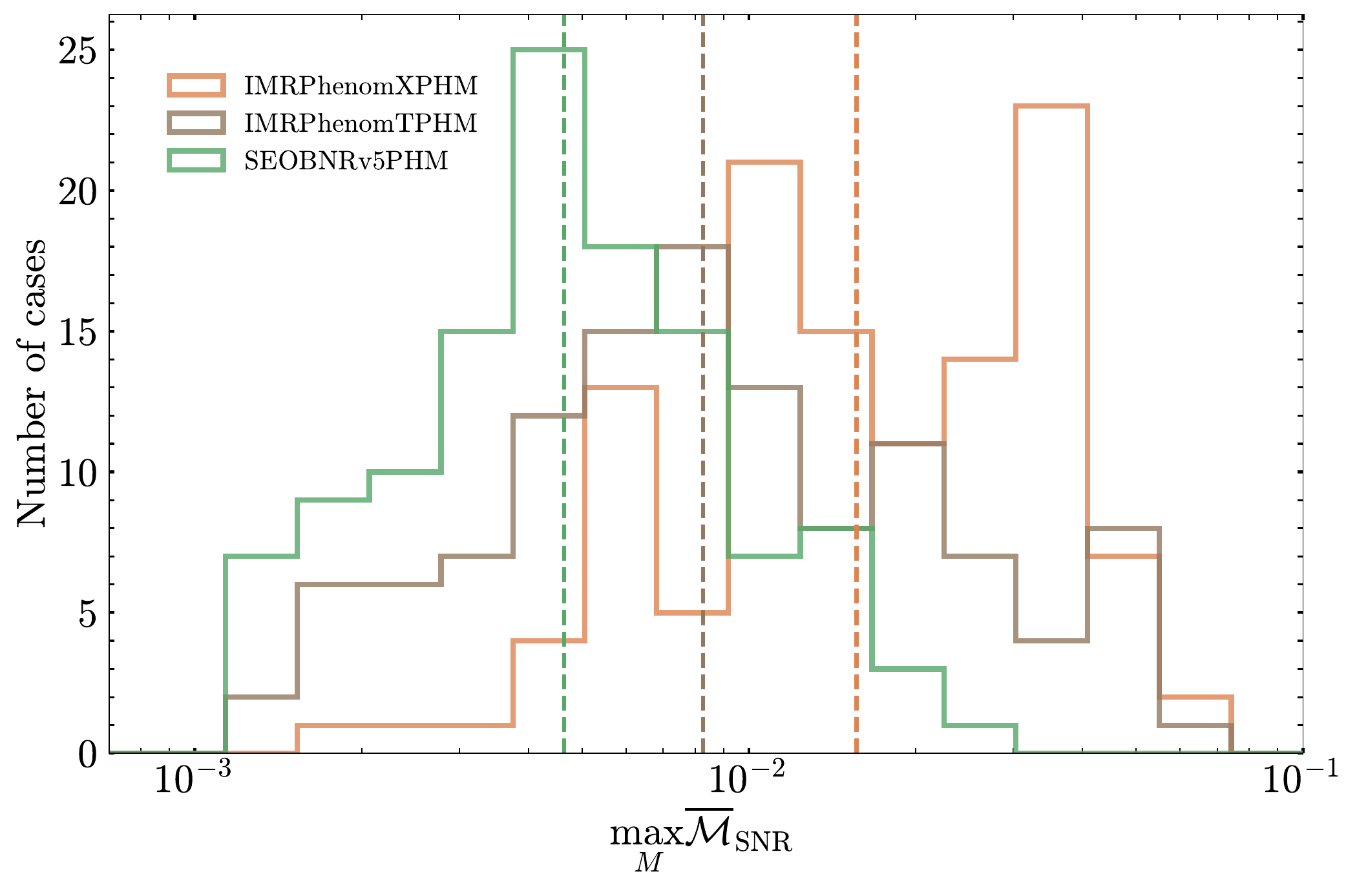}
	\caption{\textit{Left panel:} Sky-averaged SNR weighted unfaithfulness as a function of the total mass of the system $[20,300] M_\odot$, of \xphm~(orange), \tphm~(brown) and \seobfivephm~(green), against the 118 highly precessing simulations from Ref.~\cite{Ossokine:2020kjp}. \textit{Right panel:} Distribution of the maximum unfaithfulness over the total mass range for each NR simulation considered in the left plot. The vertical dashed lines indicate the median values of the distribution.}
	 \label{fig:spaghetti118_tphm}
\end{figure*}

\begin{figure*}[!]
	\includegraphics[width=\textwidth]{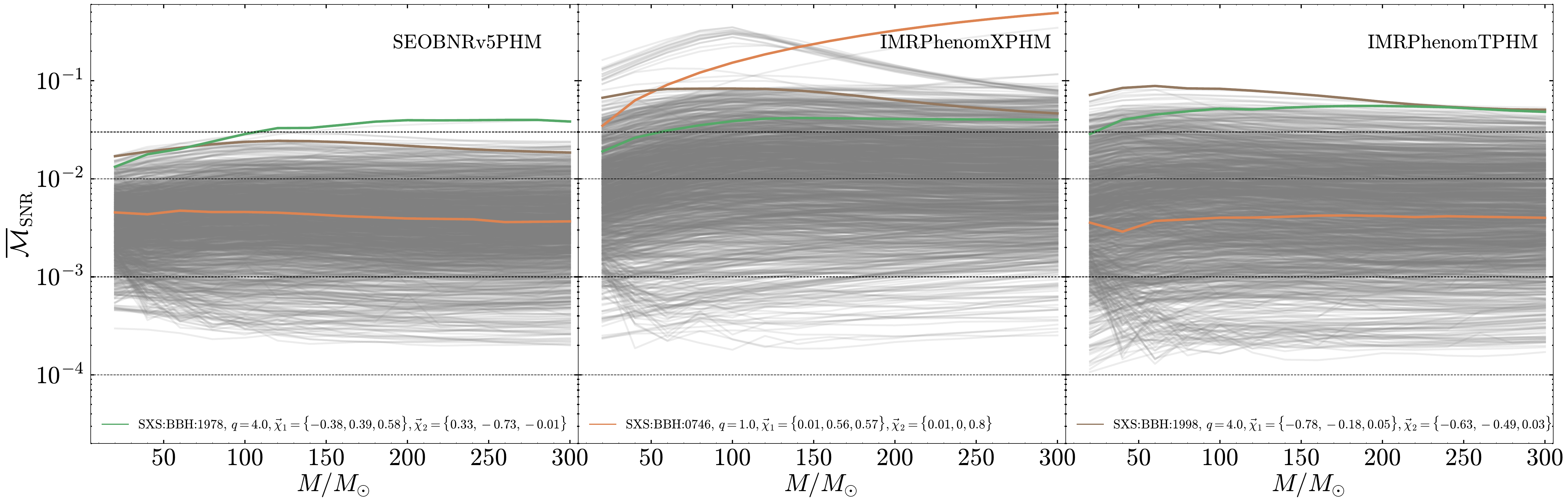}
	\caption{Sky-averaged SNR-weighted unfaithfulness as a function of the total mass of the system $[20,300] M_\odot$, of the \seobfivephm~model (left panel), and the state-of-the-art phenomenological models, \xphm~(mid panel) and \tphm~(right panel),  against 1543 precessing-spin SXS NR waveforms.}
	 \label{fig:spaghetti_tphm}
\end{figure*}

In this Appendix we contrast the accuracy against NR of the
\seobfivephm~model (and for context also the frequency-domain \xphm~model
\cite{Pratten:2020ceb} shown in Sec. \ref{sec:NRquasicircular})
with the one of the time-domain \tphm~model
\cite{Estelles:2020osj,Estelles:2020twz,Estelles:2021gvs}. We repeat
the calculation of the unfaithfulness against the catalog of NR
simulations described in Sec. \ref{sec:NRquasicircular}, both against
the set of 118 highly precessing simulations from
Ref.~\cite{Ossokine:2020kjp} (Fig. \ref{fig:spaghetti118_tphm}), and
for the full set (including the 118 highly precessing simulations) of
1543 precessing-spin SXS NR simulations
(Fig. \ref{fig:spaghetti_tphm}).

\begin{figure*}[!]
	\includegraphics[width=\linewidth]{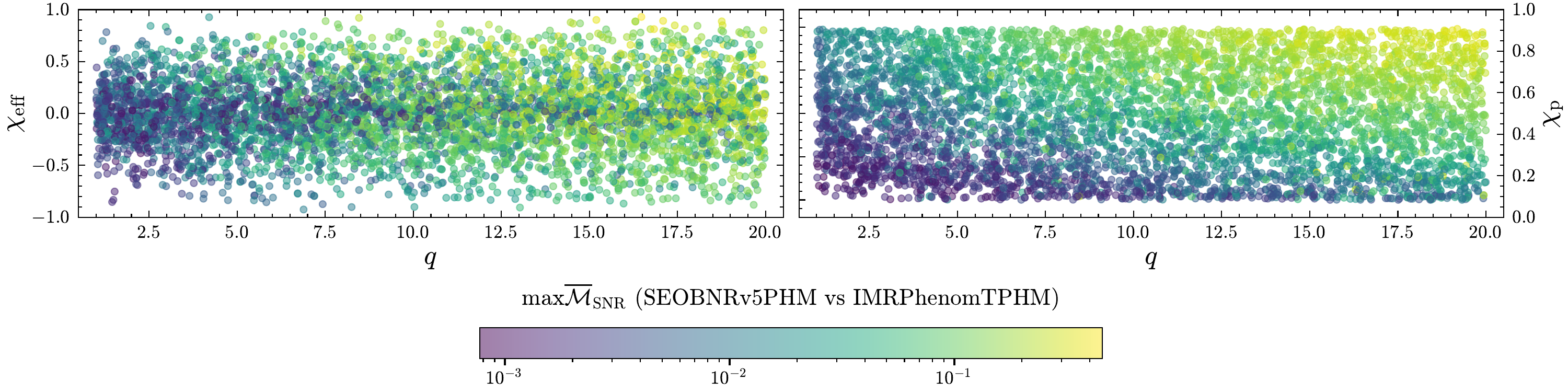}
	\caption{Maximum sky-and-polarization-averaged unfaithfulness weighted by the SNR over the total mass range $[20-300]M_\odot$ between \seobfivephm~and \tphm~for 5000 random configurations  with inclination $\iota_s = \pi/3$. The unfaithfulness grows  with increasing mass ratio and spin magnitude values, and it can reach very large values for mass ratios $q\sim 20$ and $\chi_\mathrm{p} \sim 1$.}
	 \label{fig:unf_models_tphm}
\end{figure*}

Considering the unfaithfulness against the 118 highly precessing NR
waveforms from Ref.~\cite{Ossokine:2020kjp}, we find in
Fig. \ref{fig:spaghetti118_tphm} a similar trend as in
Fig. \ref{fig:histogramMax_public118}. The \tphm~model performs better
than the \xphm~model, due to an improved description of the
precessing-spin dynamics during the inspiral and merger-ringdown,
however the lack of modeling effects due to the in-plane spin
components in the waveform causes \tphm~to still have a significant
number of cases with a maximum unfaithfulness above $3\%$ with respect
to the \seobfivephm~model. In particular, we find that the \tphm~model
has $89 \%$ $(58.5 \%)$ of cases with a maximum unfaithfulness below
$3 \%$ $(1\%)$. These numbers reduce to $72.9\%$ $(24.6\%)$ for the
\xphm~model, and they increase to $100\%$ $(85.6\%)$ for the
\seobfivephm~model. Therefore, when considering highly precessing-spin
configurations the \seobfivephm~model provides the lowest
unfaithfulness, followed closely by the time-domain phenomenological
\tphm~model, which offers an improved description of spin-precession
with respect to the frequency-domain \xphm~model.

In Fig. \ref{fig:spaghetti_tphm} we turn to a comparison against a
broader set of 1543 precessing-spin NR simulations. The \tphm~model
reaches lower values of unfaithfulness than the \seobfivephm~model for
several configurations with low precessing-spin effects, which can be
explained due to a slightly more accurate modeling of the higher order
modes in the merger-ringdown in the aligned-spin limit (see Appendix G
of Ref.~\cite{Pompiliv5} for details), but it also presents a
significantly larger number of highly precessing configurations with
unfaithfulness larger than $3\%$ with respect to the
\seobfivephm~model. Overall, the unfaithfulness of the \tphm~model is
lower than the one of the \xphm~model. More quantitatively, we find
that for \tphm, $91.6\%$ $(62.4\%)$ of cases have a maximum
unfaithfulness in the total mass range considered below $3\%$
$(1\%)$. These numbers reduce to $78.3\%$ $(38.3\%)$ for \xphm, and
increase to $99.8\%$ $(84.4\%)$ for \seobfivephm. Therefore, we find
that the \seobfivephm~model outperforms in accuracy the
phenomenological models for highly precessing-spin configurations,
while for low precessing configurations the accuracy of the models
becomes more comparable, as they rely on the accuracy of the
underlying non-precessing waveform models, which are calibrated to a
similar set of non-precessing NR waveforms.

Finally, we repeat the study of Sec. \ref{sec:ModelsComparison} and
compute the unfaithfulness between the \seobfivephm~model as the
template waveform and \tphm~as the signal, for 5000 configurations
uniformly distributed in mass ratio $q\in[1,20]$ and effective spin
parameter $\chi_p\in[0,0.99]$. Figure \ref{fig:unf_models_tphm} shows
the unfaithfulness as a function of mass ratio $q$, effective spin
parameter $\chi_\mathrm{eff}$), and effective precessing-spin
parameter $\chi_{\mathrm{p}}$. We find that for mass ratios $q<5$,
there are $99.74 \%$ ($64.5 \%$) of cases with a maximum
unfaithfulness, in total mass range $[20,300]M_\odot$, below $10 \%$
$(1\%)$, while in Sec. \ref{sec:ModelsComparison} we found that for
\xphm~these numbers decrease to $96.84 \%$ ($41.3 \%)$. The
unfaithfulness increases significantly with mass ratio and spins, with
the highest unfaithfulness values at the largest mass ratios $q \sim
20$, and effective spin precessing parameter $\chi_{\mathrm{p}} \sim
0.99$. In particular, when considering $q\leq 20$ we find that for
\tphm~there are $73.84\%$ $(30.02\%)$ cases with maximum
unfaithfulness, in the total mass range considered, below $10\%$
$(1\%)$, while these numbers decrease to $59.19\%$ $(13.45\%)$ for
\xphm~as shown in Sec. \ref{sec:ModelsComparison}. The results show
that the agreement of \seobfivephm~with the time-domain model \tphm~is
better than in the case of the frequency-domain phenomenological model
\xphm, due to the fact that the precessing-spin dynamics in \tphm~is
more accurately described than in \xphm. The existing large
differences in unfaithfulness in some regions of the parameter
space remark the necessity to populate this
region with NR waveforms in order to reduce the
systematics between models.

\pagebreak
\bibliographystyle{apsrev4-2}

\bibliography{references}

\end{document}